\documentstyle[12pt,epsf,graphics]{article}
\def\be{\begin{equation}}
\def\ee{\end{equation}}
\def\beqna{\begin{eqnarray}}
\def\eeqna{\end{eqnarray}}
\def\bear{\begin{eqnarray}}
\def\eear{\end{eqnarray}}
\def\bearst{\begin{eqnarray*}}
\def\eearst{\end{eqnarray*}}

\begin{document}

\begin{center}
{\large\bf Shortcuts in  Cosmological Branes}
\end{center}
\vspace{1ex}
\centerline{\large Elcio Abdalla, Adenauer G. Casali and Bertha Cuadros-Melgar}
\begin{center}
{Instituto de Fisica, Universidade de S\~ao Paulo\\
C.P.66.318, CEP 05315-970, S\~ao Paulo, Brazil.}
\end{center}
\vspace{6ex}

\begin{abstract}

We aim at gathering information from gravitational interaction in the
Universe, at energies where quantum gravity is required. In such a
setup a dynamical membrane world in a space-time with
scalar bulk matter described by domain walls, as well as
a dynamical membrane world  in empty Anti de Sitter space-time are
analysed.  

We later investigate the
possibility of having shortcuts for gravitons leaving the membrane and
returning subsequently. In comparison with  photons following a
geodesic inside the brane, we verify that shortcuts exist. For late
time universes they are small, but for some primordial universes they
can be quite effective. 

In the case of matter branes, we argue that at
times just before nucleosynthesis the effect is sufficiently large to
provide corrections to the inflationary scenario, especially as
concerning the horizon problem and the Cosmological Background Radiation.

\end{abstract}

\section{Introduction}
Although the Standard Model
of particle physics has been established as the uncontested theory of all
interactions down to distances of $10^{-17}$m, there are good reasons to
believe that there is a new physics arising soon at the experimental
level \cite{veneziano}. On the other hand, string theory provides an excelent 
background to solve long standing problems of theoretical high energy
physics. It is by now a widespread idea
that M-theory \cite{polchinski} can be a reasonable description
of our Universe. In the field theory limit, it is described by a solution 
of the (eventually 11-dimensional) Einstein equations with a cosmological 
constant by means of a four dimensional membrane. In this picture only 
gravity survives in the extra dimensions, while the remaining matter and 
gauge interactions are typically four dimensional.

This avenue presents us a possibility of tackling with two different
problems at the same time, namely a means of testing the up to now
far from experimental and observational data string theory, and
a theoretical framework to cosmology, whose theoretical background needs a
full understanding of quantum gravity in order to correctly deal with the
puzzling question of the initial singularity.

General Relativity always evaded quantization. Einstein gravity is a 
nonrenormalizable theory at a low loop level \cite{thooft}, and atempts to
include supersymmetry in order to cancel divergencies only postponed the
problem to a higher loop \cite{sugra}. String Theory showed a way from
its very basis of how to succesfully quantize gravity. At the same time,
and especially after the discovery of the anomaly cancellation
mechanism \cite{greenschwarz}, now even sometimes quoted as the first
superstring revolution, strings are taken as quasi unique in their
description. Moreover, with the discovery of the duality symmetry such
a uniqueness has been further enhanced \cite{duality}, leading to the
concept of a universal theory, the so called M-theory largely unknown,
describing an eleven dimensional misterious, master theory, mother of all
string descriptions on their side related by duality and containing in
their field theory limit a version of 11-dimensional supergravity
\cite{mtheory}. 

The key question however lies on what can turn a higher dimensional theory
into something realistic, describing our four dimensional world. Some
hints are known since almost a century and have been used before in
the context of supergravity. It is the Kaluza Klein approach
\cite{kaluzaklein}, where extra dimensions are compactified being
curled up in such a size that they cannot be observed in
the daily life, that in spite of being appealing, has encountered
several problems being still an active field of inquire. Still there are
proposals where some of these dimensions are large enough to be probed by
microphysics \cite{arkanihamed}.

Recently, Horava and Witten \cite{horavawitten} considered M-theory in a
rather intriguing situation. It has been argued that our universe can be
seen as a solution of a higher dimensional theory. Strings can be open or
closed. The closed string sector contains gravity and further
components. The open string sector contains matter fields isolated into
their extremities. In this proposed picture the open string extremities
containing the matter fields are restrained to live in the physical
universe, which in Horava and Witten's proposal is
described by a membrane.

Later on, Randall and Sundrum \cite{RS} described the universe as a
solution of
higher dimensional field equations with boundary conditions describing the
membrane. As a result, they obtained a space with full fledged extra
dimensions but with a warp such that the effective penetration of
gravitons into the extra dimensions is small, and the effective
gravitational interaction is observationally four dimensional. 
In this picture there is a
possibility that  gravitational fields, while propagating out of the brane,
speed up reaching farther distances in smaller time as compared to
light propagating inside the brane, a scenario that for a resident of the
brane implies shortcuts \cite{Abdalla2}---\cite{Chung}.

It is our aim in this work to further develop these ideas in the case of a
FRW brane Universe \cite{radion}. The subject was developed until now from
the point of view of the brane \cite{Abdalla1}, where all 
time dependence is embedded  in the bulk metric written in gaussian
coordinates. The price paid is the complicated form of the bulk metric and,
consequently, the complicated behaviour of geodesics in the
bulk. However, if we treat the problem from the  point of view of the
bulk, where the brane evolves in a non-trivial way in a static AdS 
background, we can construct explicitly the causal structure of null
geodesics leaving and subsequently returning to the brane.  As it turns out,
shortcuts are common, although harmless at the present days (the delay is
vanishingly small), but could be large in the era before
nucleosynthesis.

Moreover, one of the main goals of string theory nowadays is to 
prove itself able of coping with experimental evidences.  Branes have
been shown to be useful tools to understand the physics of strings and
M-theory \cite{generalbranes}. As it has 
also recently been pointed out \cite{transdimen}, brane Universes, such as 
the one described above, could imply the existence of relics of the
extra dimensions in the cosmic microwave background. Unfortunately, recent
developments with  inflation guided by a scalar field on the
brane indicate that the consistency equation is
preserved \cite{inflation}.  In this
work however, we show that if inflation took part on the brane, the
causal structure is definitely changed by those gravitational
shortcuts, possibly leading to a non-usual period of causal evolution of
scales. This could be responsible for distinct  predictions in the
cosmic microwave background structure for inflationary models.

This scenario has been proposed as an actually realizable
possibility \cite{stojkovic}---\cite{moffat}. In \cite{abdacasali} it
has been shown that in some scenarios shortcuts are very difficult to
be detected today because of 
the extremely short delay of the photon as compared to the graviton coming
from the same source.

\section{Further Motivations and Brane Cosmology}

The Planck mass is the natural scale where string effects become
important. However, it is not possible  to achieve such an energy level in
particle physics accelerators. Nonetheless, as we mentioned above,
cosmology may provide an alternative laboratory for string theory. From the
eighties several authors tried to analysed the kind of cosmology arising
from string inspired models, which are essentially general relativity in
higher dimensions together with scalar and tensor fields. In case we
also introduce the brane concept, a consistent picture of the brane
universe is achieved, and we can describe the evolution of the universe by
means of solutions of the Einstein field equations in higher dimensions
with a four dimensional membrane.

Extensive use of specific properties of strings, such as the T-duality,
may provide simple explanations of several properties of the universe,
though of course an observational check is missing. Nonetheless, a
pre big-bang scenario can be described providing a very
sophisticated physical picture of the universe. Several solutions also
indicating inflation have been obtained in what is now usually
called brane cosmology.

It is thus essential to provide means of comparing those results to some
observational data, which is the largest difficulty in string theory. The
best proposal is to use the extremely sophisticated data of the Cosmic
Microwave Background, first obtained from the pioneering COBE satelite
and more recently by the detailed analysis of the WMAP. Such observations
are opening the avenue for precision cosmology, and results with error
bars less a percent are now available. In such a case one can compare
the predictions of inflationary models with observations, and even
quantitatively distinguish between different inflationary scenarios with
a chance of favouring or discarding string inspired models of cosmology
or brane cosmology.

The definition of a scalar potential directly from string theory has been
an important issue. In \cite{dvalitye} an attempt was made in order to
derive the scalar interaction from a brane-brane interaction. This kind of
approach reviewed in \cite{quevedo} has been very recently taken over in
\cite{kalloshmaldalindeetc}. A few different and new forms for the
inflaton potential have been derived, showing that at least one can hope
to derive inflationary models directly from the properties which
caracterize string theory and its consequences, namely M-theory and branes.

\section{The junction conditions}
%





The formulation of proper junction conditions at surfaces of
discontinuity is a fundamental problem in gravitational theory.
Well-known examples are the Schwarzschild and Oppenheimer-Snyder
problems, which require the junction of the interior field of a static
or collapsing star to the exterior vacuum field. 

In newtonian theory the problem is directly solved by imposing the
standard continuity and jump conditions connecting the potential and
its first derivatives across the surface. In general relativity,
however, the gravitational potential is not only determined by the
smoothness of physical conditions but also by the smoothness of the
coordinates we are using to describe spacetime.

A pioneering work on these subjects was made in 1922 by C. Lanczos
\cite{lanczos}, on which W. Israel \cite{israel} based his landmark
formalism forty years later. Independent of these works, G. Darmois
\cite{darmois} derived junction conditions for the special case of a
boundary surface, i.e. a surface through which both the metric and the
extrinsic curvature tensor are continuous. All these ideas are now
known as the Darmois-Israel junction/thin-shell formalism, which has
found wide application in general relativity and cosmology, including
further studies of gravitational collapse, the evolution of bubbles
and domain walls in a cosmological setting, wormholes and more
recently in brane cosmology.

In this section we review the junction formalism as it was first
derived by Israel. We also derive the Darmois-Israel conditions 
from the gravitational action in the context of $D$-dimensional
theories and give an alternative approach using distribution theory.

\subsection{The Formalism}

Consider a spacetime $M$ (pseudo-riemannian manifold with signature
$(-+++)$) with metric $g_{\alpha\beta}(x^\gamma)$ in the coordinate
system $(x^\gamma)$. The absolute derivative of a smooth vector
function ${\bf A}$ defined on a curve on this spacetime parametrized by $t$
is given by
\begin{equation}
\nabla_t A^\alpha \equiv {{\partial A^\alpha }\over {\partial t}} + A^\lambda
\Gamma^\alpha _{\lambda\mu} {{dx^\mu}\over {dt}} \, .
\end{equation}

Let $\Sigma$ be a smooth hypersurface in $M$ with metric
$g_{ij}(\xi^c)$ in the coordinates $\xi^c$, which separates $M$ into
two four-dimensional manifolds $M^-$ and $M^+$, each containing
$\Sigma$ as part of its boundary.

The unit 4-normal ${\bf n}$ to this hypersurface in $M$ labels $\Sigma$ as
timelike (spacelike) for $\epsilon ({\bf n}) \equiv {\bf n}\, . \,{\bf
n} = -1 \, (1)$.

We define a natural frame of three linearly independent tangent
vectors $e^\alpha _{(i)}$ associated with the intrinsic coordinates
$\xi^i$ as
\begin{equation}
e^\alpha _{(i)} = {{\partial x^\alpha}\over{\partial\xi^i}} \, ,
\end{equation}
which gives the induced metric on $\Sigma$ as
\begin{equation}
g_{ij} \equiv {\bf e}_{(i)}\, . \,{\bf e}_{(j)} = {{\partial
x^\alpha}\over{\partial\xi^i}} {{\partial
x^\beta}\over{\partial\xi^j}}  g_{\alpha\beta} \, . 
\end{equation}

The intrinsic covariant derivative of ${\bf A}$ with respect to
$\xi^i$ is the projection of the vector $\partial {\bf A}/\partial
\xi^j$ onto $\Sigma$,
\begin{equation}
A_{i;j} = {\bf e}_{(i)}\, . \,{{\partial {\bf A}}\over {\partial \xi^j}} =
{{\partial A}\over {\partial \xi^j}} - A_h \Gamma^h _{ij} \, .
\end{equation}
We see that intrinsic covariant differentiation does not depend on the
nature of the embedding. Properties of a non-intrinsic character enter
when we consider the way in which $\Sigma$ ``bends'' in $M$. This is
measured by the variations $\partial {\bf n}/\partial\xi^i$ of the
unit normal,
\begin{equation}
\partial {\bf n}/\partial\xi^i = K_i ^j {\bf e}_{(j)} \, ,
\end{equation}
what defines the {\it extrinsic curvature 3-tensor} $K_{ij}$ of
$\Sigma$,
\begin{equation}
K_{ij} = e_{(j)}\, . \,{{\partial {\bf n}}\over {\partial\xi^i}} \equiv
{{\partial x^\alpha}\over {\partial\xi^i}} {{\partial x^\beta}\over
{\partial\xi^j}} \nabla_\alpha n_\beta = -n_\gamma \left( {{\partial
^2 x^\gamma}\over {\partial\xi^i \partial\xi^j}} + \Gamma^\gamma
_{\alpha\beta} {{\partial x^\alpha}\over {\partial\xi^i}} {{\partial
x^\beta}\over {\partial\xi^j}} \right) \, .
\end{equation}

Let $K_{ij} ^-$, $K_{ij} ^+$ be the extrinsic curvatures of $\Sigma$
associated with its embeddings in $M^-$, $M^+$. If $K_{ij} ^- \not=
K_{ij} ^+$, $\Sigma$ is called a singular hypersurface of order one,
{\it surface layer} or {\it thin shell}. If $K_{ij} ^- =
K_{ij} ^+$, $\Sigma$ is called a hypersurface of higher order or {\it
boundary surface}.

The Darmois conditions for the joining of $M^-$ and $M^+$ through
$\Sigma$ are \cite{darmois}
\begin{eqnarray}
\lbrack g_{ij}\rbrack &=& 0 \, , \label{darmois1}\\
\lbrack K_{ij}\rbrack &=& 0 \, , \label{darmois2}
\end{eqnarray}
where $[X]= X^+ - X^-$ is the jump of $X$ through the hypersurface
$\Sigma$.

A boundary surface satisfies both equations, while a thin shell only
satisfies (\ref{darmois1}). Condition (\ref{darmois2}) as it stands is
ambiguous since the orientation of the 4-vector field ${\bf n}$ has
not been specified. The Israel formalism will require the normals in
$M$ to point from $M^-$ to $M^+$. We should stress that the majority
of the existing literature deals with spherical symmetry where the
direction of the normal is clear, but in more complicated cases great
care must be taken.

The Israel formulation \cite{israel} of thin shells follows from the
Lanczos equation \cite{lanczos}
\begin{equation}
[K_{ij}] - g_{ij}[K_a ^a] = -8\pi G S_{ij} \, ,
\end{equation}
or equivalently
\begin{equation}\label{lanc}
[K_{ij}] = -8\pi G (S_{ij}-{1\over 2}g_{ij} S) \, ,
\end{equation}
where $S_{ij}$ is the surface stress-energy tensor of $\Sigma$. 

By regarding a thin shell as the limit of a layer of uniform finite
thickness $\epsilon$ as $\epsilon \rightarrow 0$, we can give a
heuristic justification for the name {\it surface stress-energy
tensor}. 

Let $\Sigma^-$, $\Sigma^+$ be the two boundary surfaces
separating the finite layer from the regions $M^-$, $M^+$. In gaussian
coordinates the equations of $\Sigma^-$ and $\Sigma^+$ are $x^1=0$,
$x^1=\epsilon$. 

In terms of the extrinsic curvature tensor and in the gaussian system
of coordinates the Ricci tensor can be written as
\begin{equation}
^4 R_{ij} = {{\partial K_{ij}}\over {\partial x^1}} + Z_{ij} \, ,
\end{equation}
where
\begin{equation}
Z_{ij} = {^3R_{ij}} -KK_{ij} + 2 K_i ^p K_{pj} \, .
\end{equation}
Integration of the Einstein field equations
\begin{equation}
R_{\alpha\beta} = -8\pi G (T_{\alpha\beta} - {1 \over 2}
g_{\alpha\beta} T) \, ,
\end{equation}
through the layer gives
\begin{equation}
-8\pi G \int ^\epsilon _0 (T_{ij} - {1 \over 2} g_{ij} T) dx^1 =
 [K_{ij}] + \int ^\epsilon _0 Z_{ij} dx^1 \, .
\end{equation}
In the limit $\epsilon \rightarrow 0$ for fixed $K_{ij} ^-$, $K_{ij}
^+$, $K_{ij}$ remains bounded inside the layer. Hence, the integral of
$Z_{ij}$ tends to zero. Comparing this result with (\ref{lanc}) we see
that
\begin{equation}
S_{ij} = \lim_{\epsilon \rightarrow 0}\int^\epsilon _0 T_{ij} dx^1 \, .
\end{equation}
Thus, $S_{ij}$ is the integral of Einstein's energy tensor through the
thickness of the layer.

\subsection{Matching Conditions from the Action}

We will consider now the problem of junction conditions in the context
of $D$-dimensional theories \cite{chre}. In order to do this we will
introduce the concept of {\it domain wall}. In a $D$-dimensional
spacetime a domain wall can be defined as an extended object with
$D-2$ spatial dimensions, which divides the spacetime in different
domains. Here we will use this term to make reference to any $D-2$
brane moving in $D$ dimensions. 

Let $M$ be a $D$-dimensional manifold containing a domain wall
$\Sigma$, which splits $M$ into two parts $M^-$ and $M^+$. The metric
must be continuous everywhere, while its derivatives must be
continuous everywhere except on $\Sigma$. We will denote $\Sigma^\pm$
as being the two sides of $\Sigma$.

Varying the Einstein-Hilbert action in $M^\pm$ gives \footnote{For
  convenience we use units such that $8\pi G=1$}
\begin{equation}\label{eh-var}
\delta S_{EH} = - {1 \over 2} \int_{\Sigma^\pm} d^{D-1}x \sqrt{-h}
\, g^{MN} n^P (\nabla_M \delta g_{NP} - \nabla_P \delta g_{MN} ) \, ,
\end{equation}
where $n_M$ is the unit normal pointing into $M^\pm$ and the induced
metric on $\Sigma^\pm$ is given by the tangential components of the
projection tensor $h_{MN} = g_{MN} -n_M n_N$. If we contract the
quantity in parenthesis with $n^M n^N n^P$, it vanishes, so we can
replace $g^{MN}$ by $h^{MN}$.

The expression (\ref{eh-var}) contains a normal derivative of the
metric variation, which is discontinuous across $\Sigma$ according to
our initial hypothesis, therefore, the contributions from $M^\pm$ will not
necessarily cancel out. In this way, it is necessary to add a
Gibbons-Hawking boundary term \cite{gibbonshawking} on each side of the domain
wall to cancel out this term,
\begin{equation}
S_{GH} = - \int_{\Sigma^\pm} d^{D-1} x \sqrt{-h}\, K \, ,
\end{equation}
with $K$ being the trace of the extrinsic curvature of $\Sigma$,
{\it i.e.} $K=h^{MN} K_{MN}$ where $K_{MN} = h^P _M h^Q _N \nabla_P
n_Q$.

The variation of this new term is
\begin{equation}
\delta S_{GH} = -\int_{\Sigma^\pm} d^{D-1} x \sqrt{-h}\, (\delta K +
{1 \over 2} K h^{MN} \delta g_{MN}) \, ,
\end{equation}
where
\begin{equation}
\delta K = -K^{MN} \delta g_{MN} - h^{MN} n^P (\nabla_M \delta g_{NP}
- {1\over 2} \nabla_P \delta g_{MN}) + {1\over 2} K n^P n^Q \delta
g_{PQ} \, .
\end{equation}
Thus, the total variation is 
\begin{eqnarray}\label{totvar}
\delta S_{EH} + \delta S_{GH} &=& \int_{\Sigma^\pm} d^{D-1} x
\sqrt{-h}\, \left[ {1\over 2} h^{MN} n^P \nabla_M \delta g_{NP} +
  K^{MN} \delta g_{MN} - \right. \nonumber \\
&&\left. - {1\over 2} K n^M n^N \delta g_{MN} - {1\over 2} K h^{MN} \delta
g_{MN} \right] \, .
\end{eqnarray} 
Notice that the first term in the rhs of (\ref{totvar}) can be written as
\begin{equation}
h^{MN} n^P \nabla_M \delta g_{NP} = \bar\nabla_M (h^{MN}n^P \delta
g_{NP}) + K n^M n^N \delta g_{MN} - K^{MN} \delta g_{MN} \, ,
\end{equation}
where $\bar\nabla$ is the covariant derivative associated to
$h$. We substitute this result into (\ref{totvar}) to
get
\begin{equation}
\delta S_{EH} + \delta S_{GH} = {1\over 2} \int_{\Sigma^\pm} d^{D-1} x
\sqrt{-h}\, (K^{MN} -Kh^{MN}) \delta g_{MN} \, .
\end{equation}

The domain wall has an action given by
\begin{equation}
S_{dw} = \int_\Sigma d^{D-1} x \sqrt{-h}\, L_{dw} \, ,
\end{equation}
whose variation is
\begin{equation}
\delta S_{dw} = \int_\Sigma d^{D-1} x \sqrt{-h}\, t^{MN} \delta g_{MN}
\, ,
\end{equation}
where $t^{MN} \equiv {2\over{\sqrt{-h}}} {{\delta S_{dw}}\over{\delta
    h_{MN}}}$ is tangential to the domain wall such that we can
    replace $\delta h_{MN}$ by $\delta g_{MN}$.

The variation of the total action $S=S_{EH} + S_{GH} +
S_{dw}$ gives the Darmois-Israel conditions
\begin{equation}
[K_{MN} - K h_{MN}] = -t_{MN} \, ,
\label{junccond25}
\end{equation}
where the brackets stand for the jump of $K$ through the domain wall
$\Sigma$.

\subsection{Junction Conditions from Distribution Theory}

The Darmois-Israel conditions describe the motion of the domain wall
through the bulk. However, sometimes we describe a
static brane. In this case the Darmois-Israel conditions reduce to
some relations between the energy density and pressure on the brane
and the coefficients of the bulk metric. An alternative derivation of
these relations can be done using distribution theory as follows.

Let us consider as an example a 5-dimensional bulk metric of the form
\cite{bdl1,bdl2} 
\begin{equation}\label{metric}
ds^2 _{(5)} = -n^2(t,y) dt^2 + a^2(t,y) \gamma_{ij} dx^k dx^j +
b^2(t,y)dy^2 \, ,
\end{equation}
where $\gamma_{ij}$ represents a maximally symmetric metric on the
3-brane located in $y=0$ with $k=-1,0,1$ parametrizing the spatial
curvature. 

The stress-energy tensor appearing in the Einstein's equations $G_{AB}
=\kappa^2 _{(5)} {\cal T}_{AB}$ can be written as
\begin{equation}
{\cal T}_{AB} = \hat T_{AB} + T_{AB} \, ,
\end{equation}
where $\hat T_{AB}$ is the stress-energy tensor of the matter on the
bulk (and possibly other branes) which we do not need to specify here,
and $T_{AB}$ corresponds to the matter content on the brane which can
be expressed quite generally as 
\begin{equation}
T^A _B = \frac {\delta(y)} {b} diag (-\rho, p,p,p, 0)\quad .
\end{equation}
The energy $\rho$ and the pressure $p$ are independent of the position
on the brane in order to recover standard cosmology on the brane.

From the metric (\ref{metric}) the Einstein's tensor components are
found to be
\begin{eqnarray}\label{einstensor}
\tilde G_{00} &=& 3 \left\{ {\dot a \over a} \left( {\dot a \over a} +
{\dot b \over b} \right) - {n^2 \over b^2} \left( {{a''}\over a} +
{{a'}\over a} \left( {{a'}\over a} - {{b'}\over b} \right) \right) +
k{n^2 \over a^2}\right\} \, , \label{G00}\\
\tilde G_{ij} &=& {a^2 \over b^2} \gamma_{ij} \left\{ {{a'}\over a}
\left( {{a'}\over a} + 2 {{n'}\over n} \right) - {{b'}\over b} \left(
{{n'}\over n} + 2 {{a'}\over a} \right) + 2 {{a''}\over a} +
{{n''}\over n} \right\} + \nonumber \\
&& + {a^2 \over n^2} \gamma_{ij} \left\{ {\dot a \over a} \left( -
{\dot a \over a} + 2 {\dot n \over n} \right) -2 {\ddot a \over a} +
{\dot b \over b} \left( -2 {\dot a \over a} + {\dot n \over n} \right)
- {\ddot b \over b} \right\} -k\gamma_{ij} \, , \label{Gij}\\
\tilde G_{05} &=& 3 \left( {{n'} \over n} {\dot a \over a} + {{a'}\over
a} {\dot b \over b} - {{\dot a'} \over a} \right) \, , \\
\tilde G_{55} &=& 3  \left\{ {{a'}\over a} \left( {{a'}\over a} +
{{n'}\over n} \right) - {b^2 \over n^2} \left( {\dot a \over a} \left(
{\dot a \over a} - {\dot n \over n} \right) + {\ddot a \over a}
\right) - k{b^2 \over a^2} \right\} \, ,
\end{eqnarray}
where a prime stands for derivative with respect to $y$ and the dot
means derivative with respect to $t$. 

In order to have a well defined geometry the metric must be
continuous across the brane. However, its derivatives with respect to
$y$ can be discontinuous in $y=0$. Thus, there is a Dirac delta
function in the second derivatives of the metric with respect to
$y$. In general we will have
\begin{equation}
a''= \hat a'' + [a'] \delta (y) \, ,
\end{equation}
where $\hat a''$ stands for the non-distributional part of the second
derivative of $a$ (the ordinary second derivative), and $[a']$ is the
jump of the first derivative across $y=0$ defined by
\begin{equation}
[a'] = a'(0^+) - a'(0^-) \, .
\end{equation}
The resulting terms with a delta function appearing in the
Einstein tensor must match the distributional part of the
stress-energy tensor in order to satisfy the Einstein's equations.
Comparing the Dirac delta functions in the components (\ref{G00}) and
(\ref{Gij}) of the Einstein's tensor we obtain the relations
\begin{eqnarray}\label{israel-5-static}
{{[a']}\over {a_0 b_0}} &=& - {{\kappa_{(5)} ^2} \over 3} \rho \, ,
\nonumber \\ \\ 
{{[n']} \over{n_0 b_0}} &=& {{\kappa_{(5)} ^2} \over 3} (3 p+ 2\rho)
\, ,  \nonumber 
\end{eqnarray}
where the subscript $0$ means that the metric coefficients take their
values on the brane.

\section{The Scenarios}

We shall consider two kinds of scenarios. First, a brane wall model
based on the exact solution of Einstein Equations and boundary conditions,
where the brane is a so-called domain wall embedded in a space-time
containing a singularity (a type of black hole) and a cosmological
constant. Subsequently we consider a membrane of the
Friedmann-Robertson-Walker type.  

\subsection{Brane Wall Model}

In the first case we have a scenario described by the gravitational
action in a D-dimensional bulk with a scalar field (the bulk dilaton), 
a domain wall potential and a Gibbons-Hawking term \cite{gibbonshawking},
\be\label{action}
S = \int_{bulk} d^D x \sqrt{-g} \left( {1 \over 2} R - {1 \over 2} (\partial
  \phi )^2 -V(\phi) \right) - \int_{dw} d^{D-1} x \sqrt{-h} ([K] +
  \hat V (\phi) ) \, ,
\ee
where $\phi$ is the bulk dilaton, $K$ is the extrinsic curvature, $V(\phi)$ 
and $\hat V (\phi)$ are bulk and domain wall potentials respectively,
and $g$ and $h$ denote the bulk and domain wall metrics. 
The potentials are here:
\begin{eqnarray}
V(\phi) &=& V_0 e^{\beta \phi} \, , \label{vbulk} \\
\hat V (\phi) &=& \hat V_0 e ^{\alpha \phi} \, . \label{vbrane}
\end{eqnarray}

We consider the bulk metric as being static and invariant under rotation,
\begin{equation}\label{metricwall}
ds^2 = -U(r) dt^2 + U(r)^{-1} dr^2 + R(r)^2 d\Omega_k ^2 \, ,
\end{equation}
where $d\Omega_k ^2$ is the line element on a $D-2$ dimensional space
of constant curvature depending on a parameter $k$. We can also consider
the brane to have a static metric, in which case the solution of the bulk
would be more complicated. The above metric is supposed to have a 
mirror symmetry $Z_2$ with respect to the domain wall. Such a symmetry 
will be used in order to impose the Darmois-Israel conditions. 
The variation of the total action (\ref{action}) including the
Gibbons-Hawking term leads directly to 
\begin{equation}\label{israeldarmois}
K_{MN} = - {1 \over {2(D-2)}} \hat V (\phi) h_{MN} \quad .
\end{equation}
The extrinsic curvature can be computed as
\begin{equation}\label{excurv}
K_{MN} = h^P _M h^Q _N \bigtriangledown_{_P} n_Q \quad ,
\end{equation}
where the unit normal, which points into $r<r(t)$, is
\begin{equation}\label{normal1}
n_M = {1 \over \sqrt{U-{{\dot r^2}\over U}}} (\dot r \, , \, -1 \, ,
\, 0 \, ... \, , \, 0) \, .
\end{equation}
Here a dot means derivative with respect to the bulk time $t$.

The $ij$ component of (\ref{israeldarmois}) can be written as
\begin{equation}\label{ij}
{{R'}\over R} = {{\hat V(\phi)} \over {2(D-2)U}} \sqrt{U-{{\dot
r^2}\over U}} \; ,
\end{equation}
while the $00$ component is
\begin{equation}\label{00}
\left({{R'}\over R} \right) ^{-1} \left( {{R'} \over R} \right)
^\prime = {{\hat V'(\phi)} \over {\hat V(\phi)}} - {{R'}\over R} \; .
\end{equation}
Here a prime denotes derivative with respect to the extra
coordinate $r$.

The equation of motion for the dilaton obtained from the action 
(\ref{action}), together with  (\ref{00}), can be simultaneously solved
with the Ansatz (\ref{metricwall}), leading to \cite{chre}
\begin{eqnarray}
\phi(r) &=& \phi_\star - {{\alpha (D-2)} \over {\alpha^2 (D-2) + 1}}
\log r \, , \label{phi} \\
R(r) &=& (\alpha^2(D-2) +1) C \hat V_0 e^{\alpha \phi_\star}
r^{1\over {\alpha^2(D-2)+1}} \, , \label{R}
\end{eqnarray}
where $\phi_\star$ and $C$ are arbitrary integration constants. 

The motion of the domain wall is governed by the $ij$ component of the
Darmois-Israel conditions (\ref{ij}). That equation can be written in terms of
the brane proper time $\tau$ as 
\begin{equation}\label{dwmotion}
{1\over 2} \left({{dR}\over{d\tau}} \right) ^2 + F(R) =0 \; .
\end{equation}

The induced metric on the domain wall is Friedmann-Robertson-Walker 
and (\ref{dwmotion})
describes the evolution of the scale factor $R(\tau)$. This equation
is the same as that one for a particle of unit mass and zero energy
rolling in a potential $F(R)$ given by
\begin{equation}\label{potential}
F(R) = {1 \over 2} U R'^2 - {1 \over {8(D-2)^2}} \hat V^2 R^2 \; .
\end{equation}
Notice that the solution only exists when $F(R) \leq 0$.

From the induced domain wall metric we find the relations between the time
parameter on the brane ($\tau$) and in the bulk ($t$) as given by
\bearst 
dt = {\sqrt{U+\left({{dr}\over{d\tau}}\right)^2}\over U} d\tau \; ,
\eearst 
so that
\begin{equation}\label{eq3}
\dot r \equiv {{dr}\over {dt}} = {{dr}\over {d\tau}} {{d\tau}\over {dt}} =
{{dr}\over {d\tau}} {U \over \sqrt{U+ \left({{dr}\over{d\tau}
}\right)^2}} \; ,
\end{equation}
where ${{dr}\over{d\tau}} = {{dR}\over{d\tau}} \left({{dR}\over
{dr}}\right)^{-1} $ can be obtained from (\ref{dwmotion}). Equation
(\ref{eq3}) describes the motion of a domain wall in the static
background as seen by an observer in the bulk.

Consider two points on the brane. In general, there are more than
one null geodesic connecting them in the D-dimensional spacetime. The
trajectories of photons must be on the brane and those of gravitons
may be outside. We consider the shortest path for both photons and gravitons.
For the latter, the geodesic equation is the same as the one considered in 
\cite{Abdalla2}, since the bulk metric is static:
\begin{equation}\label{geo}
\ddot r_g + \left( {1 \over r_g} - {3 \over 2} {{U'}\over U}\right) \dot
r_g^2 + {1\over 2} U \, U' - {U^2 \over r_g} = 0 \; .
\end{equation}
Again a dot means derivative with respect to the bulk time $t$.

The solutions of (\ref{eq3}) and (\ref{geo}) in terms of the bulk
proper time $t$ were obtained by means of a {\bf MAPLE} program. In
the next section we 
discuss the possibility of shortcuts in the cases of the various 
solutions describing different Universes defined by the domain wall solution.

\subsection{Cosmological Brane Model}

Later, we consider a scenario where the bulk is a purely Anti-de-Sitter
space-time of the form \cite{ida},
\be
ds^{2} = h(a) dt^2 - \frac{da^{2}}{h(a)} - a^2d\Sigma^2,
\label{bulkmetric}
\ee
with $ h(a) = k + \frac{a^{2}}{l^2}$, $l\sim 0.1mm$ is the Randall-Sundrum
lenght scale \cite{RS} and $d\Sigma^2$ represents the metric of the three
dimensional spatial sections with  $k= 0\; , \;\pm 1$,
\[
d\Sigma^2 = \frac{dr^{2}}{1-kr^2} + r^{2}[d\theta^{2} +
\sin^{2}(\theta)d\phi^{2}]. 
\]

The brane is localized at $a_{b}(\tau)$, where $\tau$ is the proper time 
on the brane. The unit vector normal to the brane is defined as (overdot
and prime superscript denote differentiation with respect to  $\tau$ and
$a$ respectively) 
\be
n = \dot{a}_{b}(\tau)dt - \dot{t}(\tau)da\quad . \label{normal2}
\ee
The normalization of $n$ implies the relation between the bulk time
$t$ and the brane time $\tau$,
\be
h(a_b)\dot{t}^{2} - \dot{a_b}^{2}h^{-1}(a_b) = 1
\label{nn}
\ee
and also  the usual FRW  expression for the distance
on the brane ,
\be
ds^2 = h_{ij}dx^i dx^j = d\tau^{2} - a_b^{2}d\Sigma^{2}\quad .
\label{branemetric}
\ee

We also need the second fundamental form
$K_{i j} = e^{\alpha}_{i}e^{\nu}_{j}\nabla_{\alpha}n_{\nu}$, given by
\bear
K_{rr} &=& - \frac{a_bh}{(1-kr^{2})}\dot{t} \quad , \label{krclasseII}\\
K_{\theta\theta} &=& - a_br^{2}h\dot{t}\quad
,\label{kthetaclasseII} \\
K_{\phi\phi} &=& - a_br^{2}\sin^{2}(\theta)h\dot{t}\quad
\label{kphiclasseII}\\
K_{\tau\tau} &=&   \frac{1}{h\dot{t}}\Bigl(\ddot{a}_b + \frac{h'}{2}
\Bigr)\quad . 
\label{ktauclasseII}
\eear

We now use  the Darmois-Israel conditions
for a $Z_{2}$ symmetric configuration (\ref{junccond25}), 
\be
K_{ij} = \frac{1}{2}\kappa_{(5)}^{2}\Bigl(S_{ij} -
\frac{1}{3}h_{ij}S\Bigr)\quad , 
\ee
with an isotropic distribution of matter,
\be
S_{ij} = \epsilon_T u_{i}u_{j} - p_T(h_{ij} - u_{i}u_{j})\quad .
\ee
Thus, the following relations hold \cite{Kraus},\cite{BCG}, \cite{bdl1}
\bearst
\frac{d \epsilon_T}{d \tau} &=& - 3\frac{\dot{a}_b}{a_b}
(\epsilon_T + p_T)\quad ,
\label{energy} \\
\frac{\dot{a}_{b}^2}{a_{b}^2} + \frac{h}{a_b^{2}} &=&
\frac{\kappa_{(5)}^{4}\epsilon_T^{2}}{36}\quad .
\eearst

Following \cite{Csaki2} and \cite{Cline}, we introduce an intrinsic
 non-dynamical energy density $\epsilon_0$ defined by means of $
 \epsilon_T = \epsilon_0 + \epsilon $, $p_{T} = -\epsilon_{0} + p$ where
 $\epsilon$ and $p$ correspond to the brane matter.  Thus, the junction
 equations imply the usual energy conservation on the brane, 
\bear
\frac{d \epsilon}{d \tau} &=& - 3\frac{\dot{a}_b}{a_b}(\epsilon + p)\quad 
\label{energy2}
\eear
and the  modified Friedmann equation \cite{bdl2},
\be
H^{2} = \Bigl(\frac{\dot{a}_{b}}{a_{b}}\Bigr)^2 =
 \frac{\Lambda_4}{3} + \frac{1}{M_{Pl}^2}\Bigl(\frac{\epsilon}{3} +
 \frac{\epsilon^2}{6\epsilon_{0}}\Bigr)  -
 \frac{k}{a_{b}^{2}}\quad , \label{Friedmann}
\ee
with the hierarchy
\be
M_{PL}^{-2} = \frac{\kappa_{5}^{4}\epsilon_{0}}{6}\quad  \label{hierarquia}
\ee
and the cosmological constant on the brane,
\be
\frac{\Lambda_{4}}{3} = \Bigl(\frac{\kappa_{5}^{4}\epsilon_{0}^{2}}{36} -
\frac{1}{l^{2}}\Bigr) \quad . 
\ee

The present density of the Universe is  
\[
\epsilon(0) = \Omega_0 \epsilon_{c} = \Omega_{0} 3M_{PL}^{2}H_{0}^{2}\quad,
\]
where $\Omega_0$ is the ratio between the density and the critical density of
the Universe. Aiming at the energy conservation, the Friedmann equation is
\be
H^{2} = \frac{\Lambda_4}{3} +\Omega_{0}
H_0^2\frac{a_{b0}^{q}}{a_{b}^{q}}\Bigl(1  +
\frac{\Omega_{0}}{4(1+\Lambda_4l^2/3)}\frac{L_{c}^{q}}{a_{b}^{q}}\Bigr) -
\frac{k}{a_b^{2}}\quad, 
\label{Friedmann2}
\ee
with $ L_{c}^q = a_{b0}^{q}l^2H_0^{2}$.

We thus verify that there exist three phases in the evolution of the Universe.
When $a_{b}>>L_{c}$, the linear term in the energy density prevails in the
Friedmann equation, leading to the standard cosmology. Because
$H_0l\sim 10^{-29}$ this happens in a redshift of $a_{b0}/a_{b} \sim
10^{15}$,  much earlier than the nucleosynthesis.  For  $a_{b}<<L_{c}$
the Universe expands at a slower pace as compared to the standard model,
$a_b \propto \tau^{1/q}$, and the quadratic term is the prevailing one.
In an intermediate era where $a_{b} \sim L_{c}$, both phases coexist.

We see that all the cosmological information is already known:  the
position of the brane in the extra dimension, $a_{b}(\tau)$, is just the
scale factor of the FRW metric and  the junction conditions imply that the
cosmological evolution of the brane is obtained by the usual energy
conservation (\ref{energy2}) and the modified Friedmann equations
(\ref{Friedmann}).  However, in this work  we are particularly interested
in the evolution of the brane with respect to the bulk. This relation can
be obtained  using (\ref{nn}) and transforming
from the time of the brane to the time of the bulk. Thus, the position of
the brane can also be treated as a function of the bulk proper time
satisfying the equation 
\be
\frac{d a_{b}}{dt} = \frac{d a_{b}}{d\tau}\frac{d \tau}{dt} = \dot{a}_{b}(\tau)
\frac{h(a_{b})}{\sqrt{h(a_{b}) + \dot{a}_{b}(\tau)^2}}\quad . \label{bulk}
\label{timetime}
\ee

\section{Solutions of the Geodesic Equation for a Domain Wall}
\subsection{Type I Solutions} 
We define the type I brane solutions as those for which  $\alpha=\beta=0$.
Consequently, the potentials become cosmological constants. The solution 
also has a constant dilaton $\phi=\phi_0$. A simple rescaling in the 
metric leads us to 
\begin{equation}\label{metric1}
ds^2 = -U(R)dt^2 + U(R)^{-1} dR^2 +R^2 d\Omega_k ^2 \; ,
\end{equation}
with
\begin{equation}\label{u1}
U(R) = k -2MR^{-(D-3)} - {{2V_0} \over {(D-1)(D-2)}} R^2 \; ,
\end{equation}
which corresponds to  a topological black hole solution in D dimensions 
with a cosmological constant.

As discussed in \cite{chre}, if the domain wall has positive energy 
density ($\hat V_0 >0$), the relevant part of the bulk spacetime is 
$R<R(\tau)$, which is the region containing the singularity. If it
has negative energy density, the relevant part is
$R>R(\tau)$, which is non-singular unless the wall reaches $R=0$.

\begin{figure*}[htb!]
\begin{center}
\leavevmode
\begin{eqnarray}
\epsfxsize= 5.5truecm\rotatebox{-90}
{\epsfbox{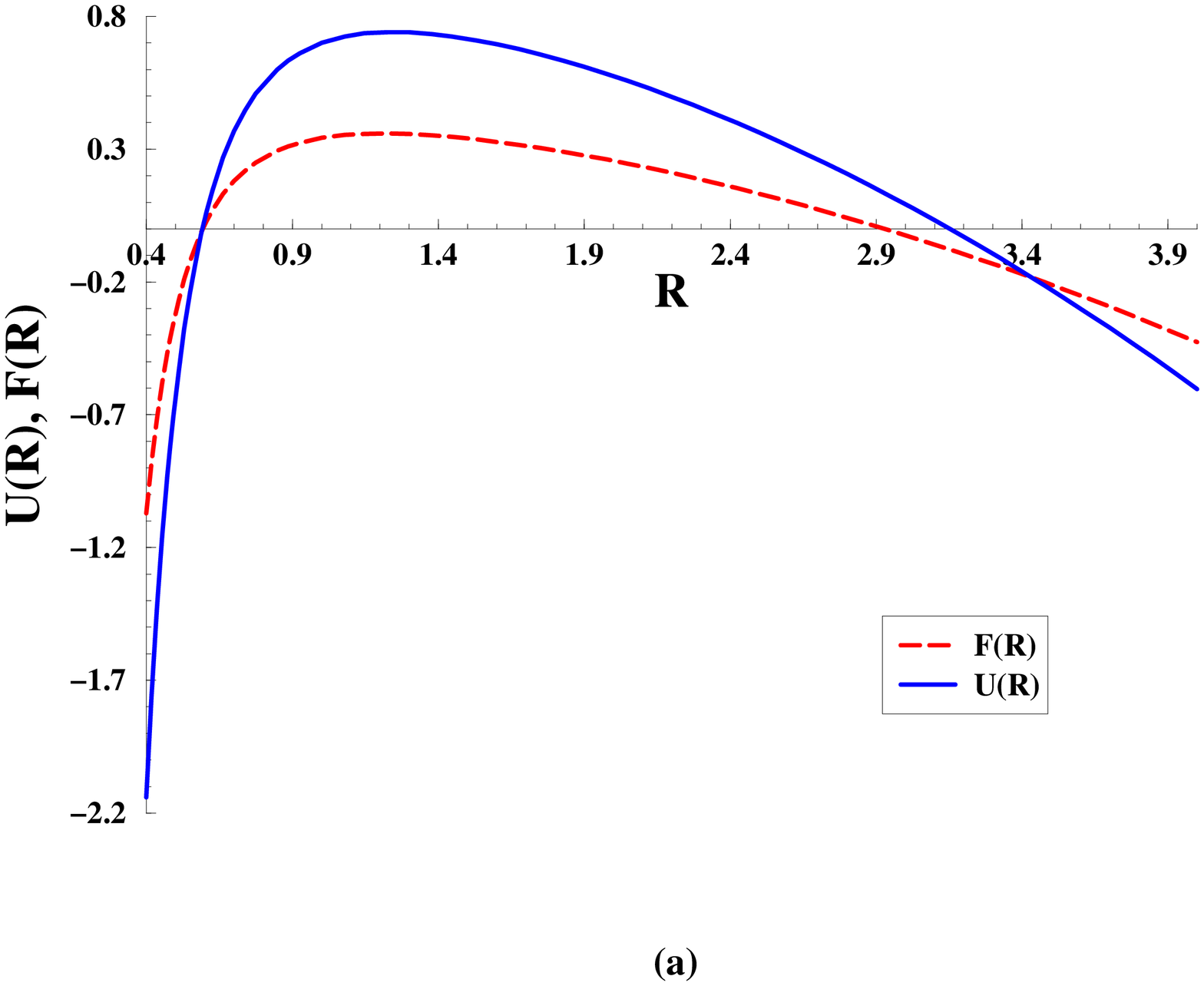}}\nonumber
\epsfxsize= 5.5truecm\rotatebox{-90}
{\epsfbox{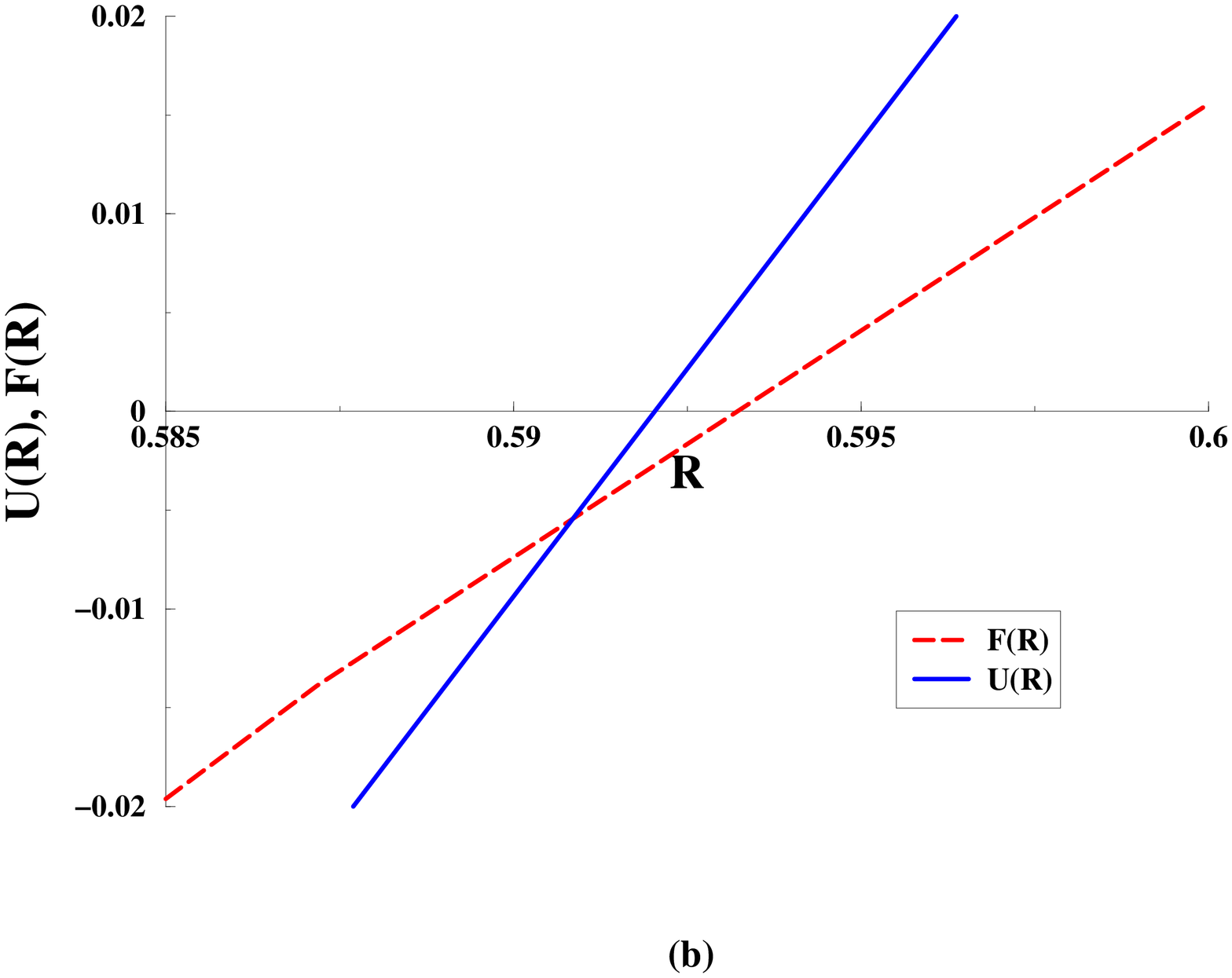}}\nonumber
\end{eqnarray}
\caption{(a) $U(R)$  and $F(R)$  for type I solutions with
$M=1/10$, $V_0=1$ and $\hat V_0=\pm 1$, (b) Zoom of the event horizon region.} 
\label{T1-U}
\end{center}
\end{figure*}
\begin{figure*}[htb!]
\begin{center}
\leavevmode
\begin{eqnarray}
\epsfxsize= 5.5truecm\rotatebox{-90}
{\epsfbox{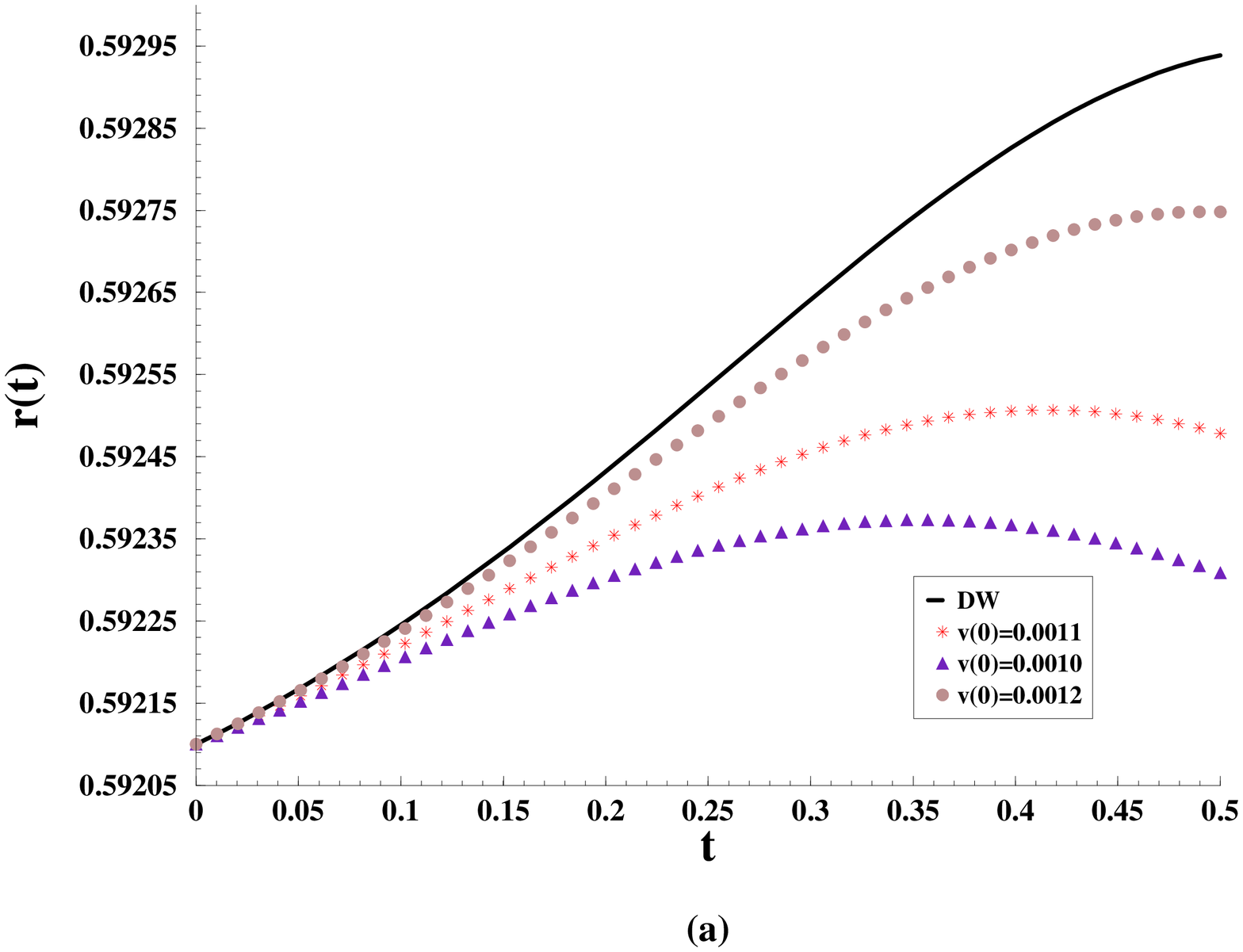}}\nonumber
\epsfxsize= 5.5truecm\rotatebox{-90}
{\epsfbox{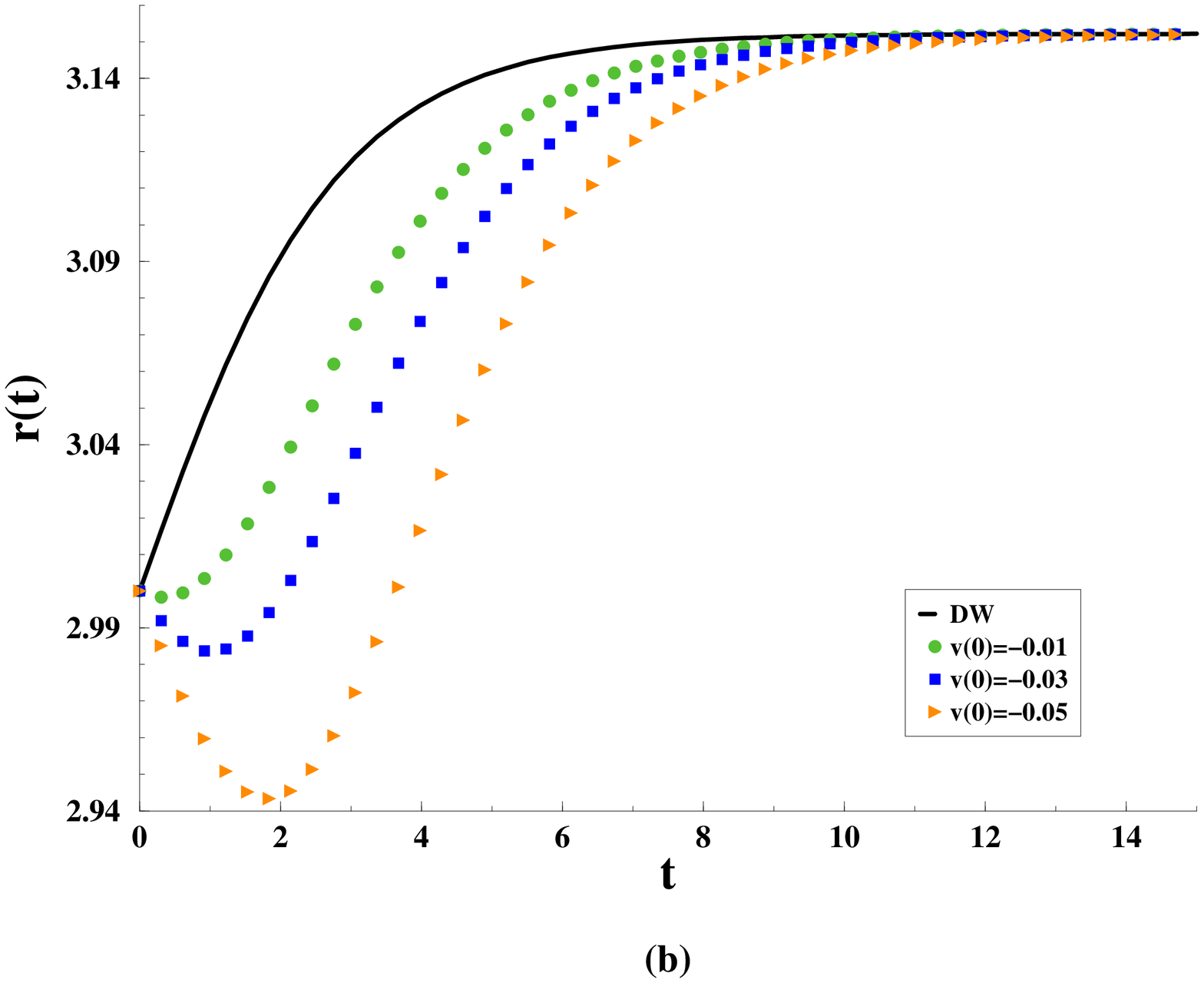}}\nonumber
\end{eqnarray}
\caption{Domain wall motion  and geodesics for type I solutions with
$M=1/10$, $V_0=1$ and $\hat V_0=1$ in (a) region I and, (b) region II.}
\label{T1-braneb}
\end{center}
\end{figure*}
\begin{figure*}[htb!]
\begin{center}
\leavevmode
\begin{eqnarray}
\epsfxsize= 5.5truecm\rotatebox{-90}
{\epsfbox{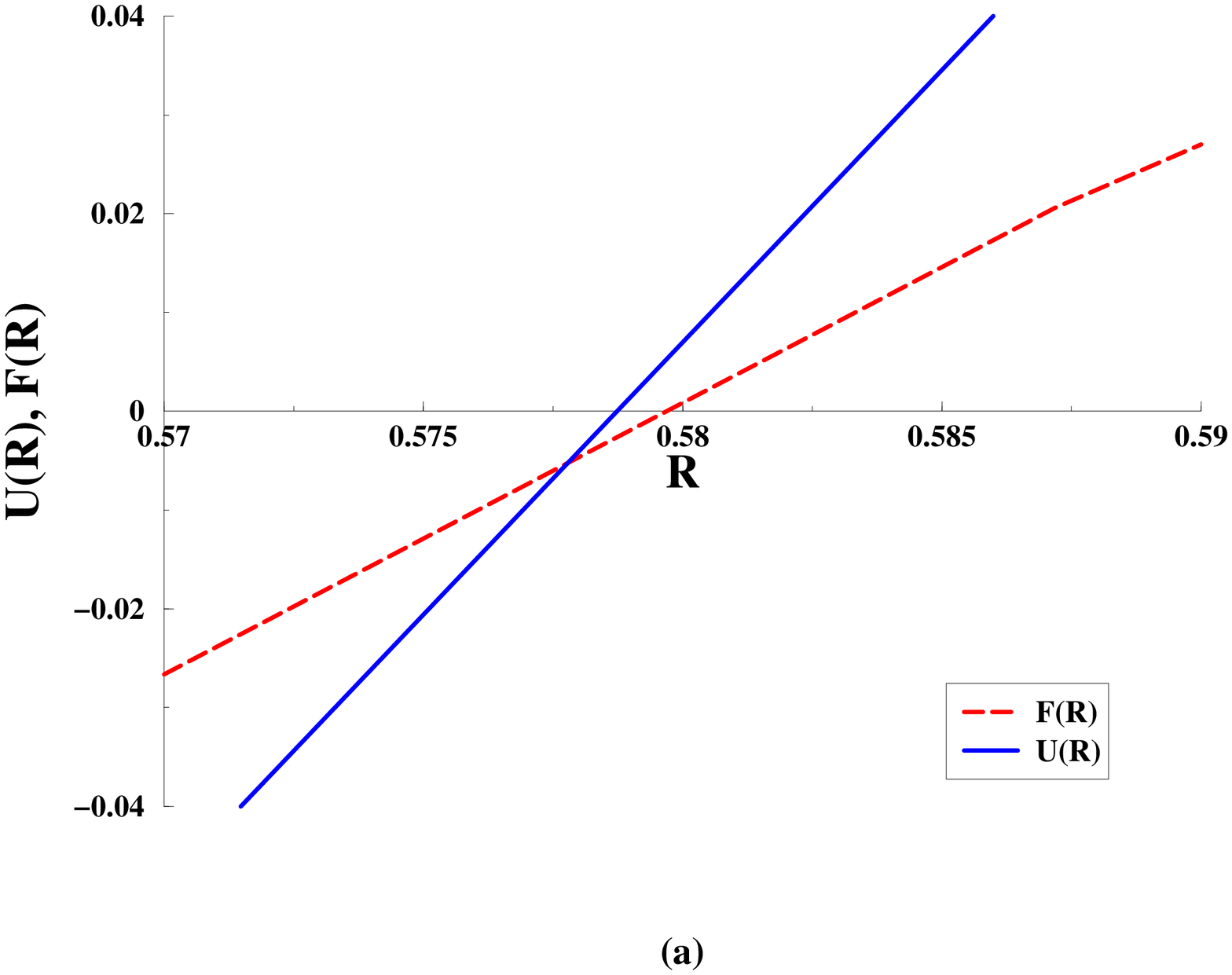}}\nonumber
\epsfxsize= 5.5truecm\rotatebox{-90}
{\epsfbox{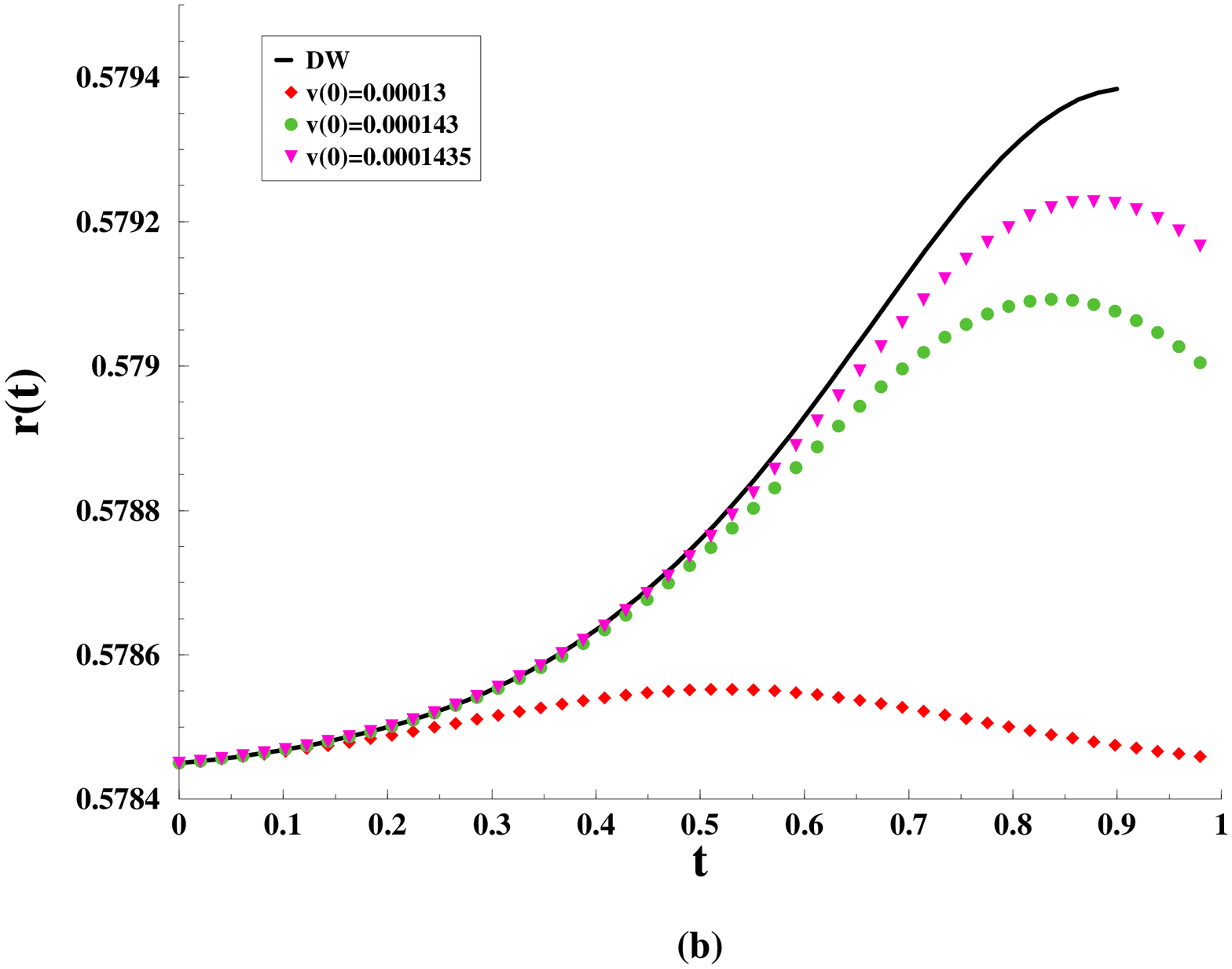}}\nonumber
\end{eqnarray}
\caption{
(a) Zoom of the region where (\ref{condition1}) holds from the graph of
$U(R)$ and $F(R)$ with $\hat \Lambda <0$ and $M>0$ for type I 
solutions, (b) Domain wall motion  and geodesics for type I solutions
with $M=1/10$, $V_0=-1$ and $\hat V_0=1$.}
\label{T1-U3}
\end{center}
\end{figure*}
\begin{figure*}[htb!]
\begin{center}
\leavevmode
\begin{eqnarray}
\epsfxsize= 5.5truecm\rotatebox{-90}
{\epsfbox{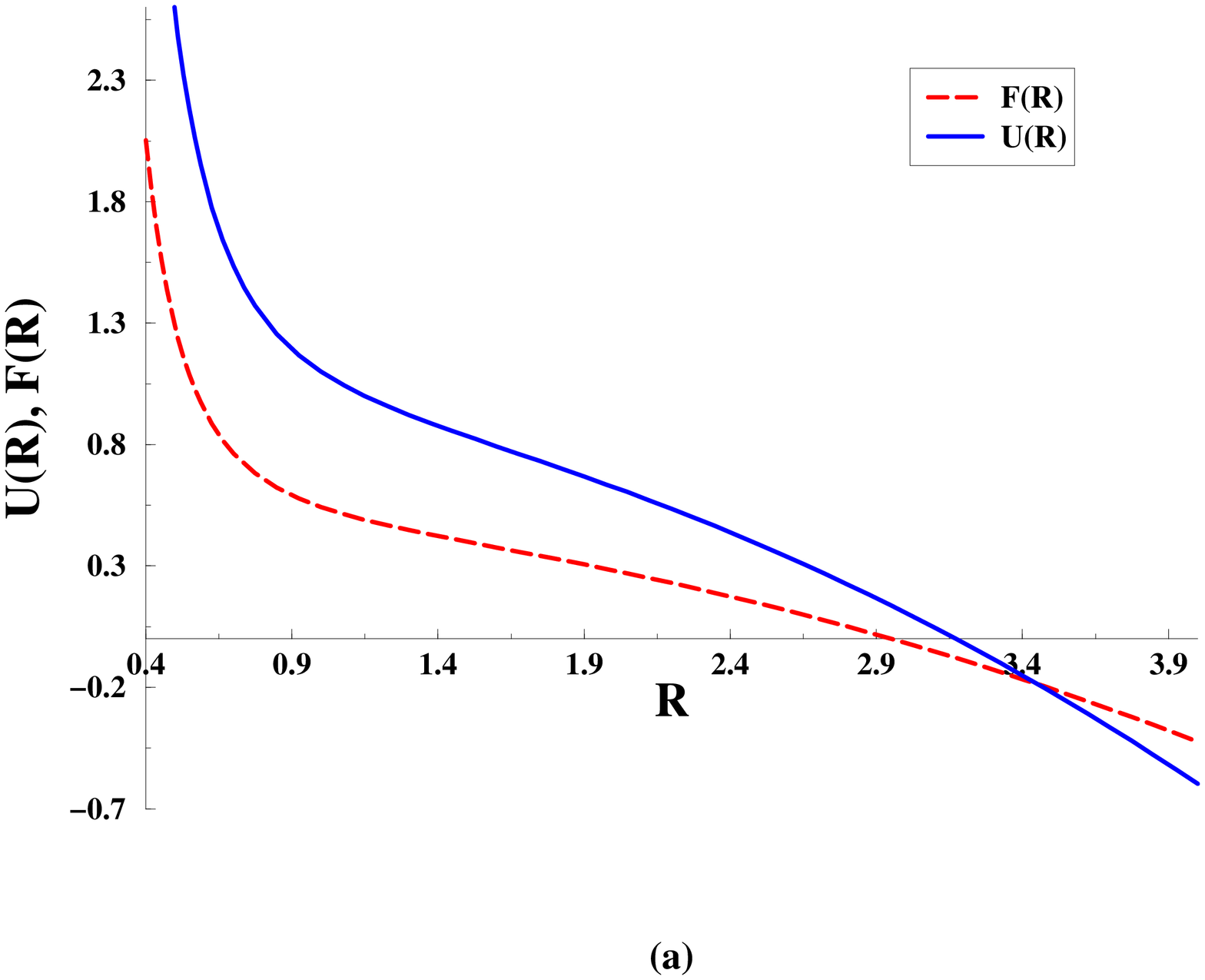}}\nonumber
\epsfxsize= 5.5truecm\rotatebox{-90}
{\epsfbox{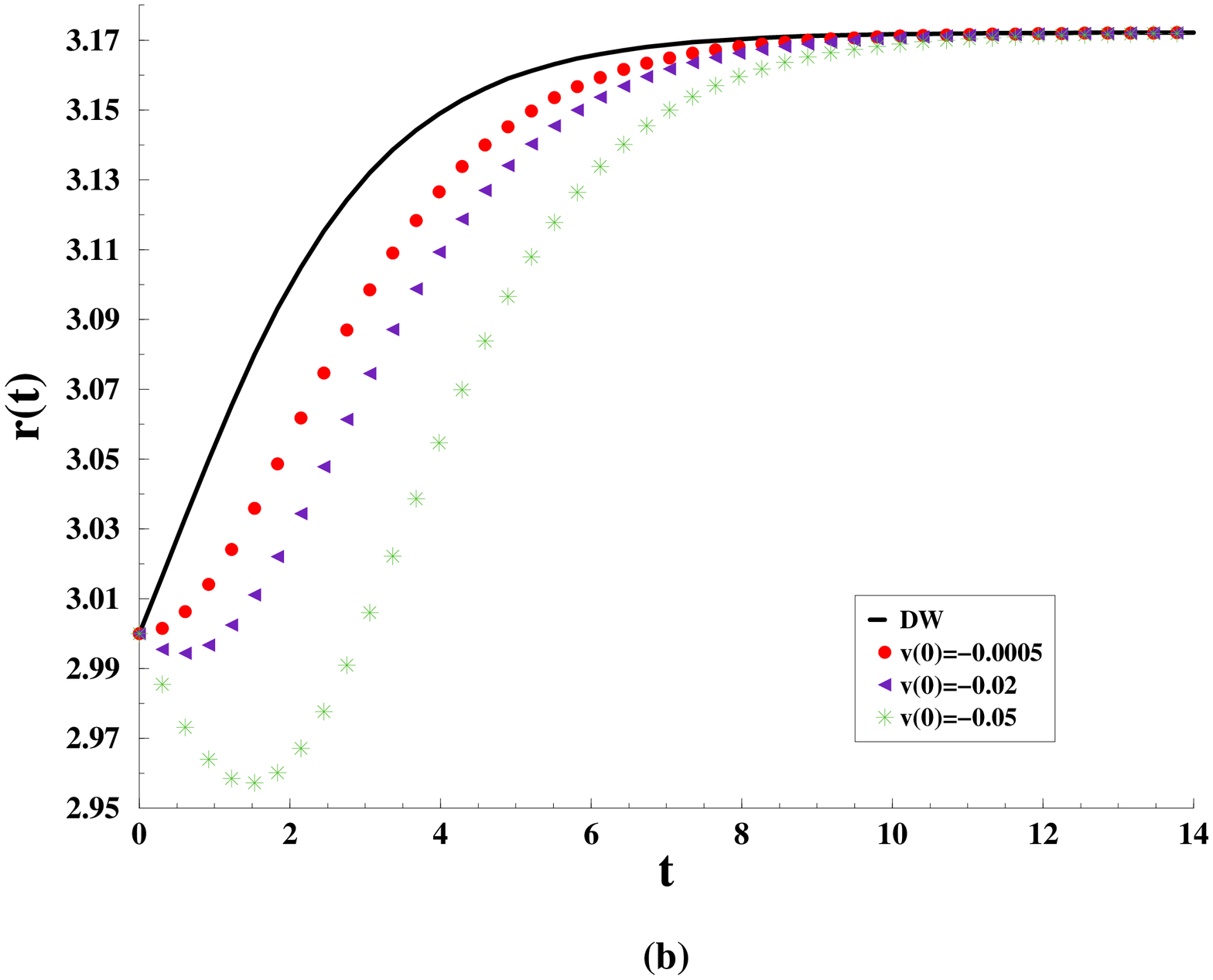}}\nonumber
\end{eqnarray}
\caption{(a) $U(R)$  and $F(R)$  with $\hat \Lambda >0$
and $M<0$ for type I 
solutions, (b) Domain wall motion and geodesics for $M=-1/10$, $V_0=1$
and $\hat V_0=1$.}
\label{T1-U2}
\end{center}
\end{figure*}
\begin{figure*}[htb!]
\begin{center}
\leavevmode
\begin{eqnarray}
\epsfxsize= 5.5truecm\rotatebox{-90}
{\epsfbox{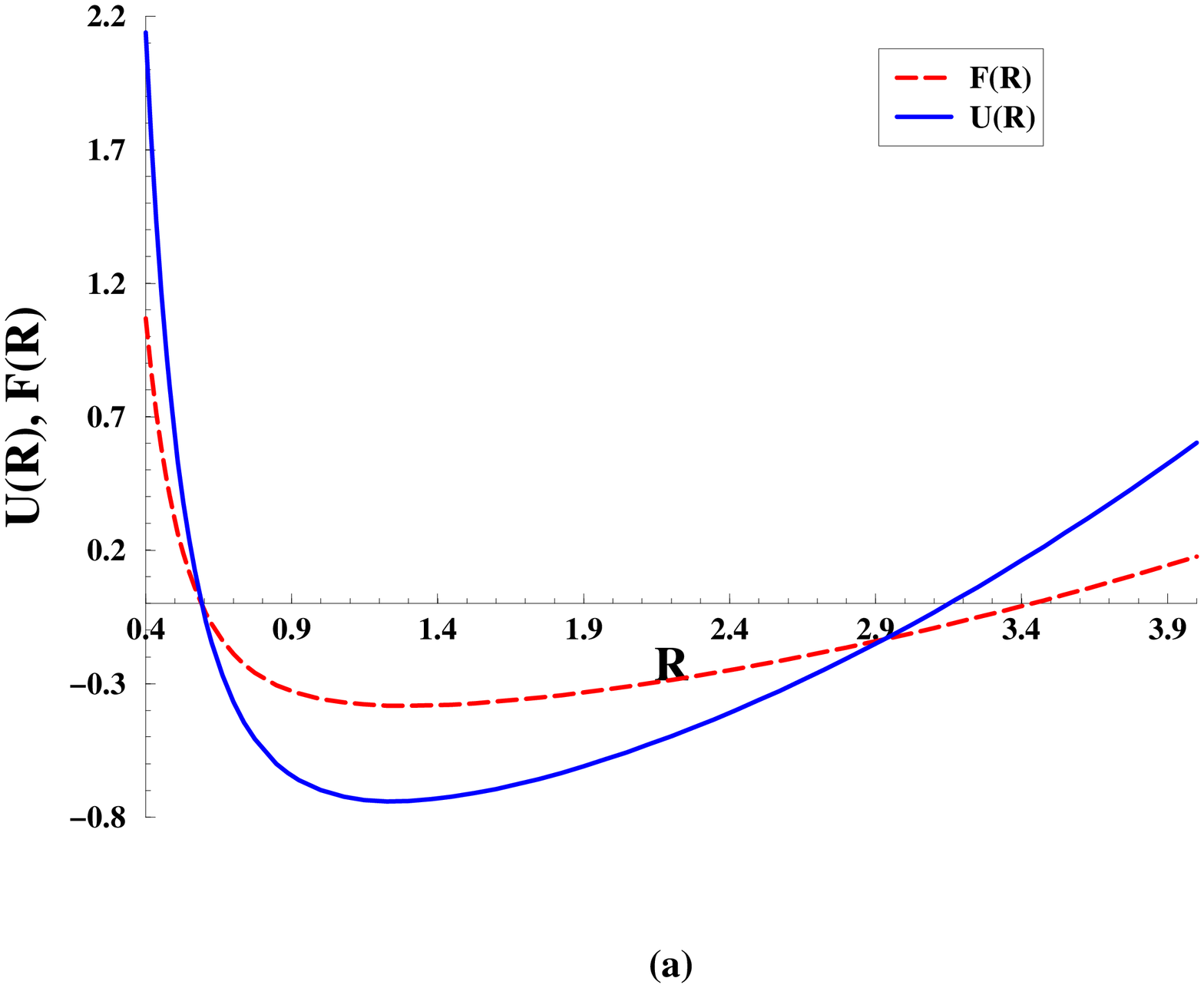}}\nonumber
\epsfxsize= 5.5truecm\rotatebox{-90}
{\epsfbox{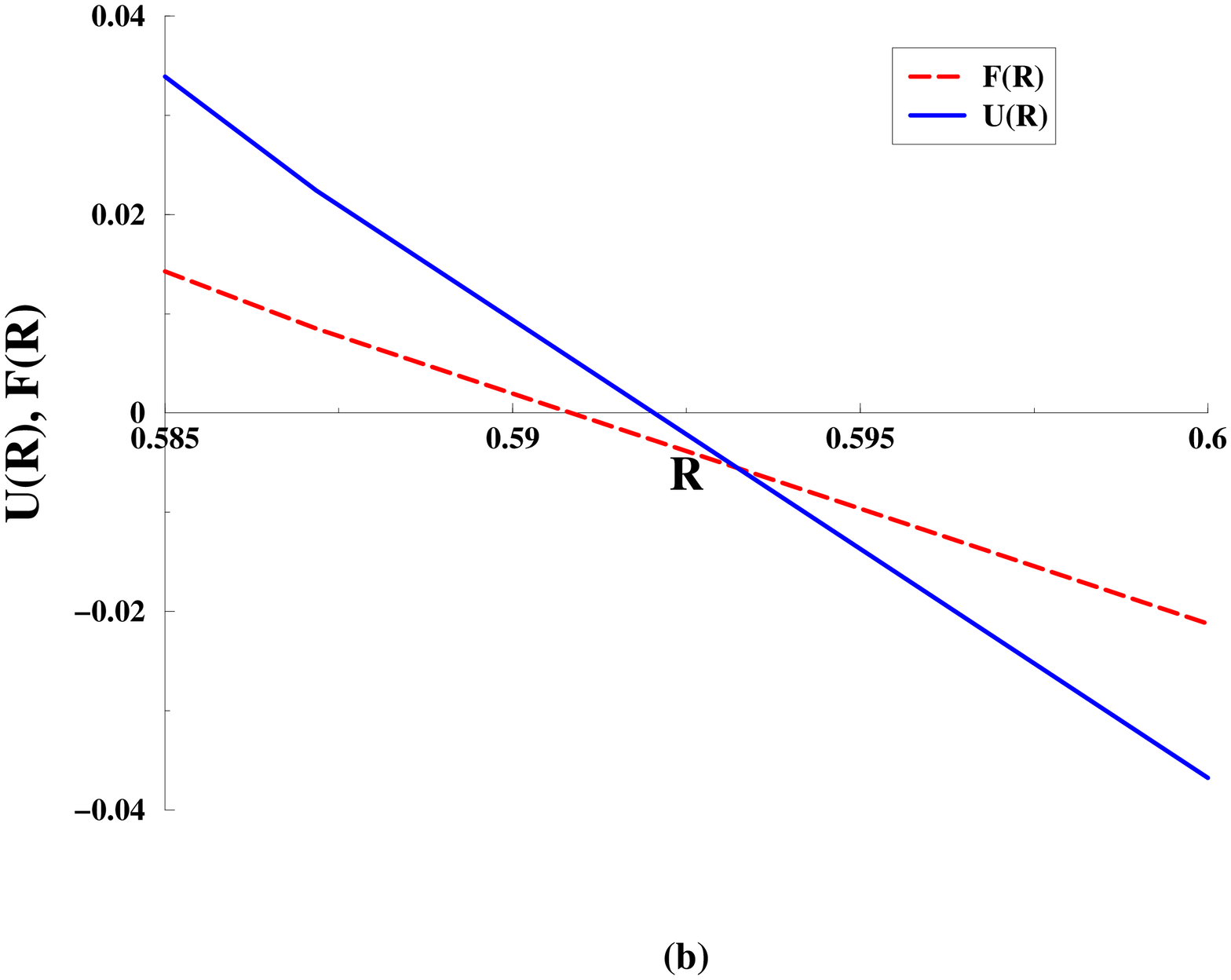}}\nonumber
\end{eqnarray}
\caption{(a) $U(R)$  and $F(R)$  with $\hat \Lambda <0$
and $M=-1/10$ for type I 
solutions, (b) Zoom of the event horizon region.}
\label{T1-U4}
\end{center}
\end{figure*}

The potential $F(R)$ ruling the evolution of the scale factor is
\begin{equation}\label{f1}
F(R)= {k \over 2} - MR^{-(D-3)} - \hat\Lambda R^2 \, ,
\end{equation}
where the effective cosmological constant on the domain wall is given by
\begin{equation}\label{lambda1}
\hat\Lambda = {1\over {D-2}} \left[ {{V_0}\over{D-1}} + {{\hat V_0 ^2}
    \over {8(D-2)}} \right] \, .
\end{equation}

We shall analyze each of the four cases presented in \cite{chre}.
As we have previously stated, the equation of motion (\ref{dwmotion})
has a solution only when $F(R) \leq 0$. This is automatic if $U(R)<0$,
i.e. if $r$ is a time coordinate; therefore, we look for solutions with 
$U(R)>0$. In fact, both conditions,
\begin{equation}\label{condition1}
F(R) \leq 0 \qquad \hbox{and} \qquad U(R)>0 
\end{equation}
can coexist in some cases as we will see in what follows. In order to
ilustrate the following examples we have chosen $D=6$ dimensions.

\subsubsection{$\hat \Lambda>0$, $M>0$}
From the graph of $U(R)$ (see Fig.\ref{T1-U}) we can choose the initial 
condition for the domain wall assuming that (\ref{u1}) describes a
dS-Schwarzschild bulk with event and cosmological horizons when $M>0$
and $V_0>0$. 

We thus choose the initial condition for the domain wall inside this
region and where $r$ is a space coordinate. From
Fig.\ref{T1-U} we notice that there are two small 
regions, $r_H \leq r < 0.593$ and $2.93 \leq r < r_C $, where
(\ref{condition1}) holds. The results are shown in  
Fig.\ref{T1-braneb}. We see that for region I the geodesics follow
the domain wall for a while and then decouple falling into the event
horizon. For region II all the geodesics and the
domain wall converge to the cosmological horizon $r_C$ independently
of the value of $\hat V_0$.

\subsubsection{$\hat \Lambda<0$, $M>0$}
This is an AdS-Schwarzschild bulk. The condition 
(\ref{condition1}) is fullfilled inside a very small range as we can 
see in Fig.\ref{T1-U3}(a). However,
all the geodesics fall into the event horizon after following some path
on the brane (see Fig.\ref{T1-U3}(b)).

\subsubsection{$\hat \Lambda>0$, $M<0$}
From Fig.\ref{T1-U2}(a) we choose the initial condition for the domain
wall equation of motion inside the region where (\ref{condition1})
holds. As we can see from Fig.\ref{T1-U2}(b), the domain wall and the
geodesics converge to the cosmological horizon $r_C$. However, after
some threshold initial velocity the geodesics diverge to the timelike
naked singularity. 
\subsubsection{$\hat \Lambda<0$, $M<0$}
Here (\ref{dwmotion}) can only have a solution when $k=-1$. This
is a topological black hole in an asymptotically AdS space.
From Fig.\ref{T1-U4} we see that there is no solution fulfilling  
(\ref{condition1}) between event and cosmological horizons. 
\subsection{Type II Solutions}

\begin{figure*}[htb!]
\begin{center}
\leavevmode
\begin{eqnarray}
\epsfxsize= 5.5truecm\rotatebox{-90}
{\epsfbox{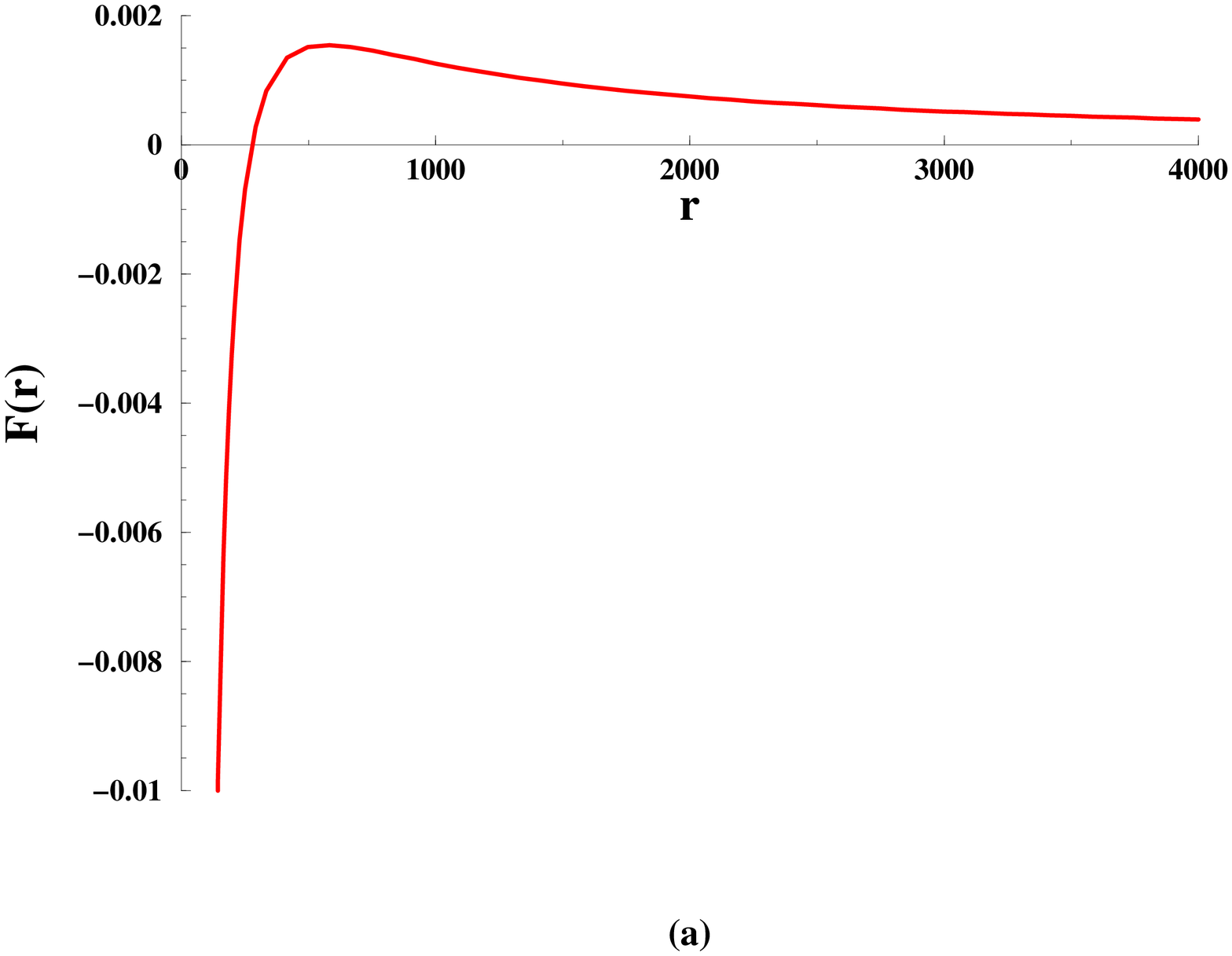}}\nonumber
\epsfxsize= 5.5truecm\rotatebox{-90}
{\epsfbox{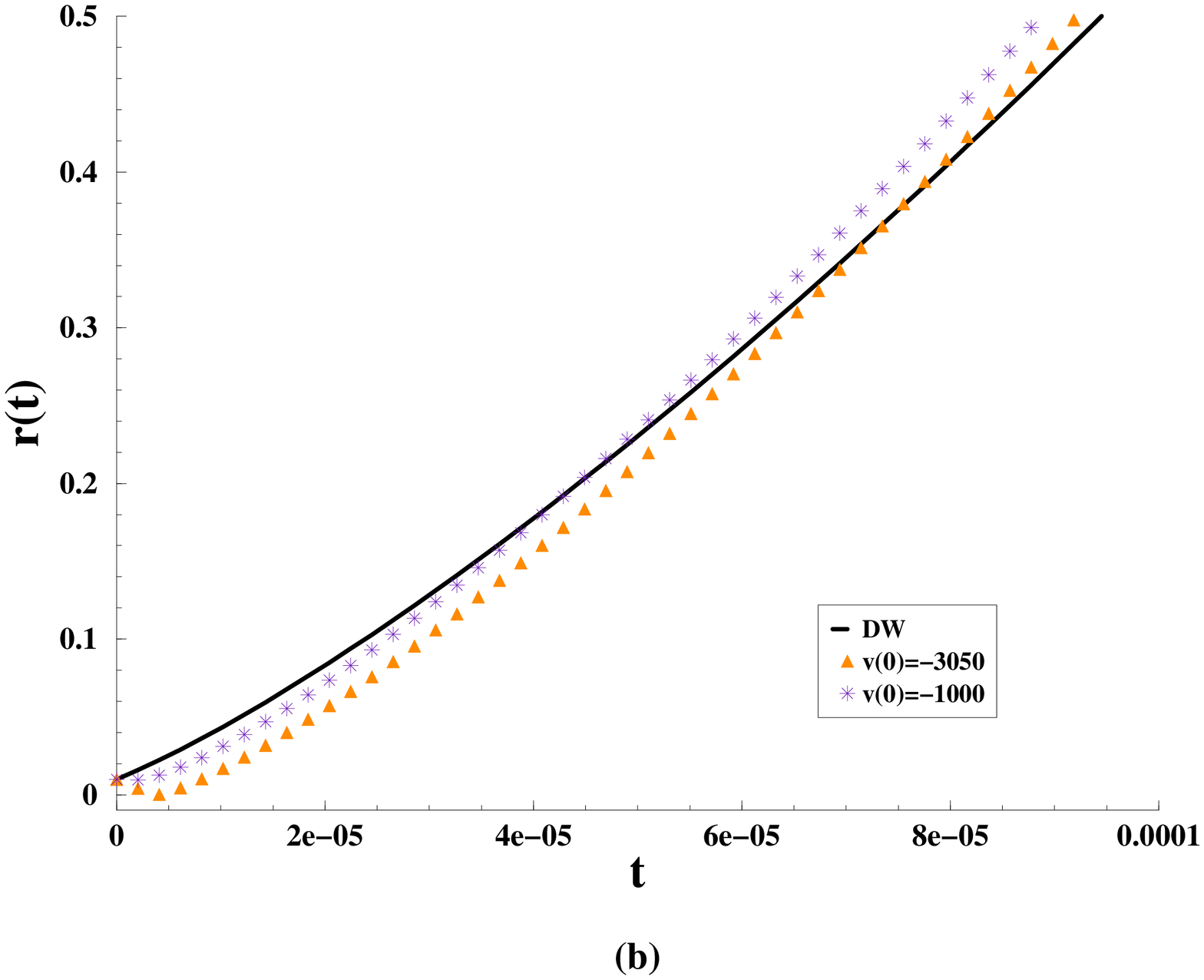}}\nonumber
\end{eqnarray}
\caption{(a) $F(r)$ with $\hat \Lambda >0$ and $M<0$ for type II 
solutions, (b) Domain wall motion  and geodesics for $V_0=1$,
$\hat V_0 =6$, $M=-10$ and $\beta=\sqrt{10}$.}
\label{T2-br1}
\end{center}
\end{figure*}
\begin{figure*}[htb!]
\begin{center}
\leavevmode
\begin{eqnarray}
\epsfxsize= 5.5truecm\rotatebox{-90}
{\epsfbox{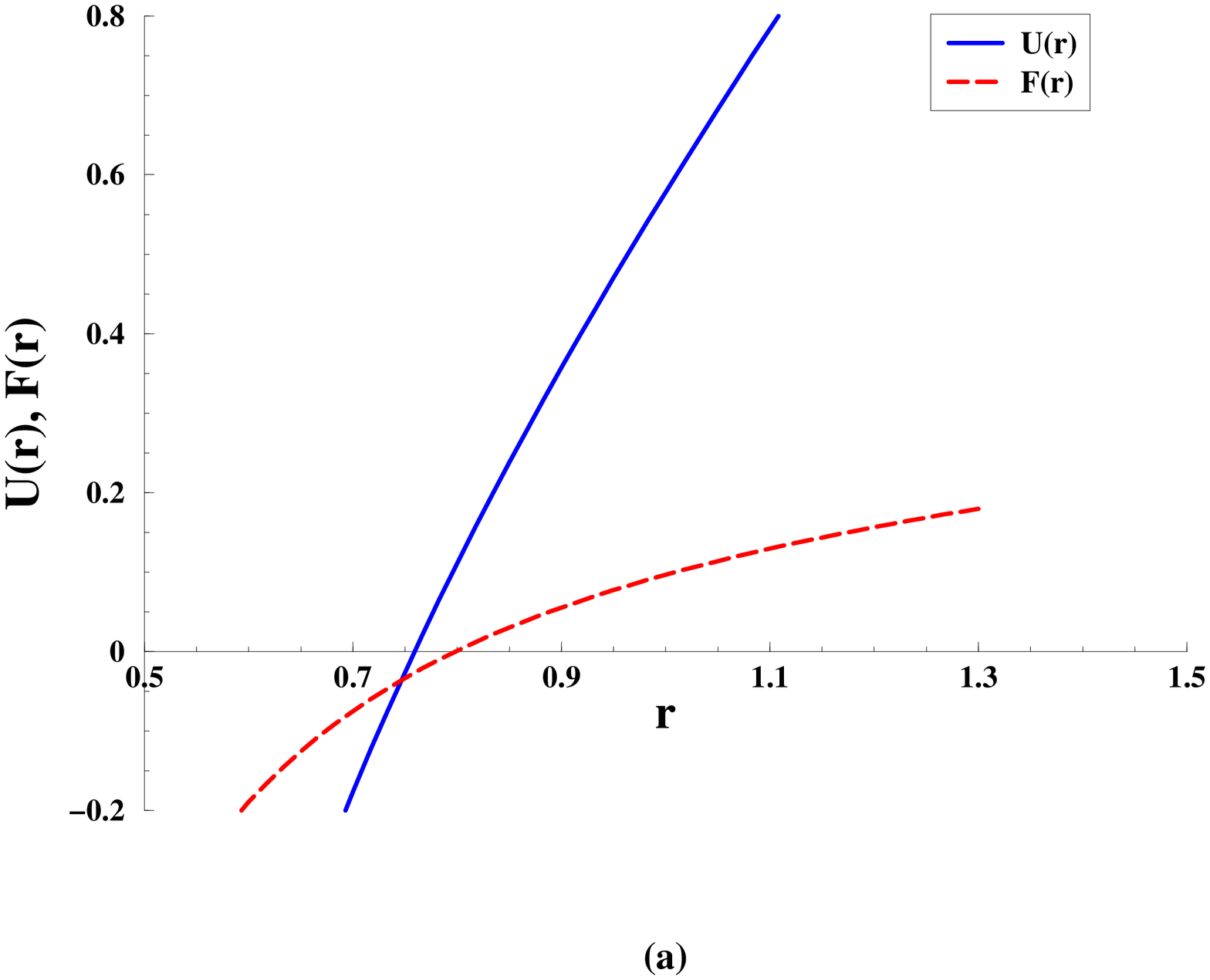}}\nonumber
\epsfxsize= 5.5truecm\rotatebox{-90}
{\epsfbox{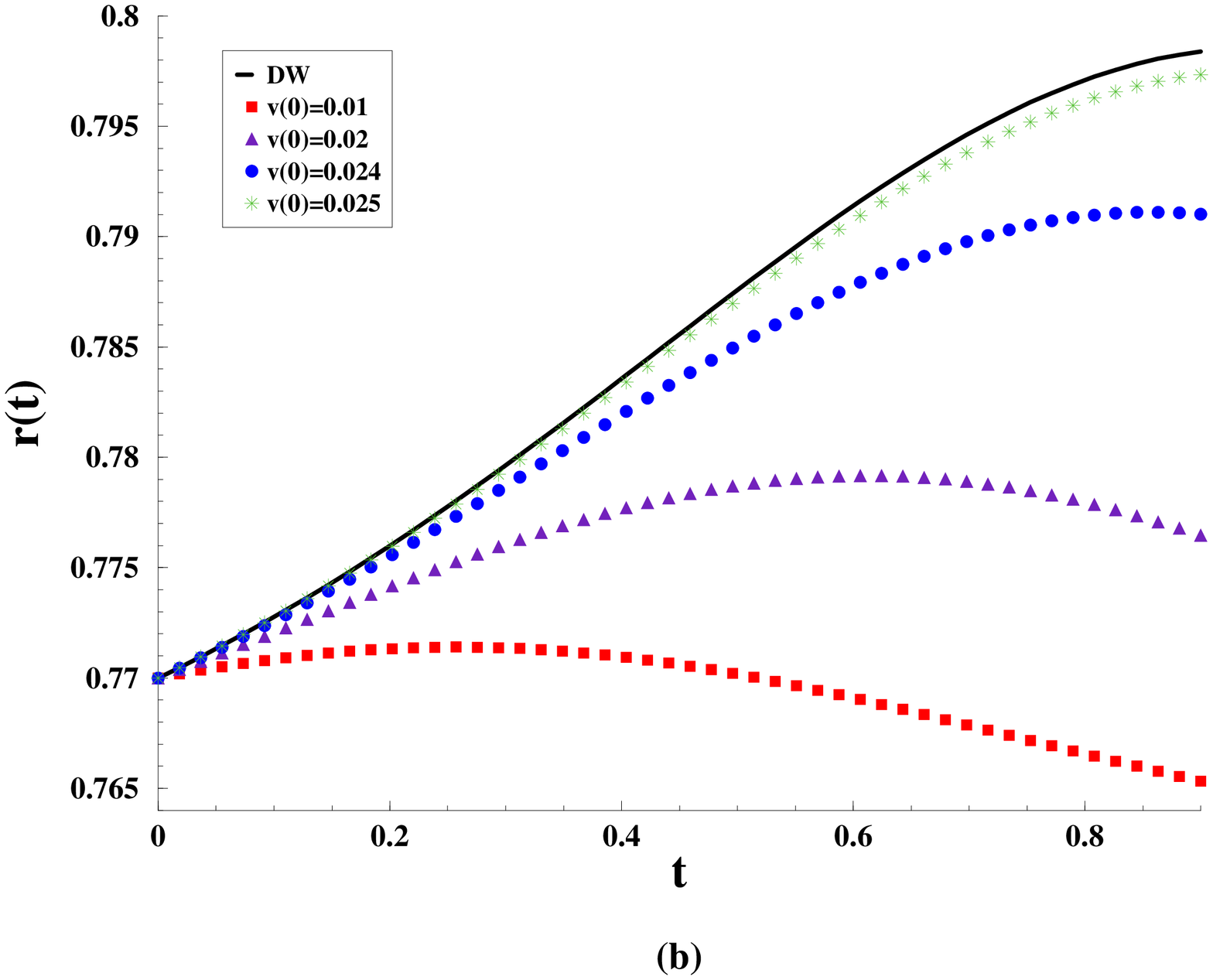}}\nonumber
\end{eqnarray}
\caption{(a) $U(r)$ and $F(r)$ with $\hat \Lambda <0$ and $M>0$ for type II 
solutions, (b) Domain wall motion  and geodesics for $V_0=-1$,
$\hat V_0 =1$, $M=1/10$ and $\beta=1/\sqrt{2}$.}
\label{T2-br2}
\end{center}
\end{figure*}
\begin{figure*}[htb!]
\begin{center}
\leavevmode
\begin{eqnarray}
\epsfxsize= 5.5truecm\rotatebox{-90}
{\epsfbox{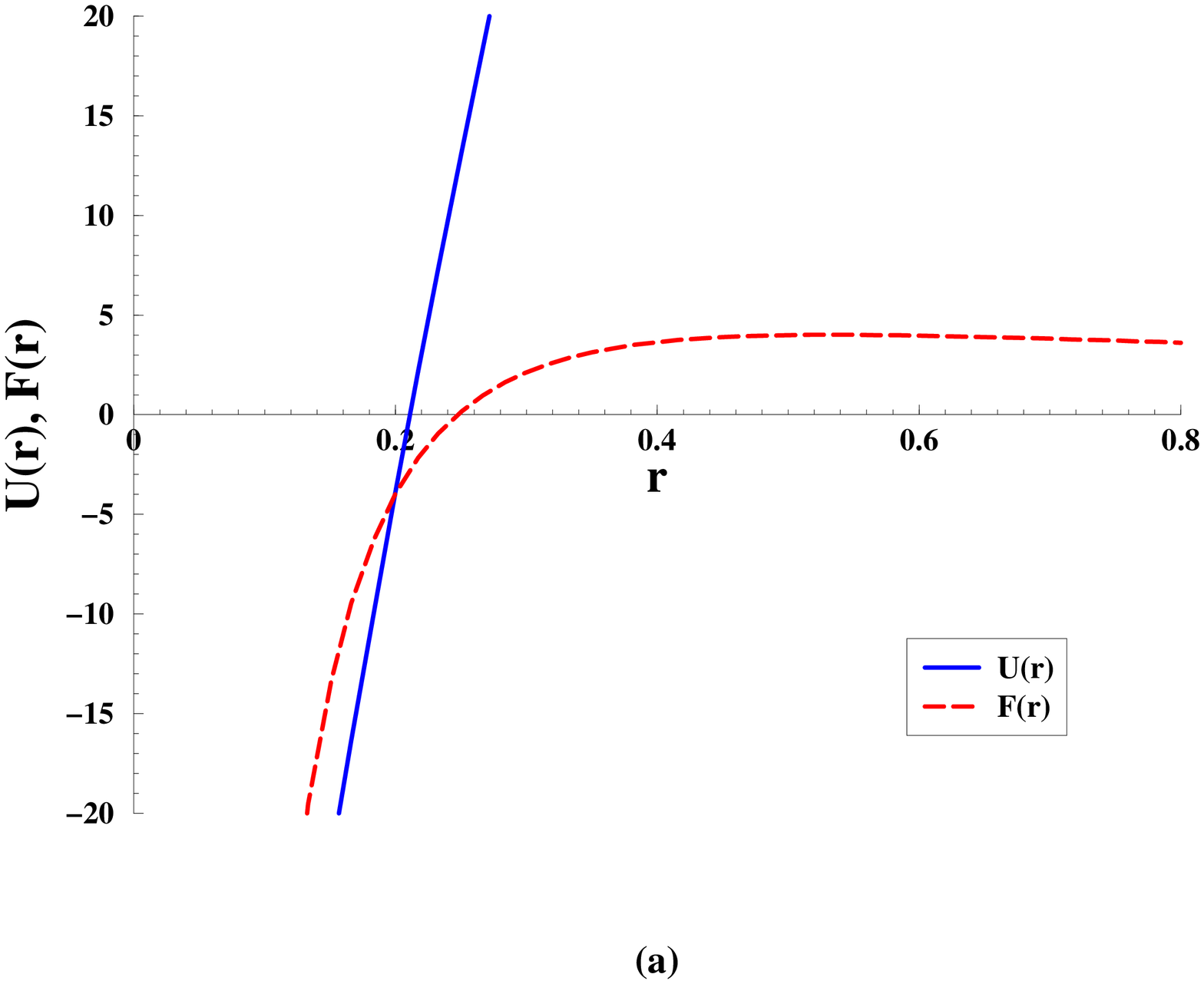}}\nonumber
\epsfxsize= 5.5truecm\rotatebox{-90}
{\epsfbox{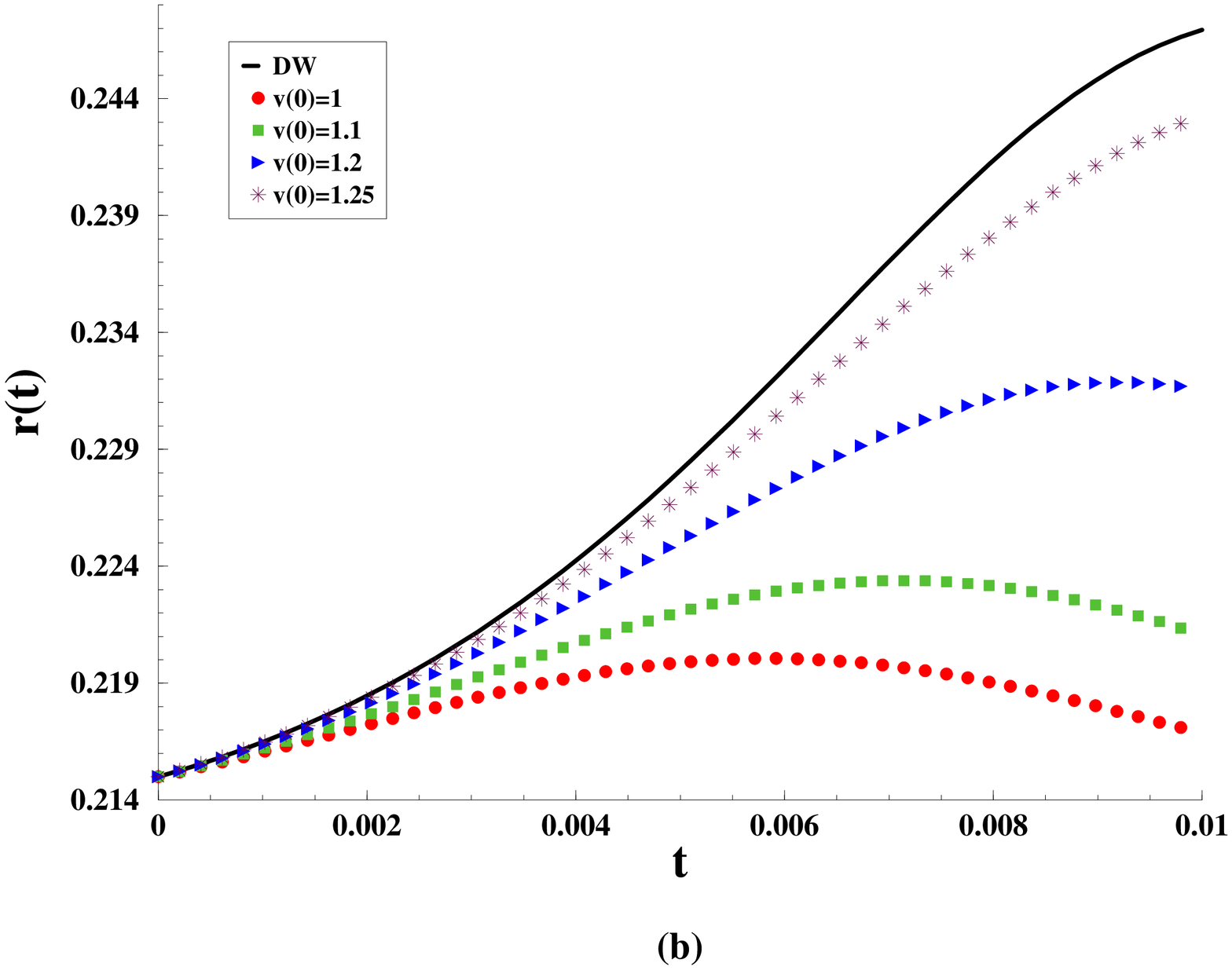}}\nonumber
\end{eqnarray}
\caption{(a) $U(r)$ and $F(r)$ with $\hat \Lambda <0$ and $M>0$ for type II 
solutions, (b) Domain wall motion  and geodesics for $V_0=-1$,
$\hat V_0 =1$, $M=10$ and $\beta=2$.}
\label{T2-br3}
\end{center}
\end{figure*}

The type II solutions have $\alpha =\beta/2$ and $k=0$. The metric is
given by
\begin{equation}\label{u2}
U(r) = (1 +b^2)^2 r^{2 \over {1+b^2}} \left( -2Mr^{-{{D-1-b^2}\over
{1+b^2}}} - {{2\Lambda}\over {(D-1-b^2)}} \right) \, ,
\end{equation}
and the scale factor is
\begin{equation}\label{R2}
R(r) = r^{1 \over {1+b^2}} \, ,
\end{equation}
where
\begin{equation}\label{lambda2}
\Lambda = {{V_0 e ^{2b \phi_0}} \over {D-2}}  \quad \quad
\hbox{and} \quad \quad b = {1 \over 2} \beta \sqrt{D-2} \quad . 
\end{equation}

The potential is given by the expression
\begin{equation}\label{F2}
F(R) = -R^{2(1-b^2)} \left( M R^{-(D-1-b^2)} + \hat \Lambda \right) \; ,
\end{equation}
where 
\begin{equation}\label{ls2}
\hat\Lambda = {e^{2b\phi_0} \over {D-2}} \left( {V_0 \over {D-1-b^2}}
+ {{\hat V_0 ^2}\over {8(D-2)}} \right) \; .
\end{equation}

There are twelve cases from which we choose those ones where $r$ is a
spatial coordinate. When $b^2<D-1$, $r$ is a spatial coordinate if
$V_0<0$. When $b^2>D-1$, $r$ is spatial if $M<0$.

We should also rewrite (\ref{condition1}) as
\begin{equation}\label{condition2}
F(r) \leq 0 \quad \hbox{and} \quad U(r)>0 \, . 
\end{equation}

\subsubsection{$\hat\Lambda>0$, $M<0$, $b^2>D-1$}

In this case $U(r)$ is always positive, whereas $F(r)$ is negative for
small $r$. From Fig.\ref{T2-br1} we see that some microscopic shortcuts
appear in the very beginning of the 
solution and after crossing the domain wall they escape to infinity.

\subsubsection{$\hat\Lambda<0$, $M>0$, $b^2<1$}

This case describes a black $(D-2)$ brane solution in AdS space. Here there is
a very small region where (\ref{condition2}) holds after 
the event horizon as we can see from Fig.\ref{T2-br2}. We show the
entire domain wall solution and we see that geodesics follow it
and then fall into the event horizon at later times.

\subsubsection{$\hat\Lambda<0$, $M>0$, $1<b^2<D-1$}

This case is also a black brane in AdS space. The region where
(\ref{condition2}) is respected is shown in Fig.\ref{T2-br3}. As in
the previous case all the geodesics follow the domain wall and at
later times fall into the event horizon.

\subsubsection{$\hat\Lambda<0$, $M<0$}

As $F(r)$ is always positive for all $b^2$, no solutions to
(\ref{eq3}) exist. 

\subsection{Type III Solutions} 

\begin{figure*}[htb!]
\begin{center}
\leavevmode
\begin{eqnarray}
\epsfxsize= 5.truecm\rotatebox{-90}
{\epsfbox{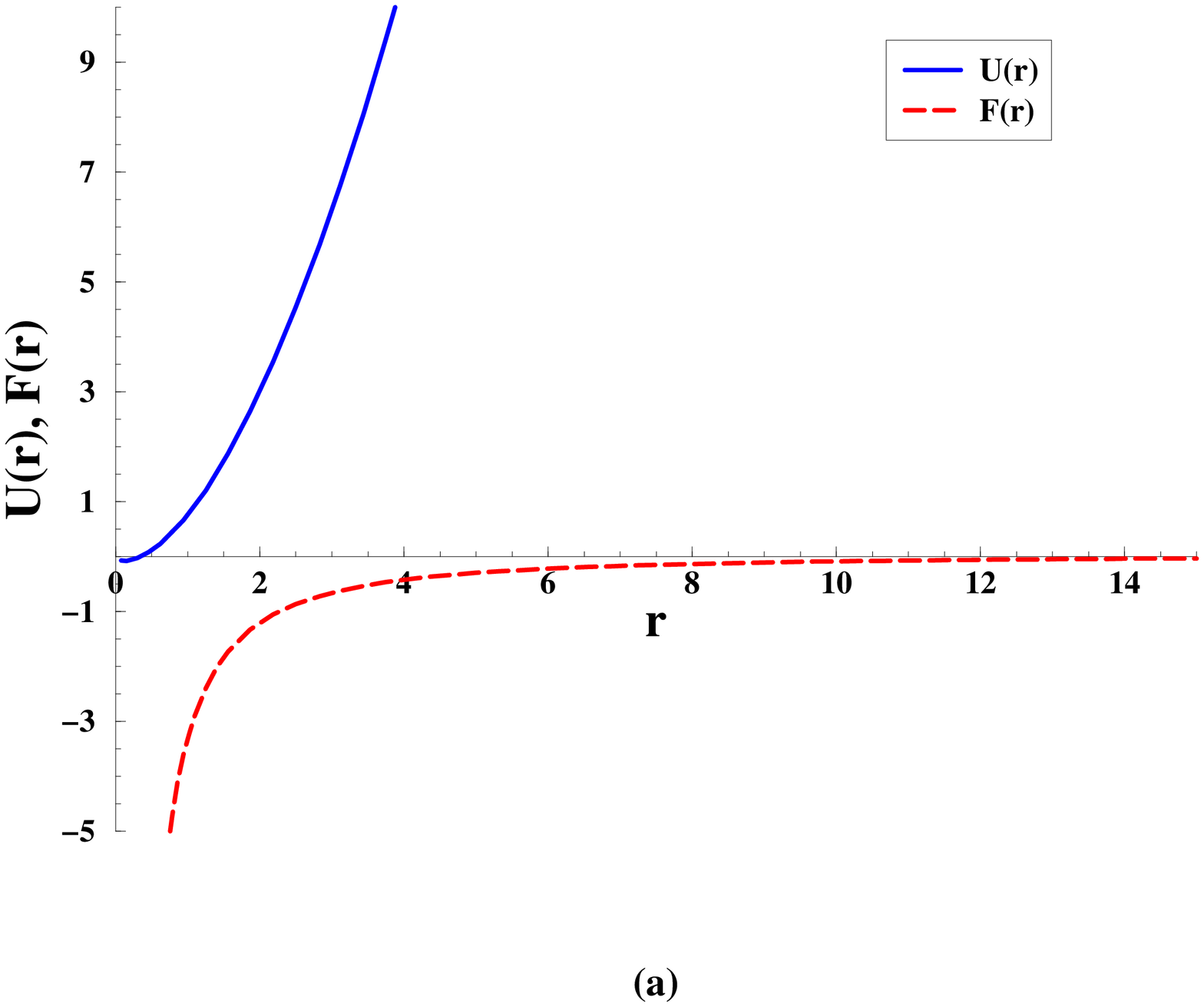}}\nonumber
\epsfxsize= 5.truecm\rotatebox{-90}
{\epsfbox{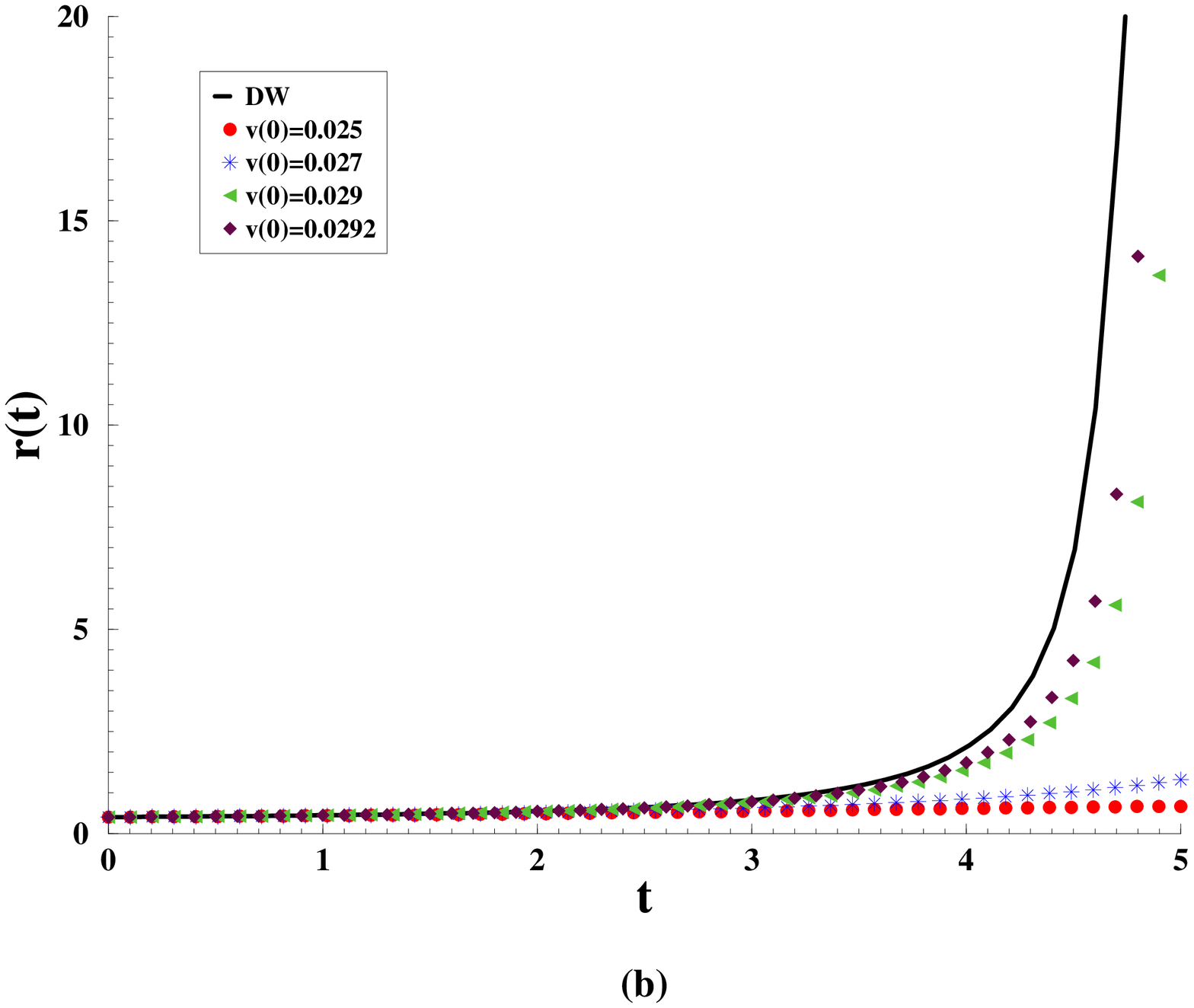}}\nonumber
\end{eqnarray}
\caption{(a) $U(r)$ and $F(r)$ for Type III solutions with $k=-1$, $M=1/10$ and
$\beta^2<{1 \over {(D-2)}}$, (b) Domain Wall motion  and
geodesics for $V_0=-1$, $\hat V_0=1$, $\phi_0=1$ and
$\beta=1/\sqrt{6}$.}
\label{T3-branea}

\begin{eqnarray}
\epsfxsize= 5.truecm\rotatebox{-90}
{\epsfbox{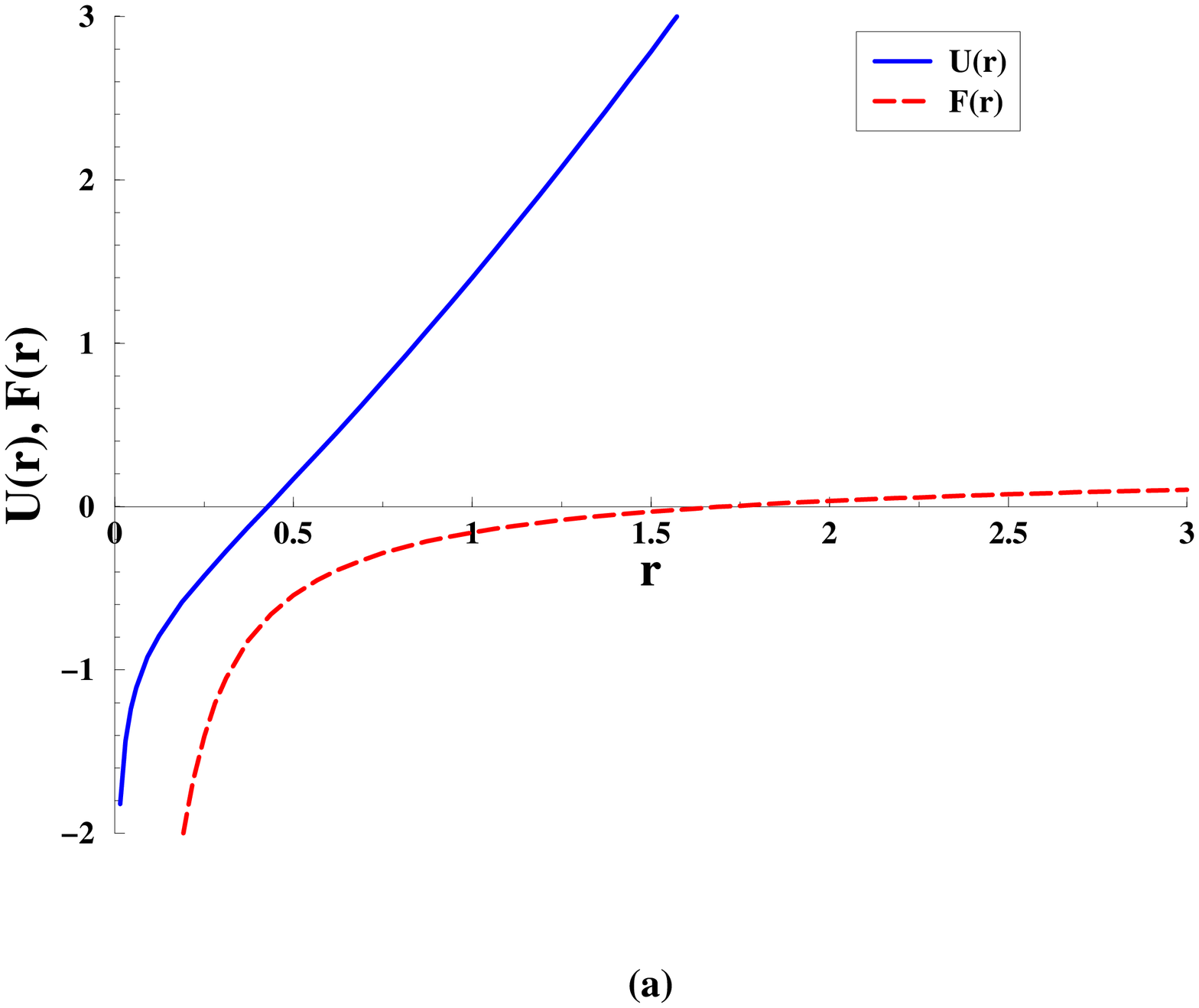}}\nonumber
\epsfxsize= 5.truecm\rotatebox{-90}
{\epsfbox{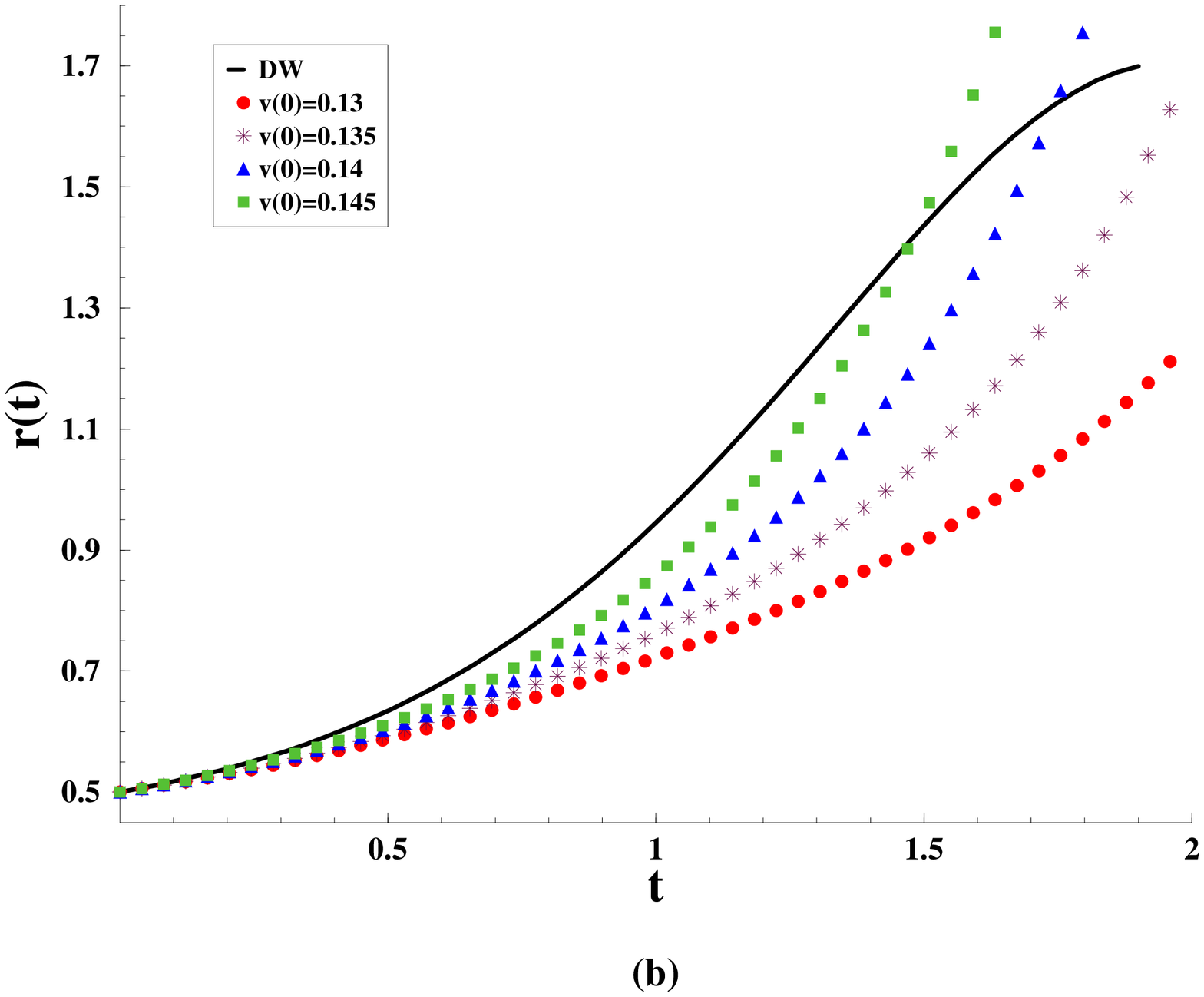}}\nonumber
\end{eqnarray}
\caption{(a) $U(r)$  and $F(r)$  for $k=-1$, $M=1/10$ and
$V_0<0$ in Type III solutions, 
(b) Domain wall motion  and geodesics with $V_0=-1$,
$\hat V_0=1$, $\phi_0=1$ and $\beta=1/\sqrt{2}$.}
\label{T3-braneb}
\end{center}
\end{figure*}
\begin{figure*}[htb!]
\begin{center}
\leavevmode
\begin{eqnarray}
\epsfxsize= 5.5truecm\rotatebox{-90}
{\epsfbox{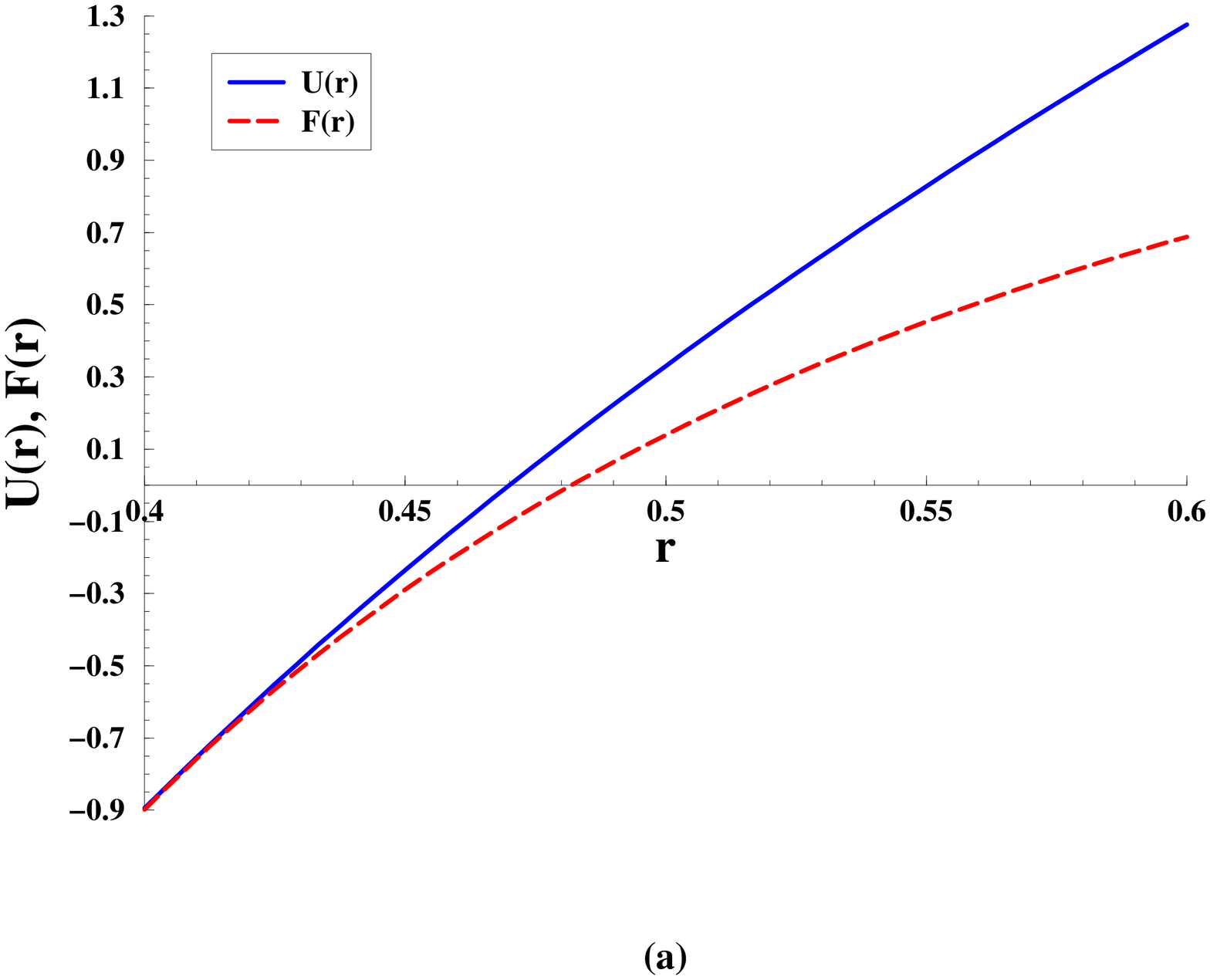}}\nonumber
\epsfxsize= 5.5truecm\rotatebox{-90}
{\epsfbox{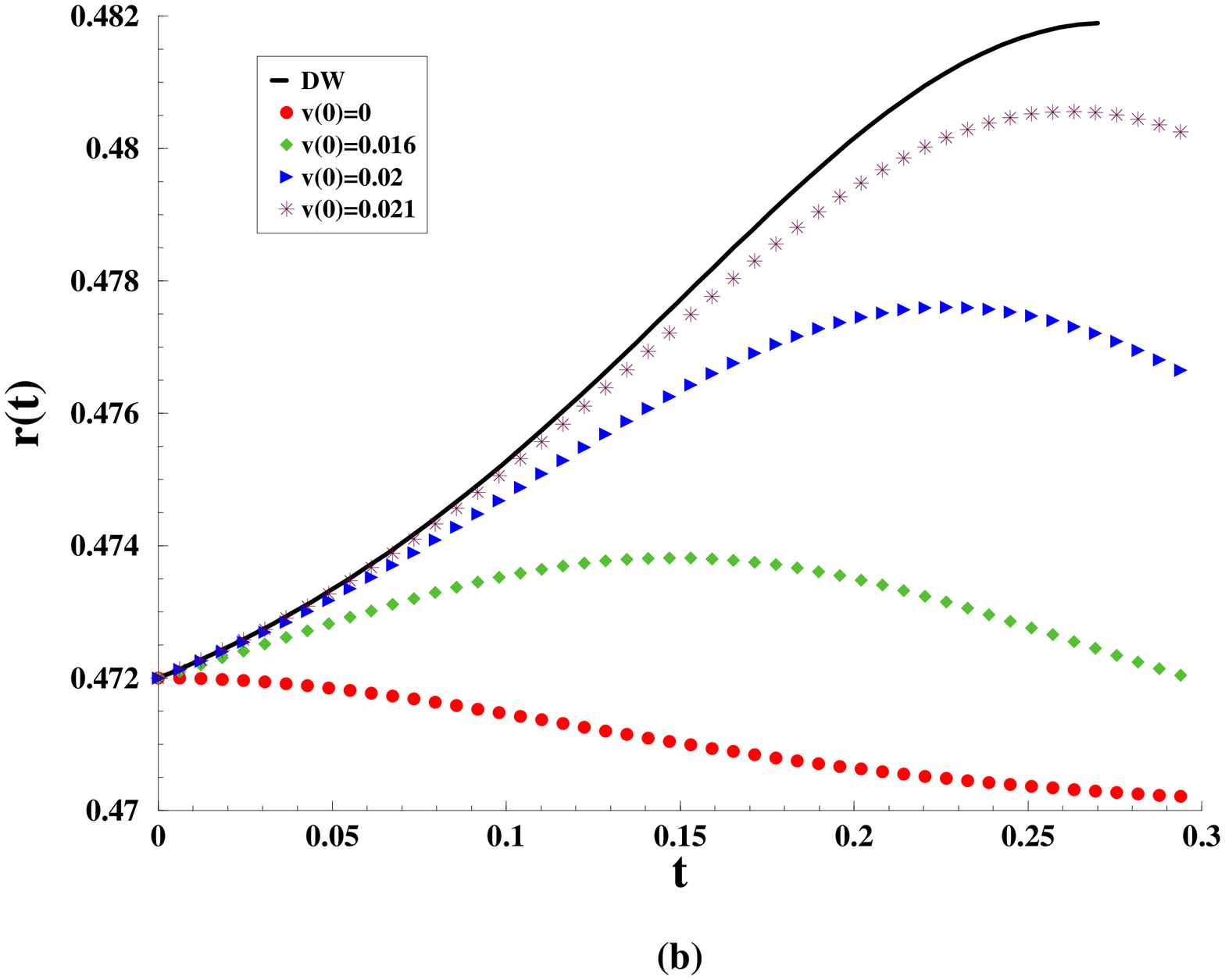}}\nonumber\\
\epsfxsize= 5.5truecm\rotatebox{-90}
{\epsfbox{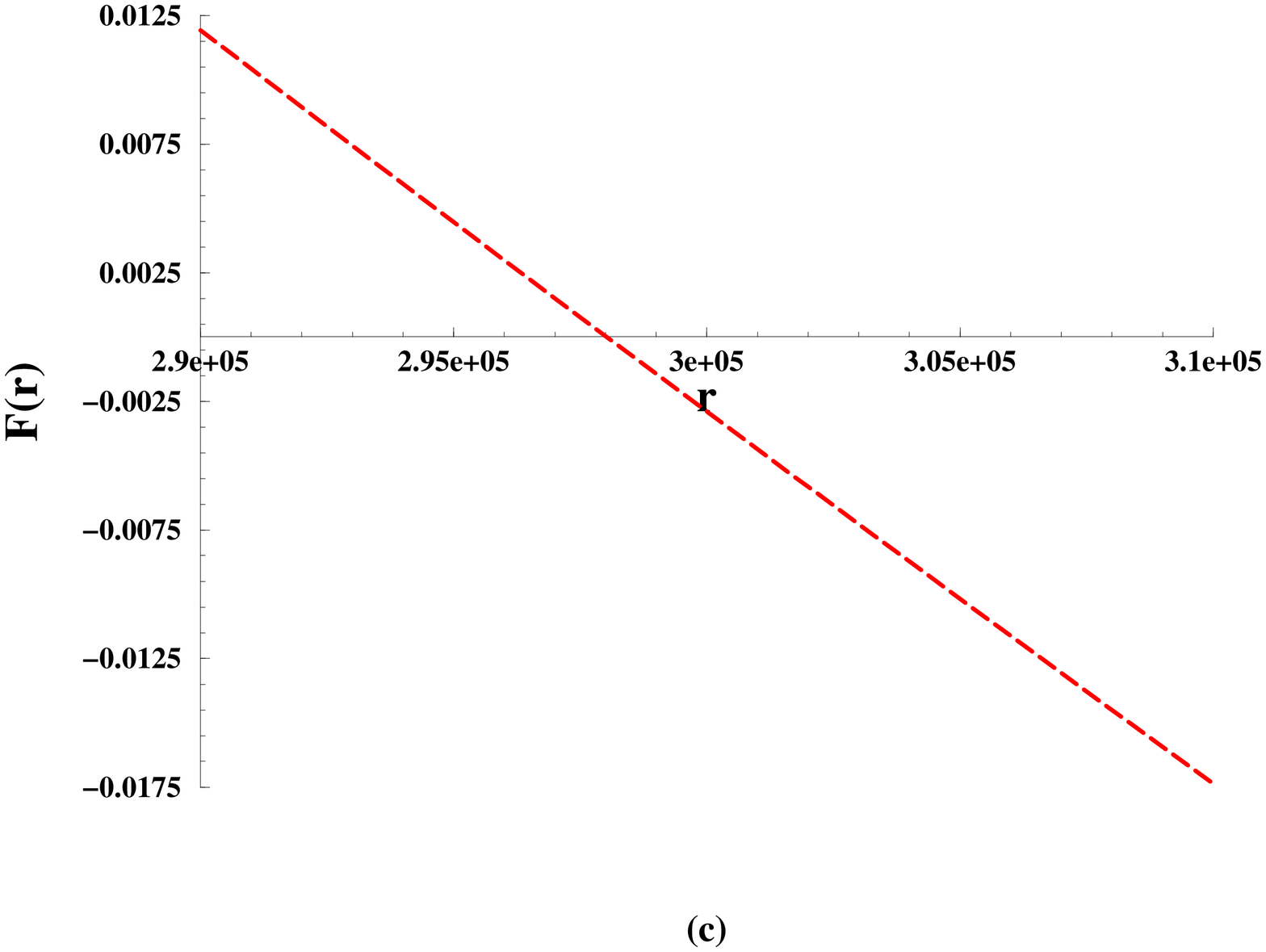}}\nonumber
\epsfxsize= 5.5truecm\rotatebox{-90}
{\epsfbox{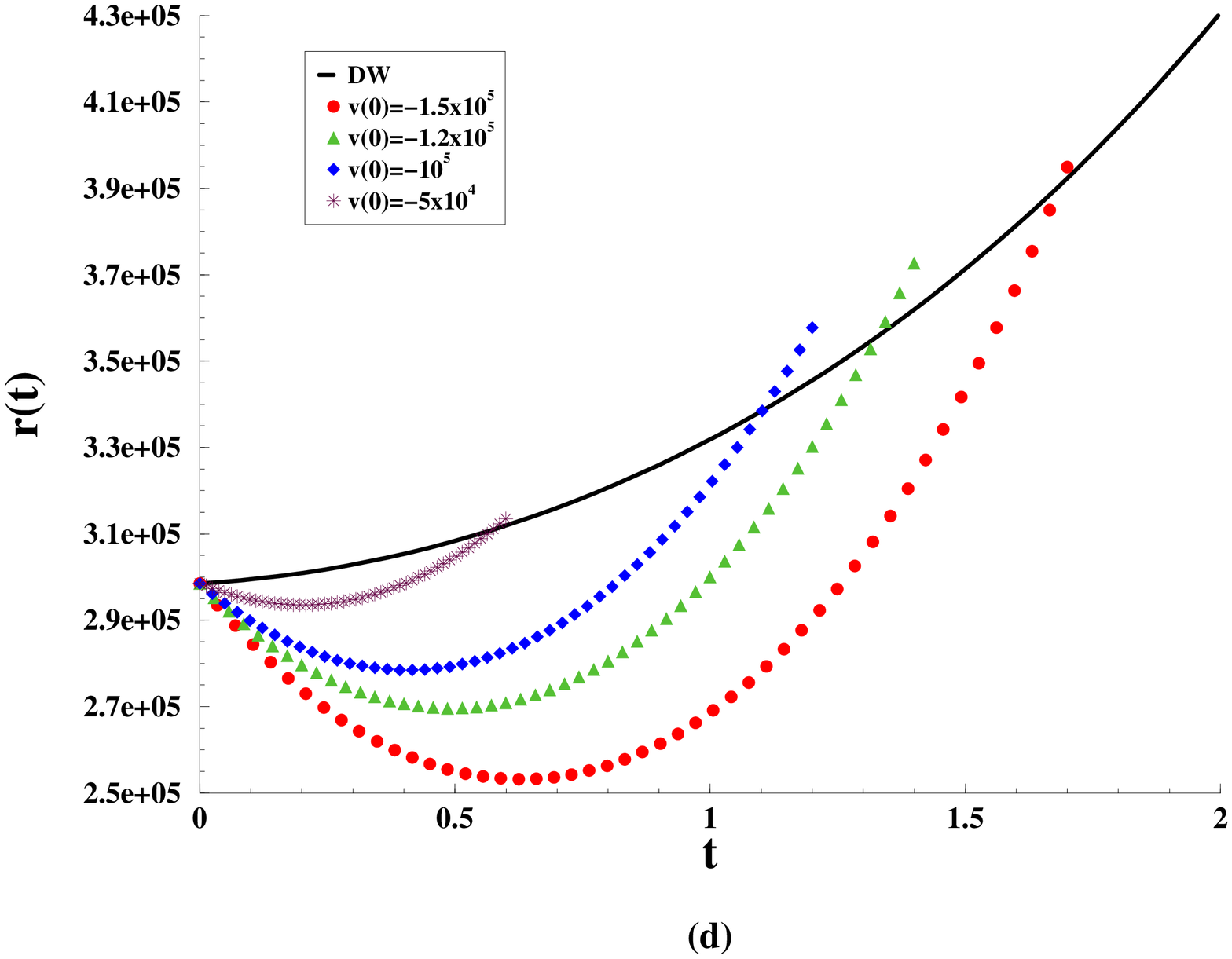}}\nonumber
\end{eqnarray}
\caption{(a) $U(r)$  and $F(r)$  for $M=1/10$ and $V_0<0$
in Type III solutions, 
(b) Domain wall motion  and geodesics with $V_0=-1$,
$\hat V_0=1$, $\phi_0=1$ and $\beta=\sqrt{5}/2$, (c) $F(r)$ in the
region of interest, (d) Domain wall motion  and geodesics under
the same conditions as (b).}
\label{T3-braned}
\end{center}
\end{figure*}

\begin{figure*}[htb!]
\begin{center}
\leavevmode
\begin{eqnarray}
\epsfxsize= 5.5truecm\rotatebox{-90}
{\epsfbox{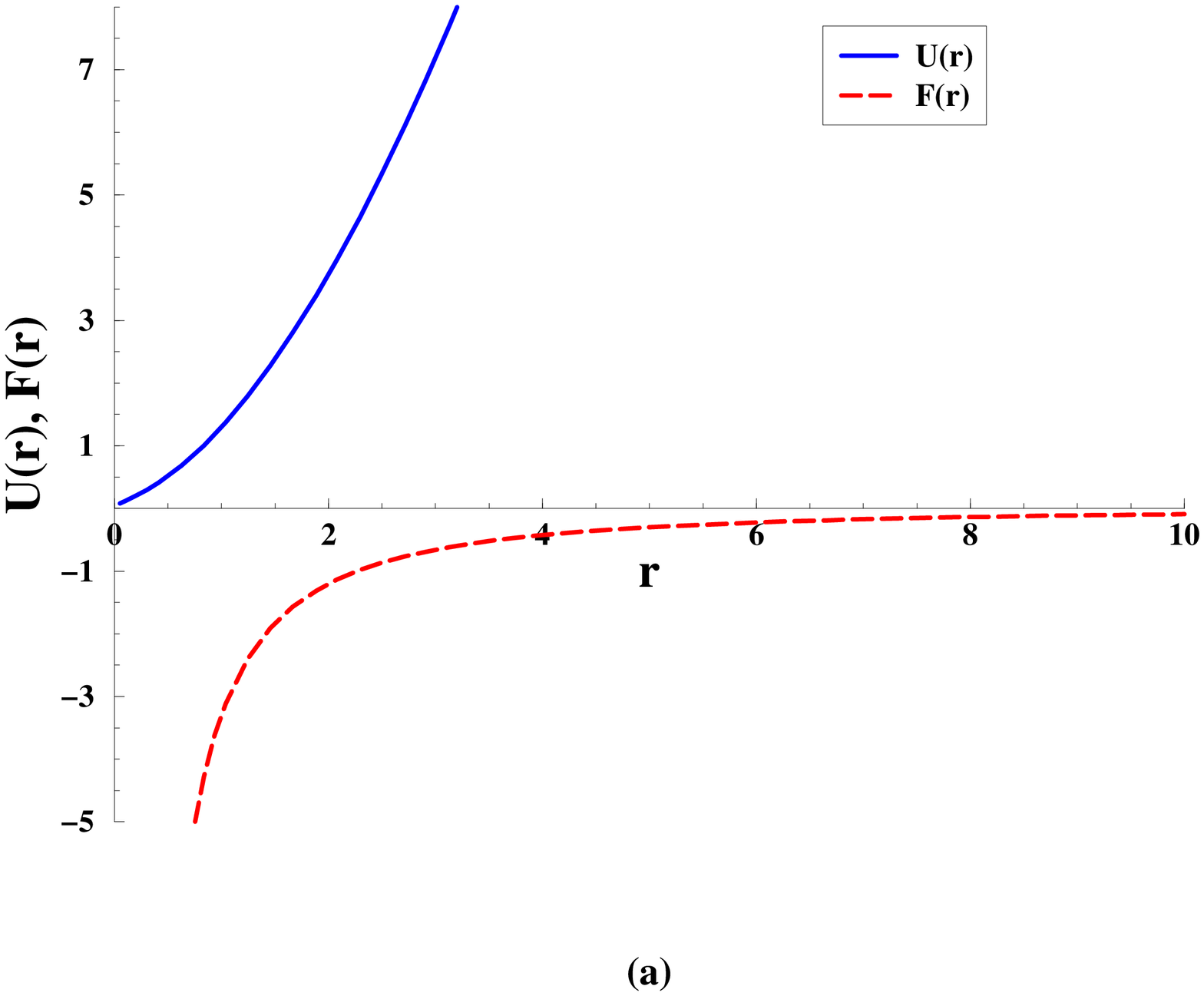}}\nonumber
\epsfxsize= 5.5truecm\rotatebox{-90}
{\epsfbox{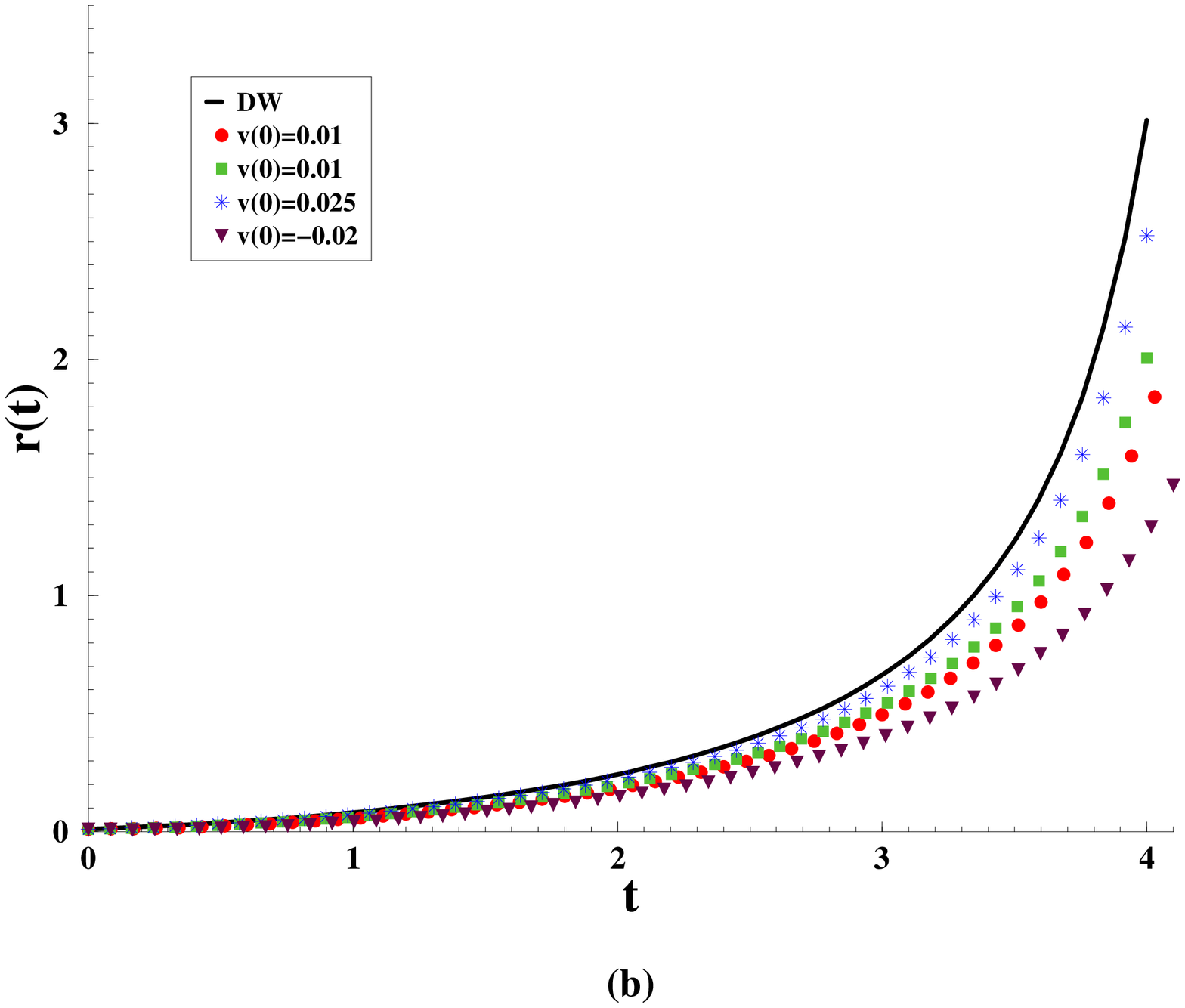}}\nonumber
\end{eqnarray}
\caption{(a) $U(r)$ and $F(r)$ for type III solutions when $k=-1$, $M=-1/10$
and $b^2<{1 \over {(D-2)}}$, (b)
Domain wall motion  and geodesics for $V_0=-1$, $\hat V_0=1$,
$\phi_0=1$ and $\beta=1/\sqrt{6}$.}
\label{T3-branee}
\end{center}
\end{figure*}
\begin{figure*}[htb!]
\begin{center}
\leavevmode
\begin{eqnarray}
\epsfxsize= 5.5truecm\rotatebox{-90}
{\epsfbox{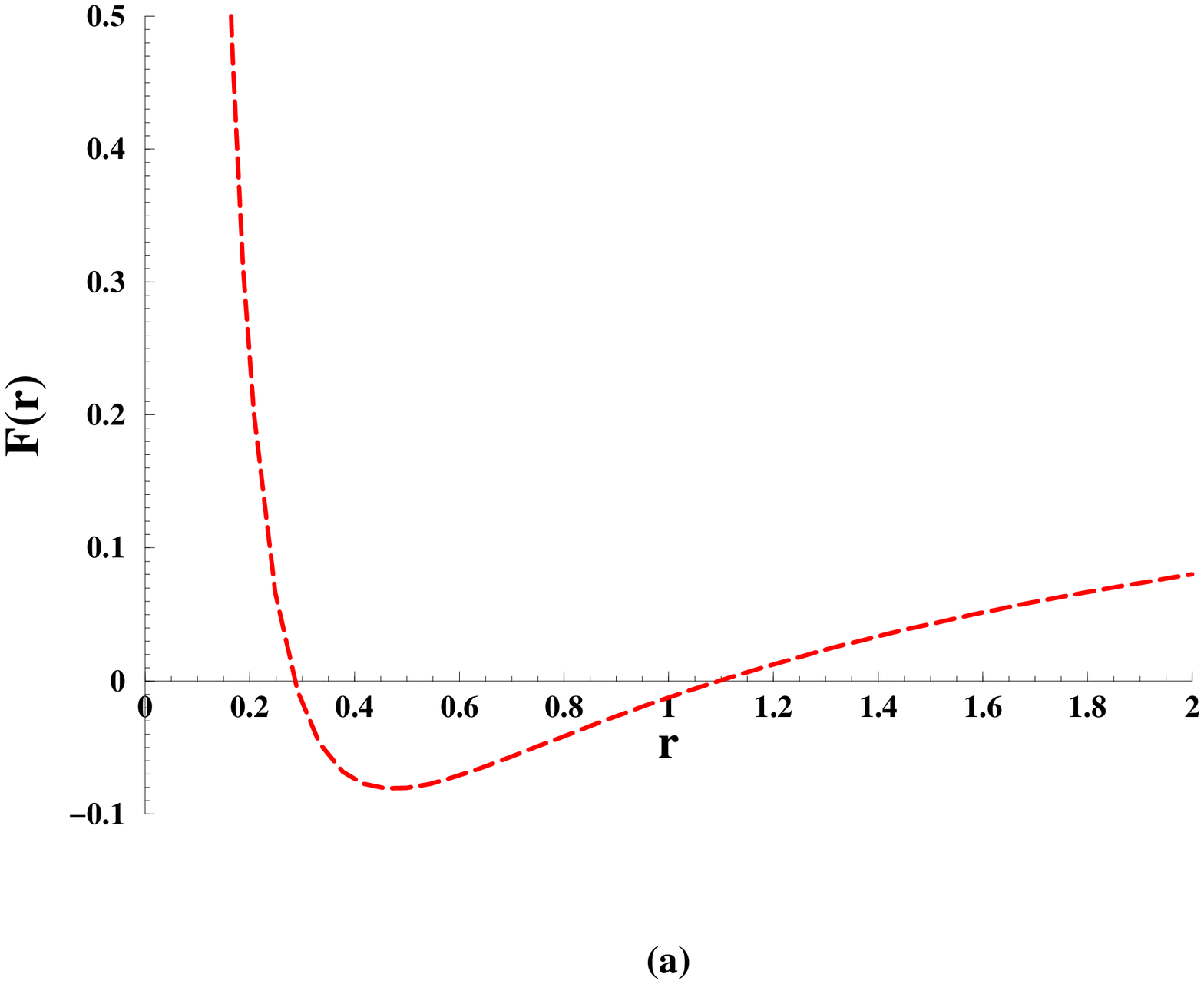}}\nonumber
\epsfxsize= 5.5truecm\rotatebox{-90}
{\epsfbox{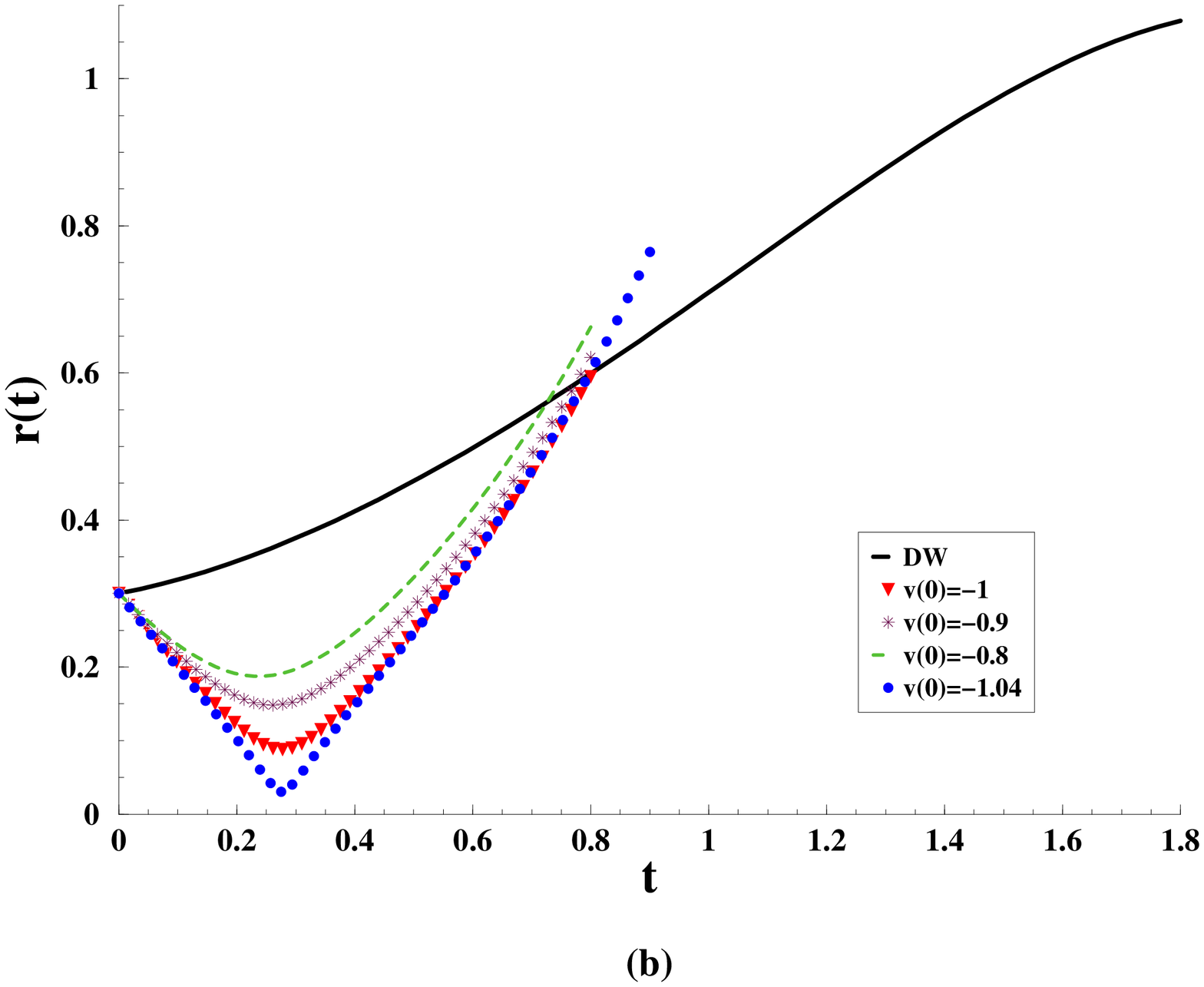}}\nonumber
\end{eqnarray}
\caption{(a) $F(r)$ for type III solutions when $k=-1$, $M=-1/10$ and ${1 \over
{(D-2)}}<b^2<1$, (b)
Domain wall motion  and geodesics for $V_0=-1$, $\hat V_0=1$,
$\phi_0=1$ and $\beta=1/\sqrt{2}$.}
\label{T3-branef}
\end{center}
\end{figure*}
\begin{figure*}[htb!]
\begin{center}
\leavevmode
\begin{eqnarray}
\epsfxsize= 5.5truecm\rotatebox{-90}
{\epsfbox{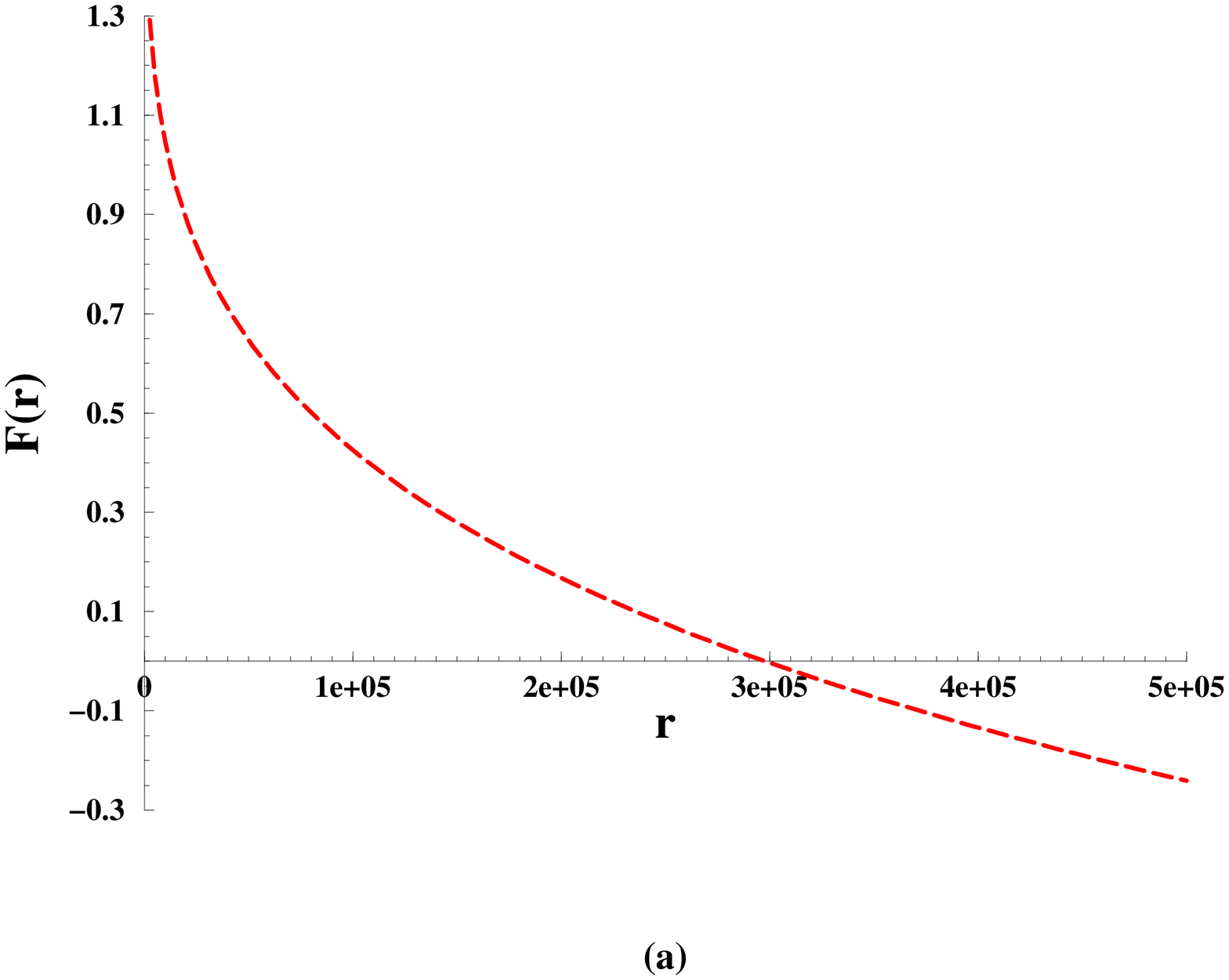}}\nonumber
\epsfxsize= 5.5truecm\rotatebox{-90}
{\epsfbox{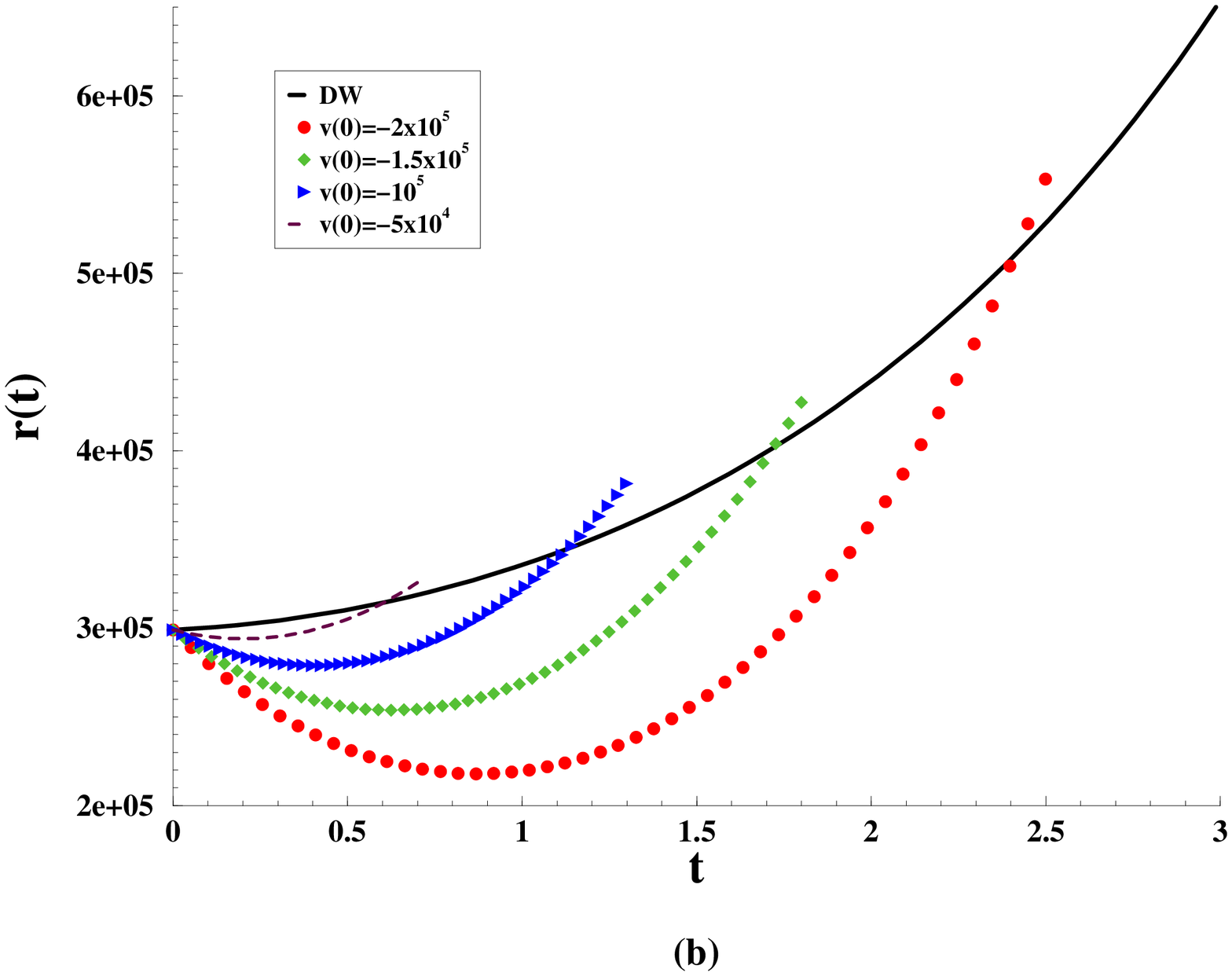}}\nonumber
\end{eqnarray}
\caption{(a) $F(r)$ for type III solutions when $M=-1/10$ and $b^2>1$, (b)
Domain wall motion  and geodesics for $V_0=-1$, $\hat V_0=1$,
$\phi_0=1$ and $\beta=\sqrt{5}/2$.}
\label{T3-braneg}
\end{center}
\end{figure*}

The type III solutions have $\alpha = {2\over {\beta(D-2)}}$.
In this case, the metric is given by
\begin{equation}\label{u3}
U(r) = (1+b^2)^2 r^{2\over {1+b^2}} \left( -2Mr^{-{1+b^2(D-3)}\over
    {1+b^2}} - {{2\Lambda} \over {(1+b^2(D-3))}} \right) \, ,
\end{equation}
and the scale factor is 
\begin{equation}\label{R3}
R(r) = \gamma r^{b^2 \over {1+b^2}} \, ,
\end{equation}
where
$\gamma = \left( {{(D-3)}\over{2k\Lambda(1-b^2)}} \right)^{1/2}
\, $.
The values of $\Lambda$ and $b$ are the same as those given in 
(\ref{lambda2}).

The potential $F(R)$ is
\begin{eqnarray}
F(R) &=& - {{(D-3)b^4} \over {2k(1-b^2)(1+b^2(D-3))}} - M \gamma^2 b^4
\left({R \over \gamma} \right) ^{-\left(D-3+{1\over b^2}\right)} -
\nonumber \\
&&- {{\hat V_0 ^2 e ^{{2\phi_0}\over b} \gamma^2} \over {8(D-2)^2}}
\left({R \over \gamma} \right) ^{-2\left({1\over b^2} -1\right)} \,
.\label{f3}
\end{eqnarray}

If $V_0>0$, $r$ turns out to be a time coordinate, while for $V_0<0$, it 
is a spatial coordinate. From the twelve cases shown in \cite{chre} we 
choose those where it is a spatial coordinate. 
The condition (\ref{condition2}) always applies.

\subsubsection{$V_0<0$, $M>0$, $b^2<{1 \over {(D-1)}}$}
This case describes a topological black hole in AdS space. From
Fig.\ref{T3-branea} we can see the region where (\ref{condition2})
holds.  
There are no shortcuts in this interval.

\subsubsection{$V_0<0$, $M>0$, ${1 \over {(D-1)}}<b^2<1$}
We again have a topological black hole in AdS space. There is a small
interval where (\ref{eq3}) has solution as we can see 
from Fig.\ref{T3-braneb}(a). Our results are shown in
Fig.\ref{T3-braneb}(b). Notice that the domain wall equation of
motion has a solution only inside the interval shown there. This means that
only a group of geodesics with initial velocity $\dot r(0)> v_c$ can
meet the domain wall after a roundabout in the bulk.

\subsubsection{$V_0<0$, $M>0$, $b^2>1$}
The black hole in AdS space appearing here has round spatial
section. In this case $U(r)$ is always positive (then $r$ is always a
spatial coordinate); however, we notice that $F(r) \leq 0$ for $r\geq
3*10^5$. We found that shortcuts
are possible for several initial velocities if $M=0$.

The case $M>0$ is shown in Fig.\ref{T3-braned}. We have two regions of
interest after the event horizon depending only on the sign of $F(r)$
since $U(r)$ is positive in this range. In the first region
all the geodesics initially follow the brane and fall into the event
horizon at later times. In the second region we have shortcuts again
for several initial velocities. 

\subsubsection{$V_0<0$, $M<0$, $b^2<{1 \over {(D-1)}}$}
Here $U(r)$ is always positive while $F(r)$ is negative in the range
shown in Fig.\ref{T3-branee}. The domain wall and the geodesics
diverge after some time near the end of the range where (\ref{eq3})
has a solution. 

\subsubsection{$V_0<0$, $M<0$, ${1 \over {(D-1)}}<b^2<1$}
In this case $U(r)$ is always positive while $F(r)$ is negative for a small 
range as seen in Fig.\ref{T3-branef}. There are several shortcuts in the 
region where the domain wall equation of motion has solution; nevertheless, 
there is a threshold velocity after which the geodesics can not
return. 

\subsubsection{$V_0<0$, $M<0$, $b^2>1$}
Now $U(r)$ is always positive and $F(r)$ will determine the
initial condition for the domain wall equation of motion. Results are
in Fig.\ref{T3-braneg}.

\subsection{Domain Wall Time and Time Delays}

The time delay between the photon traveling on the domain wall and the
gravitons traveling in the bulk \cite{abdacasali} can be calculated as
follows. Since the signals cover the same distance, 
\begin{equation}\label{timedelay}
\int {{d\tau_\gamma}\over {r(\tau_\gamma)}} = \int {{dt_g}\over
{r_g(t_g)}} \sqrt{U(r_g) - {{\dot r_g(t)^2}\over {U(r_g)}}} \, ,
\end{equation}
the difference between photon and graviton time of flight can
approximately be written as
\begin{equation}
{{\Delta\tau}\over {r}} \simeq \int_0 ^{\tau_f+\Delta\tau}
{{d\tau_\gamma}\over {r(\tau_\gamma)}} - \int_0 ^{\tau_f}
{{d\tau_g}\over{r(\tau_g)}} \, , \nonumber
\end{equation}
or in terms of the bulk time
\begin{equation}\label{delay}
\Delta\tau \simeq r(t_f) \int_0 ^{t_f} dt \left({{1}\over {r_g(t)}}
\sqrt{U(r_g)- {{\dot r_g(t)^2}\over {U(r_g)}}} - {1 \over {r(t)}}
{{d\tau}\over{dt}} \right) \, .
\end{equation}
The elapsed bulk time $t$, 
the corresponding domain wall time $\tau$ and the delays (\ref{delay})
can be computed.


\section{A Six-Dimensional Model}

We consider a six-dimensional model, such as the one constructed
by Kanti et. al. \cite{kanti}. We also search for a solution of
six-dimensional Einstein equation in AdS space of the form
\begin{equation}\label{metric6D}
ds^2 = -n^2(t,y,z) dt^2 + a^2(t,y,z) d\Sigma_{k} ^2 + b^2(t,y,z)
\left\{ dy^2 + c^2(t,y,z) dz^2 \right\}
\end{equation}
where $d\Sigma_{k} ^2$ represents the metric of the three dimensional
spatial sections with $k=-1,\,0,\,1$ corresponding to a hyperbolic, a
flat and an elliptic space, respectively.

The total energy-momentum tensor can be decomposed in two parts
corresponding to the bulk and the brane as
\begin{equation}\label{emtensor}
\tilde T^M _N = \breve T^{M(B)} _N + T^{M (b)} _N \, ,
\end{equation}
where the brane contribution can be written as
\begin{equation}\label{brane}
T^{M (b)} _N = { {\delta (z-z_0)} \over {bc}} \,diag \,(-\rho,
p,p,p,\hat p, 0) \, .
\end{equation}

An analogous development of section 3.3 for this six dimensional
model provides
the following Darmois-Israel conditions 
\begin{eqnarray}\label{israeldarmois2}
{{[\partial_z a]} \over {a_0 b_0 c_0}} &=& -{{\kappa_{(6)} ^2} \over 4}
(p-\hat p + \rho) \, ,\nonumber \\
{{[\partial_z b]} \over {b_0 ^2 c_0}} &=& -{{\kappa_{(6)} ^2} \over 4}
\left\{ \rho - 3(p-\hat p) \right\} \, ,\\
{{[\partial_z n]} \over {b_0 c_0 n_0}} &=& {{\kappa_{(6)} ^2} \over 4}
\left\{ \hat p + 3 (p+ \rho) \right\} \, . \nonumber
\end{eqnarray}

A metric of the form (\ref{metric6D}) satisfying six dimensional Einstein
equations is given by
\begin{equation}\label{bhmetric}
ds^2 = - h(z) dt^2 + {z^2 \over l^2} d\Sigma_k ^2 + h^{-1}
(z) dz^2 \, ,
\end{equation}
where here
\begin{equation}\label{spacediff}
d\Sigma_k ^2 = {{dr^2} \over {1-kr^2}} + r^2 d\Omega_{(2)} ^2 +
(1-kr^2) dy^2 \, ,
\end{equation}
and
\begin{eqnarray}
&h(z) = k + {z^2 \over l^2} - {M \over z^3}& \, ,\quad  \hbox{for
AdS-Schwarzschild bulk} \, ,\label{hsch} \\
&h(z) = k + {z^2 \over l^2} - {M \over z^3} + {Q^2 \over z^6}& \, ,\quad
\hbox{for AdS-Reissner-Nordstr\"om bulk} \, , \label{hrn} 
\end{eqnarray}
with $l^{-2} \propto -\Lambda$ ($\Lambda$ being the cosmological
constant), which describes a black hole in the bulk, located at $z=0$.  

Following \cite{csaki3}, we find a further solution by means of a
$Z_2$ symmetry inverting the space with respect to the brane
position. That is, considering a metric of the form 
\begin{equation}\label{genmetric}
ds^2 = -A^2(z) dt^2 + B^2(z) d\Sigma_{(4)} ^2 + C^2(z) dz^2
\end{equation}
and the brane to be defined at $z=z_0$, there is a solution given by
\begin{eqnarray}\label{z2sym}
&A(z), \,\, B(z), \,\, C(z)& \, ,\quad \hbox{for} \quad z\leq z_0 \,
,\nonumber\\ 
&A(z_0 ^2/z), \,B(z_0 ^2/z), \,C(z_0 ^2/z) {{z_0 ^2} \over z^2}& \, ,\quad
\hbox{for} \quad z \geq z_0 \, .
\end{eqnarray} 
The $Z_2$-symmetry corresponds to $z \rightarrow z_0 ^2/z$.

The static brane still has to obey the Darmois-Israel conditions
(\ref{israeldarmois2}), which for the metric (\ref{bhmetric}) are written as
\begin{eqnarray}\label{bhisrael}
{{[\partial_z a]} \over {a_0 ^2 c_0}} &=& -{{\kappa_{(6)} ^2} \over 4}
\rho  \, , \nonumber \\ 
{{[\partial_z n]} \over {a_0 c_0 n_0}} &=& {{\kappa_{(6)} ^2} \over 4}
(4p + 3 \rho) \, , 
\end{eqnarray}
where here 
\begin{equation}\label{a-n}
[\partial_z a] = -{2 \over l} \quad \quad \hbox{and} \quad \quad
{[\partial_z n]} = -{{h'(z_0)} \over {\sqrt{h(z_0)}}} \, .
\end{equation}

\subsection{The Shortest Cut Equation}

We consider a static version of the metric (\ref{metric6D}) with $k=0$
\begin{equation}\label{shortmetric}
ds^2 = -n^2 (z) dt^2 + a^2(z) f^2(r) dr^2 + b^2(z) dy^2 +
d^2(z) dz^2  \, ,
\end{equation}
where the graviton path is defined equating (\ref{shortmetric}) to
zero. Therefore, 
\begin{equation}\label{gravpath}
\int_{r_0} ^r f(r') dr'= \int_{t_0} ^t {{\sqrt{n^2(z) - b^2(z)
      \dot y^2 - d^2(z) \dot z^2}}\over{a(z)}} dt \equiv \int_{t_0}
      ^t {\cal L} \left[y(t), \dot y (t), z(t), \dot z (t) ; t \right] dt
\end{equation}
which naturally defines a lagrangian density. The Euler-Lagrange
equations of ${\cal L}$ define the graviton path. We first choose to
work at a constant $y$ to check on the very possibility of
(\ref{shortmetric}) allowing shortcuts. In this case the resulting
equation is simple but far from trivial,
\begin{equation}\label{zsym}
\ddot z + \left( {{a'}\over a} -2 {{n'}\over n} + {{d'}\over d}
\right) \dot z ^2 + \left( {{n n'}\over{d^2}} - {{a'}\over a} {n^2
    \over d^2} \right) = 0 \, .
\end{equation}


Notice that this case is equivalent to consider the problem in five
dimensions with the metric shown in \cite{csaki3}.

The most general case includes a $y$ dependence on the graviton path;
however, this dependence turns out to be
superflous and does not affect the $z$-equation since (\ref{zsym}) is
independently satisfied. This conclusion is not surprising if we
notice that the metric (\ref{shortmetric}) is $y$-independent.

For $k \not= 0$ cases we can also consider (\ref{zsym}) as the
shortcut equation if we assume the existence of a $y$-symmetry in our
problem. Our model represents a generalization of
\cite{csaki3}. 

\subsection{AdS-Schwarzschild Bulk}
From the Darmois-Israel conditions (\ref{bhisrael}) together with (\ref{a-n})
we have 
\begin{eqnarray}
{h \over {z_0 ^2}} &=& {{\kappa_{(6)} ^4 \rho ^2} \over {64}} \, ,
\label{jumpsch1} \\
{{h'}\over{2z_0}} &=& -{{\kappa_{(6)} ^4 \rho ^2} \over {64}}
(4\omega+3) \, , \label{jumpsch2}
\end{eqnarray}
and we can obtain the black hole mass $M$ as a function of the brane
energy density $\rho$, while $\rho$ is fixed by a {\it fine-tunning},
\begin{eqnarray}
{M\over{z_0 ^5}}&=& {2 \over 5} {k \over {z_0 ^2}} - (\omega +1)
{{\kappa_{(6)} ^4 \rho ^2} \over {40}} \, , \label{6DMQAdSSch1} \\
{{\kappa_{(6)} ^4 \rho ^2} \over {64}}&=& -{{3k}\over{z_0 ^2 (8\omega
    +3)}} - {5 \over {(8\omega +3) l^2}} \, , \label{6DMQAdSSch2}
\end{eqnarray}
where $\omega=p/\rho$.

As we saw in the previous section, the shortcuts in six dimensions are
determined from (\ref{zsym}). 

If a shortcut exists, there must be a time $t=v$ in the graviton path
when $\dot z (v)=0$ and $\ddot z(v) \geq 0$. Thus, (\ref{zsym})
evaluated at this point will give
\begin{equation}\label{zv}
\ddot z (v) + h(z_v) \left( {{h'(z_v)}\over 2} - {{h(z_v)}\over {z_v}}
\right) =0 \, .
\end{equation}

It is obvious that this minimum must be between the brane and the
event horizon $z_h$, if a horizon exists. Otherwise, there is no turning
point in the path since the graviton can not return after it goes
through the event horizon. Hence, $h(z_v) >0$. 
Thus, from (\ref{zv}) we require
\begin{equation}\label{F}
F(z_v) = {{h'(z_v)}\over 2} - {{h(z_v)}\over {z_v}} \leq 0 \quad
\hbox{for} \quad z_{h} < z_v < z_0 \, .
\end{equation}
Using (\ref{hsch}) this implies 
\begin{equation}\label{Fsch}
F(z_v) = {5\over 2} {M\over z_v^4} -{k \over z_v}\leq 0 \, .
\end{equation}
This equation has a zero in $z=z_f \not= 0$ for $k\not= 0$
$$ z_f ^3= {5 \over{2k}} M \, .$$ 
For $k=0, -1$
there is no positive root. Since the mass is positive, $F(z)>0$
everywhere preventing the coexistence of shortcuts and horizons.

On the other hand, for $k=1$ there is one real and positive root,
which must satisfy $z_f< z_0$ in order to have shortcuts. 
Taking into account (\ref{6DMQAdSSch1}) and the fact that
$\varepsilon^2$ must be positive in (\ref{6DMQAdSSch2}), \footnote{From
now on, we will denote $\varepsilon^2=\kappa_{(6)} ^4 \rho^2 /64$ in
six dimensions.}
we have a first restriction for $\omega$,
\begin{equation}\label{schom2}
\omega +1 >0 \, .
\end{equation}

%
%

The positivity of the black hole mass gives a second restriction for
$\omega$, 
\begin{equation}\label{schomega}
-1<\omega < -{3\over 4} \, ,
\end{equation}
and also for the brane position,
\begin{equation}\label{schbpos}
{{z_0 ^2}\over l^2} < - {{\omega + 3/4}\over {\omega +1 }} \, .
\end{equation}

\begin{figure*}[htb!]
\begin{center}
\leavevmode
\begin{eqnarray}
\epsfxsize= 7.0truecm\rotatebox{-90}
{\epsfbox{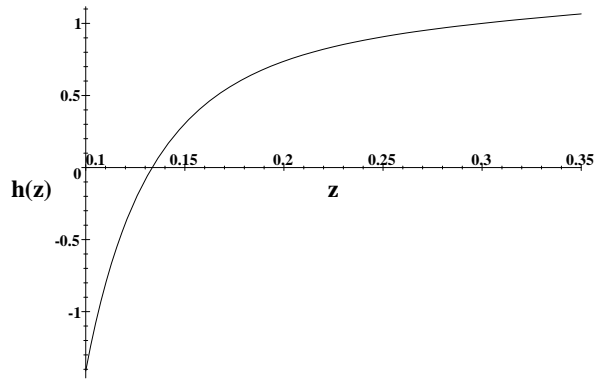}}\nonumber
\end{eqnarray}
\caption{$h(z)$ in six-dimensional AdS-Schwarzschild bulk with the
 brane located at $z=1/3$. Notice that the singularity is shielded by a
 horizon.}  
\label{6DAdSSch1}
\end{center}
\end{figure*}

\begin{figure*}[htb!]
\begin{center}
\leavevmode
\begin{eqnarray}
\epsfxsize= 7.0truecm\rotatebox{-90}
{\epsfbox{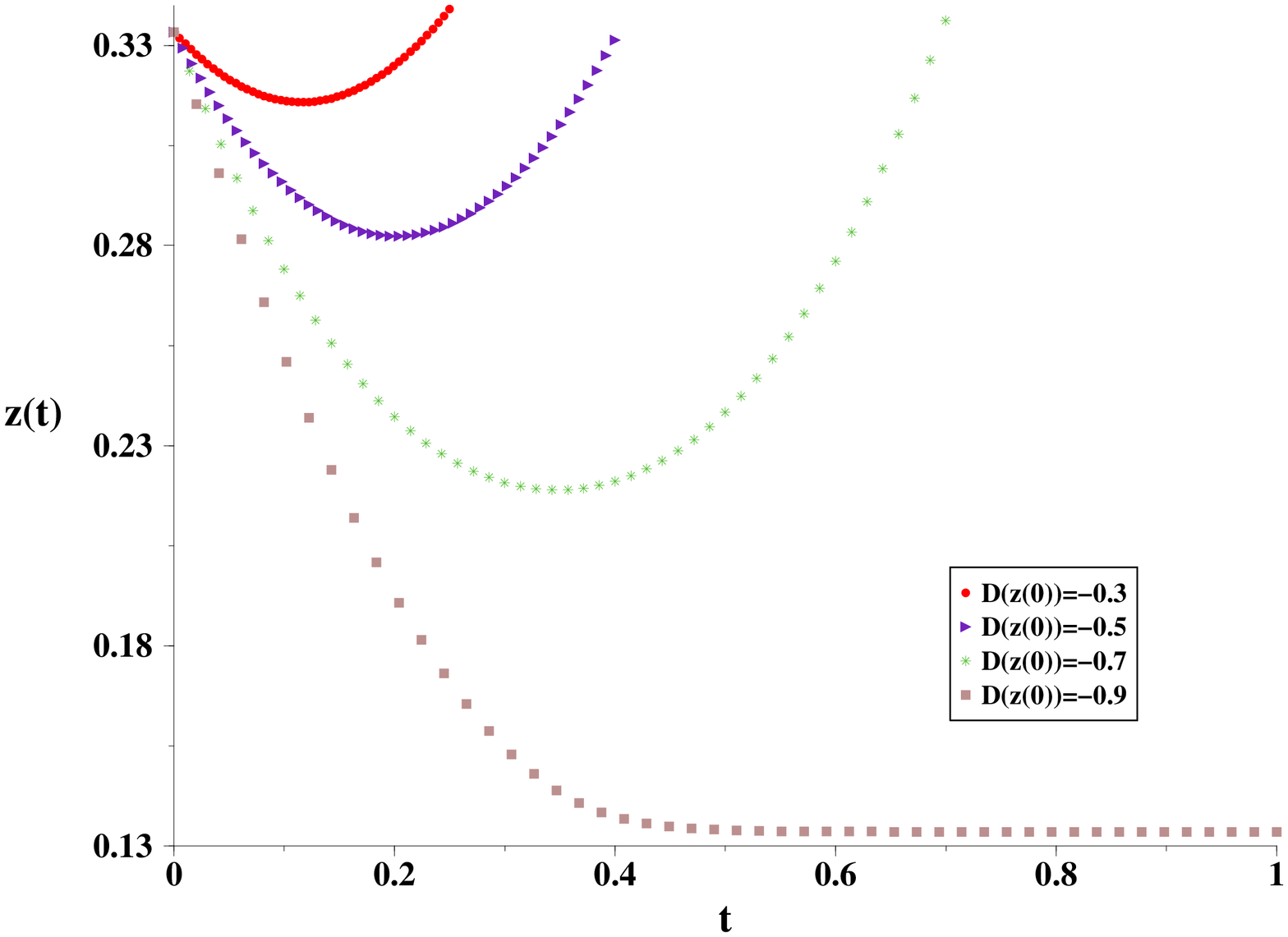}}\nonumber 
\end{eqnarray}
\caption{Shortcuts for several initial velocities in six-dimensional
 AdS-Schwarzschild bulk. Notice that there is a threshold initial velocity for
 which the graviton can not return to the brane and falls into the
 event horizon.}   
\label{6DAdSSch2}
\end{center}
\end{figure*}

If we follow both (\ref{schomega}) and (\ref{schbpos}) together with
the fine-tunning for the energy (\ref{6DMQAdSSch2}), we will have several
shortcuts in AdS-Schwarzschild bulks with shielded singularity.
In Figs. \ref{6DAdSSch1} and \ref{6DAdSSch2} we illustrate an example
with $\omega=-4/5$, $z_0=1/3$, and $l=1$. Notice in Fig.
\ref{6DAdSSch1} that the horizon appears before the brane. 

Since this case is equivalent to consider the problem in five
dimensions with $h(z)$, $M$ and $\rho$ given in \cite{csaki3},
analogous results are obtained \cite{Abdalla2}.

\subsection{AdS-Reissner-Nordstr\"om Bulk}

From the Darmois-Israel conditions (\ref{bhisrael}) we have for the black
hole mass and charge,
\begin{eqnarray}\label{6DMQAdSRN}
{M\over {z_0 ^5}} &=& {{2k}\over{z_0 ^2}} + {8\over{3l^2}} +
{{\kappa_{(6)} ^4} \over {24}} \rho^2 \omega \, , \nonumber \\
{Q^2 \over {z_0 ^8}} &=& {k\over{z_0 ^2}} + {5 \over{3l^2}} +
{{8\omega +3}\over 3} \, {{\kappa_{(6)} ^4 \rho^2} \over {64}} \, . 
\end{eqnarray}

At this stage it is convenient to carefully study the possibility
of existence of shortcuts for every value of $k$. 

As it was found in the AdS-Schwarzschild case, (\ref{F}) determines
the existence of shortcuts.
Using (\ref{hrn}) we see that (\ref{F}) has a zero in $z=z_f\not=0$
when
\begin{equation}\label{zerosF}
{5\over 2} M z_f ^3 -4Q^2 -kz_f ^6=0 \, .
\end{equation}
If $k=0$, we have a real root in
\begin{equation}\label{k0root}
z_f ^3 = {{8Q}\over{5M}} \, .
\end{equation}
If $k=1$, we have two roots in
\begin{equation}\label{k1root}
z_f ^3 = {5\over4}M \pm {1\over 4} \sqrt{25M^2-64Q^2} \, .
\end{equation}
Finally, if $k=-1$, we have
\begin{equation}\label{k-1root}
z_f ^3 = -{5\over4}M \pm {1\over 4} \sqrt{25M^2+64Q^2} \, .
\end{equation}

Notice that $F(z)$ has at most one real and positive zero if $k=0, -1$
and at most two positive zeros if $k=1$.

Analysing $h(z)$ and its derivatives 
%
%
we conclude that for positive mass there is just
one zero for $h'(z)$, and hence, at most two horizons for $h(z)$.

When there is one horizon, $h'(z)$ is negative before it and positive
after, crossing $h(z)$ at the very horizon. If there are two horizons,
$h'(z)$ vanishes at a point between Cauchy and event horizons,
being negative before this point and positive after, while $h(z)$ is
positive at all points except between both horizons. Taking into
account both the sign and zeros of these functions, $h'(z)$ crosses
$h(z)$ between the Cauchy horizon and the point at which $h'(z)$ vanishes.

Since $h'(z)/2$ has the same sign as $h'(z)$ and vanishes at the same
point, and in the same way $h(z)/z$ has the same sign of $h(z)$ and
vanishes at the same points, we conclude that, existing horizons, $F(z)$
necessarily vanishes at some point $z=z_c$ such that $0<z_c\leq z_h$. 
However, as we pointed out before, for $k=0$ or $k=-1$ there is only
one positive root of $F(z)$. As $F(z)<0$ for $z<z_c$, then $F(z)>0$
for $z>z_c$. Thus, because $z_c \leq z_h$, $F(z)>0$ for $z>z_h$
contrary to what was required in (\ref{F}). This implies that
there are no shortcuts with $k=0$ or $k=-1$ when horizons exist. 

In five dimensions the proof is very similar and we arrive to the same
conclusion.

In the case $k=1$ $F(z)$ has two real, positive and
distinct roots,
\begin{eqnarray}
r_1 ^3 = {5\over4}M - {1\over 4} \sqrt{25M^2-64Q^2} \, , \label{root1}\\
r_2 ^3 = {5\over4}M + {1\over 4} \sqrt{25M^2-64Q^2} \, . \label{root2}
\end{eqnarray}
This is the only situation where the shortcuts can coexist with a
shielded singularity. In fact, this situation necessarily requires
the second root of $F(z)$ being at some point before the brane
position $z_0$. This also implies $F(z_0)<0$.

In addition, we must have both $Q^2$ and $M$ positive.

Given the fact that we have horizons, if the brane is not between them
or at a horizon 
position, $h(z_0)>0$. Furthermore, in order to guarantee that the brane
is located after the event horizon, we also need $h'(z_0)>0$.

From the previous discussion we will have one or two
horizons if and only if $h(r_1)\leq 0$.

In summary, shortcuts in bulks with shielded singularities can occur
only if $k=1$ and also if the following conditions are supplied,

\begin{enumerate}
\item $h(z_0)>0$ and $h'(z_0)>0$ to have both horizons before the
brane.

\item $F(z_0)<0$ and $r_2<z_0$ to have shortcuts with shielded
singularity.

\item $Q^2>0$ and $M>0$, which assures the positivity of the black
hole mass and the square of the charge.

\item $h(r_1)\leq 0$ in order to have horizons.
\end{enumerate}

As we have done for the Schwarzchild case, we can  analyze each condition and impose certain restrictions on
$\omega$, $\rho^2$, and $z_0$. In short, by purely analytic considerations we conclude that shortcuts
in bulks having no naked singularities and a static brane embedded in
can only appear if $k=1$ and if the following conditions are
satisfied, \cite{Abdalla2}

\begin{enumerate}
\item We must choose $\omega$ such that $-1<\omega<-3/4$;

\item Given $\omega$, the brane must be located at a position such that
\begin{equation}\label{branepos}
{z_0 \over l} < {1 \over 2} \sqrt{- {{3+4\omega}\over {1+\omega}}} \, ,
\end{equation}
which is the same condition as AdS-Schwarzschild case (\ref{schbpos});

\item Given (\ref{branepos}), the energy $\varepsilon$ must satisfy
\end{enumerate}
\begin{equation}\label{energycond}
{1\over{32\omega^2}} \left( - {{32\omega z_0 ^2}\over l^2} + 9 +3
\sqrt{-64\omega {{z_0 ^2}\over l^2} +9 -64{{z_0 ^2}\over l^2}\omega^2}
\right) < z_0 ^2 \varepsilon^2 < {1\over {8\omega +3}} \left(-3-{{5z_0
^2}\over l^2} \right) \, .
\end{equation}

In this way, it turns out to be simple to find shortcuts in bulks with
shielded singularities.

As an example, let us choose $\omega=-9/10$. From (\ref{branepos}) we
must have $${z_0 \over l} < {\sqrt{6}\over 2} \, ,$$ then we choose $l=1$ and
$z_0=1$. 

From (\ref{energycond}) we have
$$ {{35}\over{24}} + {5\over{72}} \sqrt{41}< \varepsilon^2 < {{40}
\over{21}} \, ,$$ so we choose $\varepsilon = \sqrt{238/125}$.


\begin{figure*}[htb!]
\begin{center}
\leavevmode
\begin{eqnarray}
\epsfxsize= 7.0truecm\rotatebox{-90}
{\epsfbox{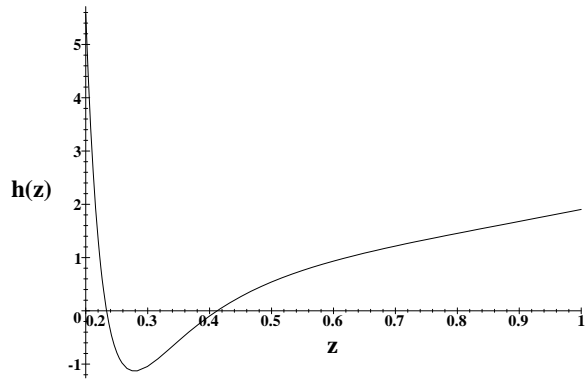}}\nonumber
\end{eqnarray}
\caption{$h(z)$ in six-dimensional AdS-Reissner-Nordstr\"om bulk with
the brane located at 
$z=1$. Notice that the singularity is shielded by two horizons.} 
\label{h(z)}
\end{center}
\end{figure*}
\begin{figure*}[htb!]
\begin{center}
\leavevmode
\begin{eqnarray}
\epsfxsize= 7.0truecm\rotatebox{-90}
{\epsfbox{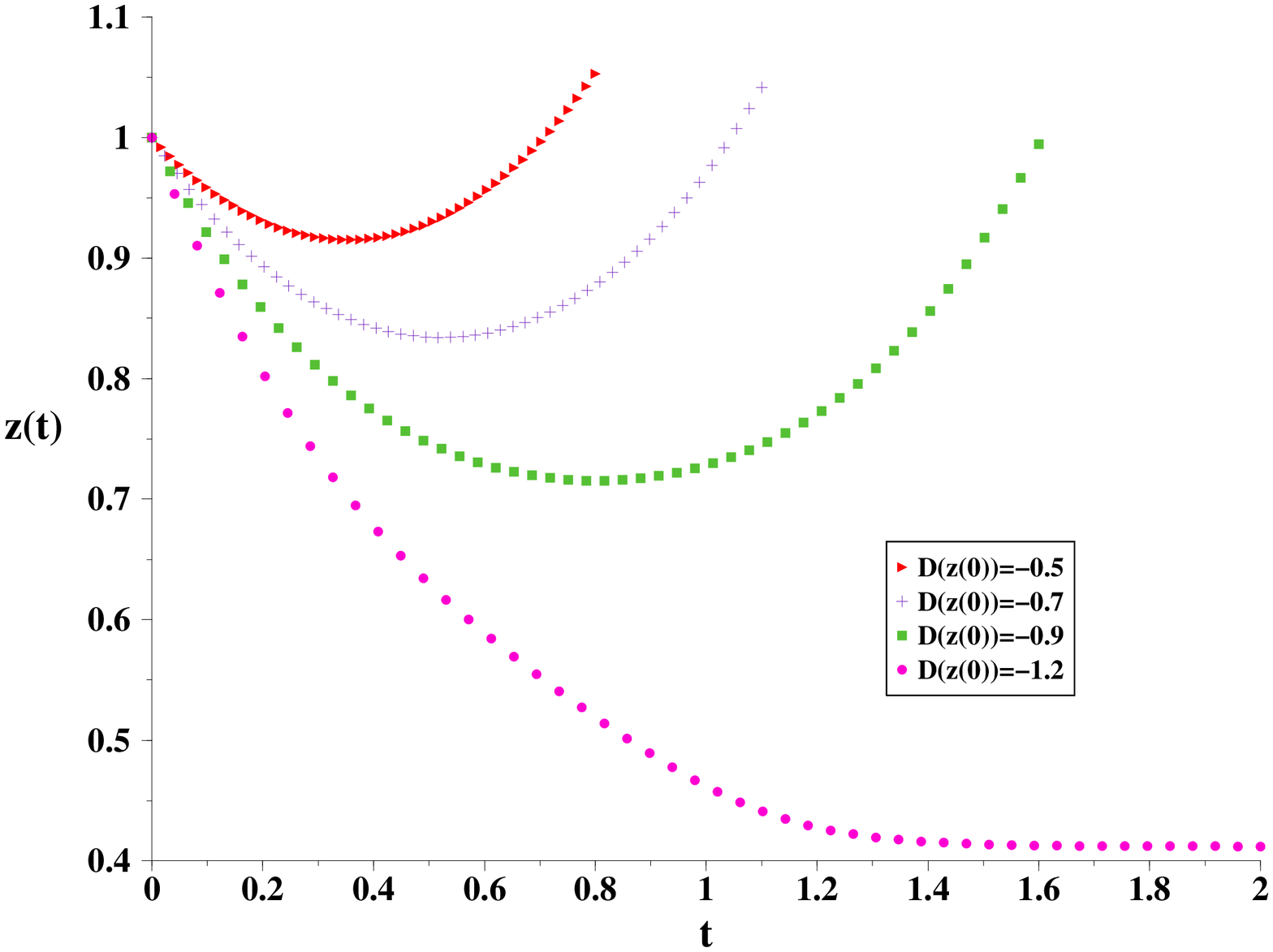}}\nonumber
\end{eqnarray}
\caption{Shortcuts for several initial velocities in six-dimensional
AdS-Reissner-Nordstr\"om bulk. Notice that there is threshold initial
velocity for which the graviton
can not return to the brane and falls into the event horizon.}  
\label{6Dsc}
\end{center}
\end{figure*}

In figure (\ref{h(z)}) we plot $h(z)$ with these
conditions. Notice that the singularity is protected by an event horizon
and the brane is at $z=z_0=1$. 

In figure (\ref{6Dsc}) we plot the graviton paths obtained from
(\ref{zsym}) under the previous conditions for a variety of initial
velocities showing that, in fact, shortcuts appear when we choose the
parameters following the complete analysis shown in this section.


The analysis in five dimensions can be performed analogously.


\section{The FRW-Brane evolving in the Bulk}

The main reason to study the evolution of the brane from the
point of view of the bulk is to simplify the analysis of gravitational
signs leaving and subsequently returning to the brane. In fact in the
static AdS backgound (\ref{bulkmetric}), the FRW brane
(\ref{branemetric}) has a 
particularly simplified  equation for a null geodesic in the bulk,
$a=a(t)$ \cite{Abdalla2},
\be
\frac{\ddot{a}(t)}{a} + \frac{\dot{a}^{2}(t)}{a^{2}}\Bigl(1 -
\frac{3h'(a)a}{2h(a)}\Bigr) + \frac{h(a)}{a}\Bigl(\frac{h'(a)}{2} -
\frac{h}{a}\Bigr)=0\quad. 
\label{nullgeodesic}
\ee

On the other hand, the evolution of the brane in the bulk,
$a=a_{b}(t)$,  at early times, $a_{b}<<L_{c}$, is dictated by
(\ref{timetime}), 
\[
 \frac{d a_{b}}{d t} =  \frac{h(a_{b})}{\sqrt{1 +
 \frac{h(a_{b})}{\dot{a}_{b}(\tau)^2}}} = h(a_{b})\Bigl( 1 + \frac{k +
 \frac{a_{b}^{2}}{l^{2}}}{\frac{\Omega_{0}^2}{4(1+l^2\Lambda_4)}
\frac{L_c^{2q}}{l^2a_{b}^{2q-2}}}\Bigr)^{-1/2}\quad ,
\]
where we used the fact that the quadratic term in the energy prevails. Thus,
\[
 \frac{d a_{b}}{d t} = h(a_{b})\Bigl( 1 +
 \frac{4(1+l^2\Lambda_4)}{\Omega_{0}^2}\frac{a_{b}^{2q}}{L_{c}^{2q}}(1 +
 kl^2/(a_b^2))\Bigr)^{-1/2}\quad ,
\]
and since the observed cosmological constant is
at most $\Lambda_4 \sim H_0^{2}$, $\Lambda_4l^2<<1$ we have for
$a_{b}<<L_{c}$, 
\be
 \frac{d a_{b}}{d t} \approx h(a_{b})\Bigl( 1 -
 \frac{2(1+l^2\Lambda_4)}{\Omega_{0}^2}\frac{a_{b}^{2q}}{L_{c}^{2q}}(1 + 
kl^2/(a_b^2))\Bigr)\quad . \label{primitivo}
\ee

Substituting the result for the evolution of the brane, $\dot{a}(t) =
h(a)$, in the geodesic equation (\ref{nullgeodesic}), 
we verify that it is satisfied. Therefore,  the trajectory of the brane
differs from the trajectory of the null geodesic by a term of the order
$\Bigl(\frac{a_{b}}{L_{c}}\Bigr)^{2q}$. 

This means that for $a_{b}<<L_{c}$, the trajectory of the brane in the
bulk is governed by
\be
\frac{da_{b}(t)}{dt} = k + \frac{a_{b}^{2}}{l^{2}}\quad . \label{branageo}
\ee

Thus, if $k=0$,
\be
a_b(t) = \frac{a_b(0)l^{2}}{l^{2} - a_b(0)t}
\label{abtran}
\ee
and, if $k=-1$,
\be
a_b(t) = \frac{2l(l+a_b(0))}{e^{2t/l}(l-a_b(0)) + l + a_b(0)} - l\quad.
\ee

In this last situation, if the Universe begins under the
Randall-Sundrum scale, $a(0)<l$, it will recolapse to the singularity in a
finite time,
\[
t = \frac{l}{2}\ln \Bigl(\frac{l+a_b(0)}{l-a_b(0)}\Bigr)\quad.
\] 
There is an event horizon when $a=l$ if $k=-1$.

In the case of an elliptic Universe,
\be
a_{b}(t) = l\tan\Bigl(\frac{t}{l} +
\tan^{-1}\Bigl(\frac{a_b(0)}{l}\Bigr)\Bigr)\quad . 
\ee
Starting at the singularity
$t_{0}=\tau_{0}=a_{b}(t_0)=0$,  we have
\be
a_{b}(t) = l\tan\Bigl(\frac{t}{l}\Bigr)\quad .
\ee 
Note that the evolution of the brane in the bulk is linear near the
initial singularity ($a(t)\sim t$ for $t<<l$), diverging at the critical
time $t_{c} = \frac{\pi}{2}l$. In fact, the behaviour of all solutions is
similar near the critical time
\bear
t_{c} &=& \frac{l^2}{a_b(0)}\quad,\nonumber\\
t_{c} &=& \frac{l}{2}\ln
\Bigl(\frac{l+a_b(0)}{a_b(0)-l}\Bigr)\quad,\nonumber \\
t_{c} &=& \frac{\pi}{2}l - l\tan^{-1}\Bigl(\frac{a_b(0)}{l}\Bigr)
\quad ,
\eear
for $k=0,\: -1,\: +1$ respectively.

As we approach the critical time,
$a_{b}(t)$ increases quickly. When $a_{b}(t)\sim L_{c}$, equation
(\ref{branageo}) is no longer valid, and the trajectory of the brane is no
longer a geodesic. Thus, for a very short period, from the point of view
of the bulk the brane undergoes a phase transition. 
Before the critical time, from the point of view of the bulk, there
is no time left for the remaining graviton geodesics to reach the
brane. 

\begin{figure*}[htb!]
\vspace{-2cm}
\begin{center}
\leavevmode
\begin{eqnarray}
\epsfxsize= 9.0truecm \rotatebox{-90}
{\epsfbox{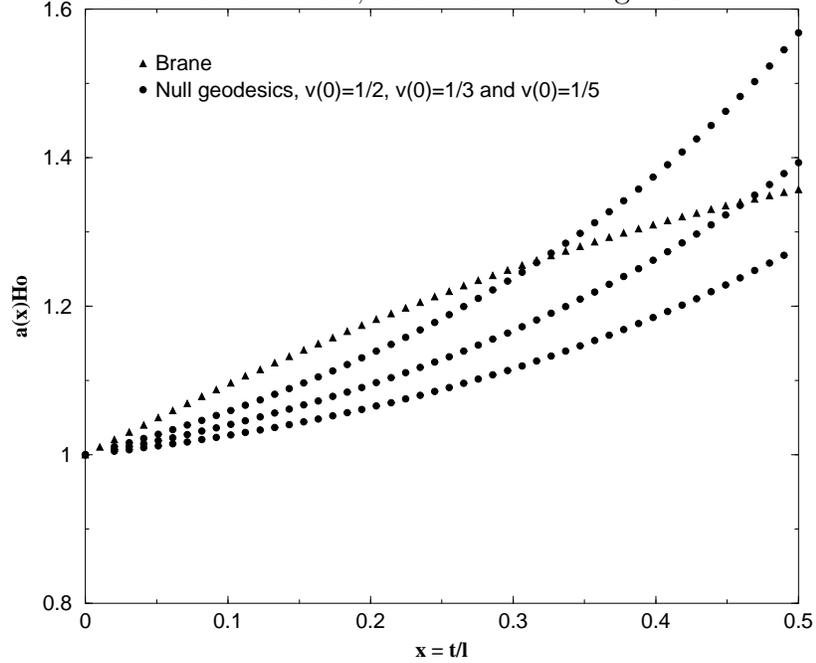}}\nonumber
\end{eqnarray}
\caption{The trajectory of the brane in the bulk for $\Omega_0=2$, matter 
dominated era (MDE) with $k=1$. Also null geodesics starting in
the brane with various initial velocities.}
\label{atual}
\end{center}
\end{figure*}

For later times the evolution of the brane is softer,
and shortcuts should appear. In fact, the numerical solutions of the
brane evolution equation (\ref{timetime}) and the null geodesics
equation (\ref{nullgeodesic}) in the bulk
indicate the presence of shortcuts in late times universes,  as
exhibited in Fig. \ref{atual}.



\subsection{The Effect of Shortcuts in Late Times Universes}

The shortcuts just found for late times Universes could be used to probe
the extra-dimensionality by the aparent violation of causality on the
brane. Thus, suppose that in $\tau=t=0$ an object could emit eletromagnetic and
gravitational waves and that we could be able to detect both signs in
times $\tau'_{\gamma}$ and $\tau'_{g}$ respectively. 
We compute now the order of magnitude of the advance in time of the
graviton. Since the signals cover the same distance on the brane,
\be
\int_{0}^{\tau'_{\gamma}} \frac{d\tau_{\gamma}}{a_{b}(\tau_{\gamma})} =
\int_{0}^{t'_{g}}\frac{dt_{g}}{a(t_{g})} 
\sqrt{h(a) - \frac{\dot{a}(t_g)^{2}}{h(a)}}\quad . \label{intgeo}
\ee
Here  $a$ denotes the coordinate defining the geodesic in the bulk, and
differs from the coordinate of the brane $a_b$. In terms of the
dimensionless parameters $y$ and $x$, $a=Ly$ and $t=Tx$ the last integral
reads 
\bear
\int_{0}^{t'_{g}} \frac{dt_{g}}{a(t_{g})}\sqrt{h(a) -
\frac{\dot{a}(t)^{2}}{h(a)}} &=& \int_{0}^{t'_{g}} \frac{dt_{g}}{l}\sqrt{1
+ \frac{l^{2}}{L^{2}y^{2}}  - 
\frac{l^{2}}{T^2}\frac{\dot{y}(x)^{2}}{y^{2} +
  y^{4}\frac{L^{2}}{l^{2}}}}\quad . 
\eear
Using the relation between the time on the brane and that in the bulk we
express the above expression in terms of the time interval of the
observer on the brane, and together with the Friedmann equations we get
%
\bearst
&&\int_{0}^{\tau'_{g}}
\frac{d\tau_{g}}{l}\frac{1}{h(a_{b})}\sqrt{\frac{y_{b}^{2}L^{2}}{l^{2}}+ 
\Lambda_4L^2y_b^2 +
\frac{\Omega_{0}}{y_{b}^{q-2}}\frac{L_{c}^{q}}{l^2L^{q-2}}}\sqrt{1 +
\frac{l^{2}}{L^{2}y^{2}}  - \frac{l^{2}}{T^2}\frac{\dot{y}(x)^{2}}{y^{2} +
y^{4}\frac{L^{2}}{l^{2}}}}  \\ 
&&\quad =\int_{0}^{\tau'_{g}} \frac{d\tau_{g}}{a_{b}(\tau_{g})}\frac{1}{1  +
\frac{l^{2}}{L^{2}y_{b}^{2}}}\sqrt{1 +l^2\Lambda_4 +
\frac{\Omega_{0}}{y_{b}^{q}}\frac{L_{c}^{q}}{L^{q}}}\sqrt{1 +
\frac{l^{2}}{L^{2}y^{2}}  - \frac{l^{2}}{T^2}\frac{\dot{y}(x)^{2}}{y^{2} +
y^{4}\frac{L^{2}}{l^{2}}}}\quad .
\eearst

If $L>>L_{c}>>l$, we obtain, at second order in  $L_{c}/L$
and $l/L$,
\bearst
\int_{0}^{\tau'_{\gamma}} \frac{d\tau_{\gamma}}{a_{b}(\tau_{\gamma})} &=& 
\int_{0}^{\tau'_{g}} \frac{d\tau_{g}}{a_{b}(\tau_{g})}\Bigr[1  +
\frac{1}{2}\frac{\Omega_{0}} 
{y_{b}^{q}}\frac{L_{c}^{q}}{L^{q}}  + \frac{1}{2}l^2\Lambda_4-
\frac{l^{2}}{L^{2}} 
\frac{1}{y_{b}^{2}} \nonumber \\ &+& \frac{1}{2}\frac{l^{2}}{L^{2}y^{2}}  -
\frac{1}{2}\frac{l^{4}}{T^2L^2}\frac{\dot{y}(x)^{2}}{y^{4}}\Bigl]\quad .
\eearst

Thus, at first order, the time difference between the photon and the
graviton is corrected in the integrand by terms of order $
\frac{L_{c}^{q}}{a_{b}^{q}}$. Today this factor is at most $10^{-58}$ and
in the time of decoupling  $10^{-46}$, showing that
the time advance of the graviton can
be safely neglected and is of no physical significance, in spite of the 
fact that the trajectory of the brane is distinctively different from the 
null geodesic.

\subsection{The Effect of Shortcuts in The Early Universe}

From the analysis developed so far we learned that the periods of
evolution of the Universe differ by the scale that defines
the physical significance of the shortcuts. When $a_{b}<<L_{c}$, the
trajectory of the brane in the bulk differs from the extreme geodesics by
$(a_{b}/L_{c})^{2q}$ and the shortcuts do not appear since the brane itself
provides the graviton geodesic.

In the period when $a_{b}>>L_{c}$, the trajectory of the brane is far from a
null geodesic and shortcuts appear, but they are not significant since the
{\it skin depth} of the graviton in the bulk is defined by the parameter
$l<<L_{c}<<a_{b}$. The difference between the time intervals of the photon and
the graviton is of the order $(L_{c}/a_{b})^{q}$.

\begin{figure*}[htb!]
\vspace{-1cm}
\begin{center}
\leavevmode
\begin{eqnarray}
\epsfxsize= 10.0truecm \rotatebox{-90}
{\epsfbox{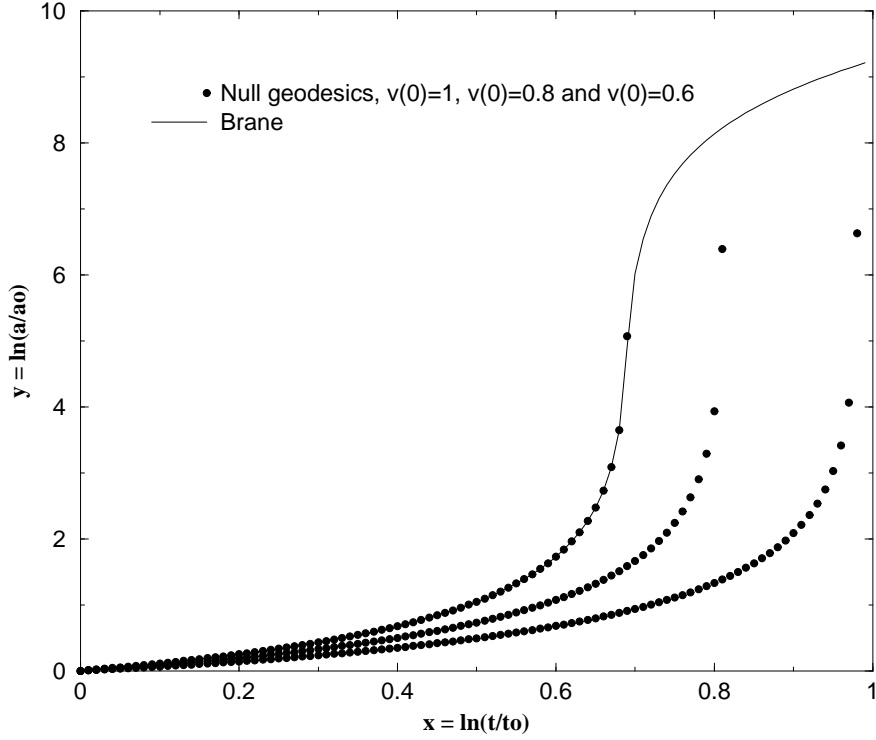}}\nonumber
\end{eqnarray}
\caption{The trajectory of the brane in the bulk leaving
$a=0.01 L_c$ in the   RDE. Geodesics in the bulk with different initial
velocities are exhibited intercepting the brane after the critical time
$t_c$.} 
\label{horizonte}
\end{center}
\end{figure*}

However, from the continuity of the evolution of the brane in the
bulk, we expect that there is also an intermediate situation
$a_{b}\sim L_{c}$
 when physically important shortcuts could appear since the evolution of
the brane is far enough from a geodesic. Indeed, in Fig.
\ref{horizonte},  rescaling the geodesics in the bulk we can observe the 
behaviour of the brane as compared to the geodesics that start on it at a
later time. It is clear that these shortcuts are
serious mediators of homogenization of the matter on the brane in the era
before nucleosynthesis \cite{Ishihara}, \cite{Caldwell}.

From the evolution of the brane in the bulk in the intermediate epoch we
thus conclude that there is a critical age  $t_{c}$, after which the 
gravitational waves leaving the brane return before the arrival of the
photons released at the same time as the gravitons. The behaviour of
the geodesic in the bulk shows that any geodesic starting on the brane at
a certain instant will be singular at a time later than the critical time,
indicating that it will return to the brane. Thus, information leaks
between regions which apparently are causally disconnected at times $t>t_{c}$.

In order to study the horizon problem we now compare, at a certain time
$\tau_{n}$ previous to 
nucleosynthesis, the graviton horizon $R_{g}$ with the
observable proper distance of the universe (from the radiation
decoupling
until today) \cite{Chung}. Since the graviton evolves in a bulk geodesic,
\bear
R_{g} &\equiv& \int \frac{dr}{\sqrt{1-kr^{2}}}  = \int \frac{dt}{a}
\sqrt{h(a) - \frac{\dot{a}^{2}}{h(a)}} = \int
\frac{d\tau}{a}\frac{\sqrt{h(a_{b}) + \dot{a}_{b}^{2}}}{h(a_{b})}
\sqrt{h(a) - 
\frac{\dot{a}^{2}}{h(a)}}\quad \nonumber \\  &=&  
\int_0^{\tau_{n}} \frac{d\tau}{a_b}\frac{1}{1 + l^{2}/a_b^2}\sqrt{1 +
\Lambda_4 l^2 + \frac{\Omega_0L_c^q}{a_b^q}\Bigl(1 +
\frac{\Omega_0L_c^q}{4a_b^q(1+\Lambda_4l^2)}\Bigr)}\times\nonumber\\
&&\times\sqrt{1 + \frac{l^{2}}{L^{2}y^{2}} -
  \frac{l^{2}}{T^2}\frac{\dot{y}(x)^{2}} 
{y^{2} + y^{4}\frac{L^{2}}{l^{2}}}} \quad ,\label{graviton1}
\eear
while on the brane, the size of the observable Universe is
$R =  \int \frac{d\tau}{a_b(\tau)}$.

We use the known results for the usual particle horizon
\[
R  = \frac{1}{\sqrt{\Omega_0}H_0a_{b0}}\int_{z(\tau)}^{z(0)}z^{-q/2}dz
=\frac{2}{(q-2)\sqrt{\Omega_0}H_0a_{b0}} (z(\tau)^{1-q/2} -
z(0)^{1-q/2})\quad , 
\]
where $z(\tau) = a_{b0}/a_{b}(\tau)$. Today, 
\[
R \sim \frac{2}{\sqrt{\Omega_0}H_0a_{b0}} \equiv R_0\quad.
\]

For the computation of the graviton horizon we work in a primordial era
before nucleosynthesis. Thus, in the Friedmann equation we can neglect the
usual cosmological term as well as the curvature term. At an epoch between
Planck era and nucleosynthesis, 
$l<< a_b << H_0$, it is  safe to neglect terms involving  $l^2/L^2$ in
(\ref{graviton1}), and we find
\[
R_g \approx 
\int_0^{\tau_{n}} \frac{d\tau}{a_b}\sqrt{1 + \frac{\Omega_0L_c^q}{a_b^q} +
\frac{\Omega_0^2L_c^{2q}}{4a_b^{2q}}}\quad .
\]

Using the Friedmann equation we get
\be
R_g = 
\frac{1}{\Omega_{0}^{1/2}H_{0}a_{b0}}\int_{z(\tau_n)}^{z(0)}
\frac{dz}{z^2} \frac{\Bigr(1 + \frac{\Omega_0H_0^2l^2}{2}z^4\Bigl)}{\sqrt{
    1 +   \frac{\Omega_0 H_0^2l^2}{4}z^4}}\quad.
\ee

The integral diverges for arbitrarly high redshifts, proving that the
horizon problem is potentially solvable. The behaviour of this integral
can be determined,
\[
R_g \approx 
\frac{l}{a_{b0}}z(0)\quad.
\]

Comparing with the size of the Universe today we have
\[
\frac{R_g}{R_0}  \sim \frac{\sqrt{\Omega_0}}{2}H_0l z(0)\quad.
\]

It looks like shortcuts are not enough to solve the horizon problem since
we would need to go back in time to $z(0) \sim (H_0l)^{-1} \sim
10^{29}$, $10^{11}$ times higher than the redshift at the Planck time
on the brane associated with the fundamental scale of gravity,
$\kappa_{5}$.

We may note, however, that there
are actually two related time scales.  In the primordial  Universe 
the brane is evolving as a part of the bulk, with velocities  close to 
that of light, and time intervals  on the brane correspond to  much 
longer intervals in the bulk. In fact, the relation between these  
scales is obtained from $t_{0}=\tau_{0}=a_{b}(\tau_0)=0$ and
\[
t  = \int_{0}^{\tau}d\tau
\frac{\sqrt{h(a_{b})+\dot{a}_{b}^{2}(\tau)}}{h(a_{b})}\quad . 
\]

For $a_{b}<<L_{c}$ we have $\tau<<l$, and one finds as a consequence
\[
t  \approx \int_{0}^{\tau}d\tau \frac{\dot{a}_{b}(\tau)}{h(a_{b})} =
\int_{0}^{a_{b}(\tau)}\frac{da_{b}}{k+a_{b}^{2}/l^{2}}\quad .
\]
This implies for the example of a closed universe with $\Omega_0=2$,
\[
t = l\arctan\Bigl(\frac{a_{b}(\tau)}{l}\Bigr)\quad .
\]

Therefore, when $a_{b}(\tau)\sim l$, that is, when the brane time is 
$\tau \sim 10^{-67}s$, the corresponding bulk time $t\sim \frac{\pi}{4}l \sim
10^{-11}$s is much larger than the Planck scale, $t >> M_{(5)}^{-1}$.

If we assume that quantization is mandatory according to the bulk Planck
energy scale, geodesics that start in the bulk with $z(0)\sim (H_0l)^{-1}$
should be sufficient to homogenize the Universe before nucleosynthesis
reaching  $R_{g}\sim 1$. In order to verify this, let us note that we
can  study analytically 
the whole causal structure of the gravitational signs for a $k=0$ Universe.

The geodesic equation, (\ref{nullgeodesic}), is quite simple for
purely AdS $k=0$ space-times
\be
\frac{1}{a^{2}}\frac{d a}{dt}  = \frac{v(0)}{a(0)^{2}} \quad.
\ee

Thus, a geodesic that starts on the brane at $a=a(0)$ and $t=0$, with
initial velocity $v(0)$, returns to it when
\[
t_r \approx \frac{a(0)}{v(0)} \quad.
\]

The expression for the gravitational horizon can also be integrated
\bear
R_{g} &=& \int_{0}^{t_{r}}\frac{dt}{a}\sqrt{h(a) - \frac{\dot{a}^{2}}{h(a)}}
= \int_{0}^{t_{r}}\frac{dt}{l}\sqrt{1 - \frac{l^{4}\dot{a}^{2}}{a^{4}}}
\nonumber \\ &=& \frac{t_{r}}{l}\sqrt{1 - \frac{v(0)^{2}l^4}{a(0)^{4}}}\quad.
\eear
Using the relation between the returning time and the initial velocity,
\bear
R_{g} = \frac{t_{r}}{l}\sqrt{1 - \frac{l^{4}}{a(0)^{2}t_r^{2}}}\quad.
\eear

In order to relate the returning time  to  the
redshift,  we must integrate the relation between time of the
bulk and time of the brane (\ref{timetime}). We already know that
from $a(0)$ to the critical period of transition a time of $t_c\sim
\frac{l^{2}}{a(0)}$ has passed. After that, the evolution is dominated by
the usual term in Friedmann equation and we can use
\bear
t_{r} &\approx& \frac{l^2}{a(0)} + \int_{L_{c}}^{a_r}l\frac{da_b}{a_b^{2}H}
\approx \frac{l^2}{a(0)} +
\int_{L_{c}}^{a_r}da_b\frac{l}{\sqrt{\Omega_0}H_{0}}a_{b}^{q/2-2}
\nonumber \\ &\approx& \frac{l^2}{a(0)} +
\frac{2l}{(q-2)\sqrt{\Omega_0}H_{0}a_{b0}}\Bigl(z_{r}^{-q/2+1} -
10^{-15}\Bigr)\quad, 
\eear
where $z_r$ must, of course, be greater than the redshift in the
transition, $z_{Lc}\sim 10^{15}$. 

Substituting back in the expression of the gravitational horizon,
\bear
R_{g} &=& \Bigl[\frac{l}{a(0)} +
\frac{2}{(q-2)\sqrt{\Omega_0}H_{0}a_{b0}}\Bigl(z_{r}^{-q/2+1} -
10^{-15}\Bigr)\Bigr]\nonumber\\ &\times& \sqrt{1 -  \Bigl[1 +
  \frac{2a(0)}{(q-2)\sqrt{\Omega_0}lH_{0}a_{b0}}\Bigl(z_{r}^{-q/2+1} -
  10^{-15}\Bigr)\Bigr]^{-2} }\quad. 
\eear

When considering the interesting situation of a high initial redshift $z(0) =
a_{b0}/a(0) > (H_0l)^{-1}$, this expression can be approximated by
\bear
R_{g} \approx \frac{l}{a(0)}\sqrt{\frac{4a(0)}{(q-2)\sqrt{\Omega_0}lH_{0}a_{b0}}\Bigl(z_{r}^{-q/2+1} - 10^{-15}\Bigr) }\quad.
\eear

Comparing with the size of the horizon today we find
\bear
\Bigr(\frac{R_{g}}{R_{0}}\Bigl)_{z_r} \approx
\sqrt{\sqrt{\Omega_0}\frac{lH_0}{(q-2)}}\sqrt{z(0)\Bigl(z_{r}^{-q/2+1} -
  10^{-15}\Bigr)} \quad. 
\label{completehorizon}
\eear

Thus, as we have previously noted, if sufficiently large
redshifts were available, the graviton horizon in a past epoch could be
larger than the present size of the observable Universe.

We argue, however, that those high redshifts could be
available. Indeed, if inflation takes place on the brane, high
redshifts could be present in the beginning of the
inflationary epoch.

 Denoting the redshift when
inflation  ends by $z_e$, if the size of the present Universe, $R_0$, is
expected to be in causal contact during inflation, we must reach at
least a redshift in the beginning of inflation, $z(0)$, that solves
\[
\frac{a_{b0}R_{0}}{z(0)} = H^{-1}(z_{e}) \quad.
\]

The unusual results in brane-world cosmology are expected if inflation
ends before the transition time, when the quadratic term in Friedmann
equation dominates. In this case  we get
\be\label{result}
\frac{a_{b0}R_{0}}{z(0)} = \frac{1}{\Omega_{0}H_{0}^{2}l z_e^{4}}
\quad \quad \hbox{and} \quad \quad
z(0)  =  2\sqrt{\Omega_{0}}H_{0}l z_e^{4}\quad. 
\ee

If $H(z_e)l>>1$, it is simple to
note, from equation (\ref{timetime}), that the evolution of the brane
in the bulk is not altered during inflation. Thus, we can substitute the
result(\ref{result}) in the complete expression for the causal 
gravitational horizon (\ref{completehorizon}),
\bear
\Bigr(\frac{R_{g}}{R_{0}}\Bigl)_{z_r} \approx
\sqrt{\frac{2\Omega_0}{(q-2)}} H_{0}l z_e^{2}\sqrt{\Bigl(z_{r}^{-q/2 + 1}
  - 10^{-15}\Bigr)} \quad. 
\eear

This equation tells us that  in the time of
nucleosyntesis, when $z_r\sim 10^{10}$,  $R_{g}/R_{0}\sim 1$ for a model
with inflation ending just 
before what would be the Planck era with $z_e \sim 10^{17}$ (where
$z_{Pl} \sim 10^{18}$). This proves that successfully inflationary
models ending before the transition time necessarily  make relevant
changes in the causal structure of the universe. 

\begin{figure*}[htb!]
\vspace{-1cm}
\begin{center}
\leavevmode
\begin{eqnarray}
\epsfxsize= 10.0truecm \rotatebox{-90}
{\epsfbox{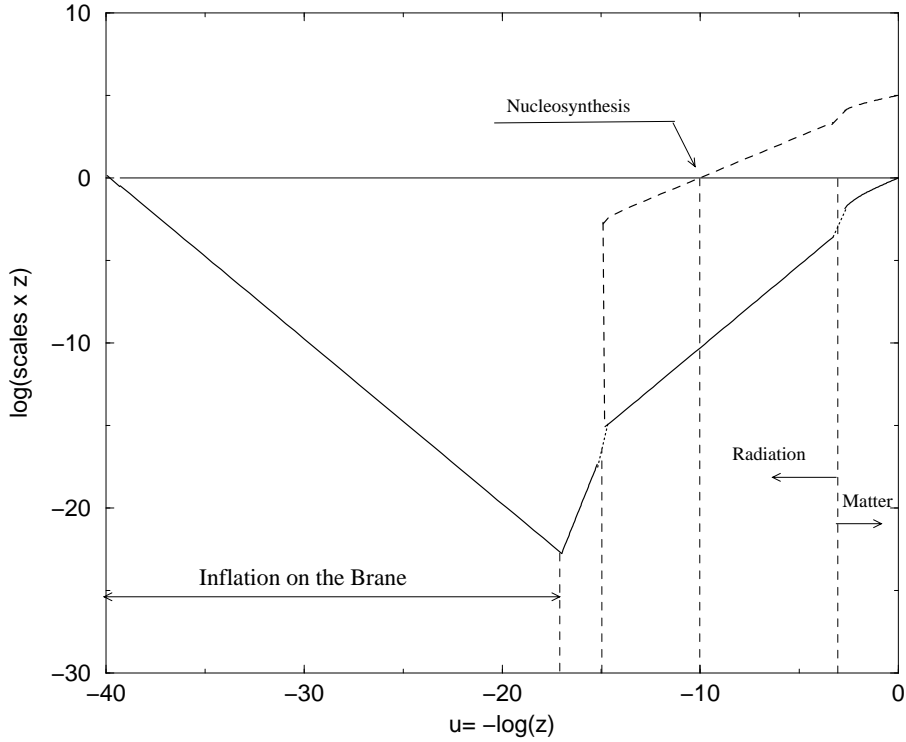}}\nonumber
\end{eqnarray}
\caption{We plot in log-log scale the evolution of the fraction of
graviton horizon and the present observable size,
$R_{g}/R_0$ (dashed line) and the same for the Hubble horizon
$a_{b}^{-1}H^{-1}/R_{0}$ (solid line), for an inflationary model in the
brane that ends after Planck 
time ($u_{PL}=-\log(z_{PL}) = -18$),  and before transition time
($u_{TR} = -15$),  in $u_{end} = -\log(z_{end}) = -17$. The present
scale would be under the de-Sitter horizon if redshifts like
$u(0)=-39$ were available. This imply in strong modifications of the
causal structure for gravitational signs after the transition time.}
\label{inflation}
\end{center}
\end{figure*}

We are able to sketch the gravitational horizon for this kind of
configuration. In Fig. \ref{inflation} we show the behaviour of the
Hubble horizon in comoving coordinates $H^{-1}a_{b}^{-1}/R_{0}$ and the
graviton horizon $R_g/R_0$ for an 
inflationary model ending just before the Planck era, $z_e=10^{17}$ and
producing the necessary number of e-folds to solve the horizon
problem. The fraction of the Universe in causal
contact by gravitational signs in the nucleosynthesis epoch is just
the present horizon $R_{0}$. Today, the gravitational horizon would be
$10^{5}R_0$.

\section{Conclusions}

We have shown that gravitational shortcuts in three  braneworld models are
common and there are many consequences.

In the brane wall
model, where the Universe is replaced
by a domain wall, we have proved that shortcuts may exist and above
all abundant, which is a necessary condition in order to solve the
homogeneity problem. The model shows interesting results as 
the delay of the time of flight inside the brane that can be comparable
with the time of flight of the graviton itself. This lends further
support for a thermalization via graviton exchange through the extra
dimensions.

In the brane static models, the AdS-Schwarzschild and
AdS-Reissner-Nordstr\"om bulks also
open up the possibility of having 
shortcuts provided both the spatial section has positive curvature and a
set of strong restrictions on the brane intrinsic tension must be satisfied.
Moreover, its location in the bulk has to be respected. 
It is interesting to notice that despite the fact that the charge
contributes 
to have a negative $F(z_0)$ and thus facilitates the existence of
shortcuts, there are more restrictive conditions for the energy coming
from $Q^2>0$ and from the horizons equation  which do not
appear in the uncharged case. In this way, the results favor the existence of
shortcuts in bulks with shielded singularities with the same conditions for
$\omega$ and $z_0$ as the AdS-Schwarzschild case and also impose what
is basically a fine-tunning in the energy that already exists in the
uncharged model directly from the junction conditions.

Finally, studying the shortcut problem in the cosmological  braneworld model  from the point of
view of the bulk, we have explicitly shown that shortcuts are indeed
common in late time Universes, though they are extremelly small and the
time advance of the graviton can be safely neglected. However, we have
also learned that gravitational signs may  leave and 
subsequently return to the brane even in early universes. We
have shown that those shortcuts exist  and that the new  scale of the
model, $l$, implies in a minimum time scale for the reception of those
signs by an observer on the brane. Before that critical time the brane
itself evolves like a null-geodesic in the bulk. If high initial
redshifts were available the shortcuts just found could 
solve the horizon problem without inflation. More important
however,  may be the
effect of those shortcuts in an inflationary epoch on the brane.
   
Braneworld  models incorporate two changes in the cosmology, namely, the
modified Friedmann equation and the possibility of leaking of gravity in
the extra dimension. Using the first of those modifications it was shown
that, remarkably, the consistency equation is mantained in the brane-world
formalism when the inflation is guided by a scalar field minimally coupled
on the brane \cite{inflation}. This consists in bad news for those who
expect that brane cosmological configurations could probe the extra
dimensionality of our Universe.

However, we have shown that with an
inflationary epoch in the brane evolution the causal
structure of the universe could be strongly modified. This could be a
sign of an unusual evolution of the perturbations  from the
time they cross the de Sitter horizon, $H^{-1}$, during inflation,
through the time they became causally connected again. In this case,
there could be  distinct
predictions for the  microwave background radiation structure even with
the same consistency equation  during inflation. 
Thus, further investigation on the dynamics of perturbations in
inflationary brane-world models may prove
useful to probe the dimensionality of space-time.

\bigskip
{\bf Acknowledgements:} This work has been supported by Funda\c c\~ao
de Amparo \`a Pesquisa do Estado de S\~ao Paulo {\bf (FAPESP)} and Conselho
Nacional de Desenvolvimento Cient\'\i fico e Tecnol\'ogico {\bf
(CNPq)}, Brazil. 


\end{document}